\documentclass[onecolumn,amsmath,amssymb,nofootinbib,12pt]{article}
\pdfoutput=1 

\usepackage{jheppub} 
\usepackage[permil]{overpic}
\usepackage{bbold} 

\usepackage{amstext}

\newcommand{\arXiv}[1]{\href{http://www.arXiv.org/abs/#1}{#1}}

\def\ie{{\it i.e.},\ }
\def\eg{{\it e.g.},\ }

\def\a{\alpha}
\def\b{\beta}
\def\d{\delta}

\def\eps{\epsilon}
\def\l{\lambda}
\def\g{\gamma}
\newcommand\gm{\gamma_\mt{max}}

\def\s{\sigma}
\def\t{\tau}
\def\vp{\varphi}
\def\D{\Delta}
\def\G{\Gamma}

\def\O{{\cal O}}

\def\]{\right]}
\def\[{\left[}

\def\({\left (}
\def\){\right )}

\newcommand{\del}{\partial}

\newcommand{\be}{\begin{equation}}
\newcommand{\ee}{\end{equation}}
\newcommand{\bea}{\begin{eqnarray}}
\newcommand{\eea}{\end{eqnarray}}
\newcommand{\ba}{\begin{eqnarray}}
\newcommand{\ea}{\end{eqnarray}}
\newcommand{\beal}{\begin{align}}
\newcommand{\eeal}{\end{align}}
\newcommand{\beq}{\begin{equation}}
\newcommand{\eeq}{\end{equation}}
\newcommand{\beqa}{\begin{eqnarray}}
\newcommand{\eeqa}{\end{eqnarray}}
\newcommand{\beqar}{\begin{eqnarray*}}
\newcommand{\eeqar}{\end{eqnarray*}}

\newcommand{\reef}[1]{(\ref{#1})}

\newcommand{\mt}[1]{\textrm{\tiny #1}}
\def\<{\langle}
\def\>{\rangle}
\newcommand{\tr}{\mathop{\mathsf{tr}}\nolimits}
\newcommand{\rank}{\operatorname{rank}}
\def\gibbs{\rho_\beta}

\def\dg{\delta g}
\def\gf{{\mathfrak g}}
\def\dgf{\delta{\mathfrak g}}

\def\ins{{\rm in}}
\def\out{{\rm out}}
\def\shell{{\rm shell}}

\newcommand\work{W_{\alpha,\beta}}
\newcommand\Work{W_{\alpha,{\rm R}}}
\newcommand{\br}{\beta_\mt{R}}
\newcommand{\rhoR}{\rho_\mt{R}}
\newcommand{\odd}{{\mathcal O}_\D }

\preprint{NSF-ITP-18-010}

\title{\boldmath Holographic second laws of black hole thermodynamics}

\author[a]{Alice Bernamonti,}
\author[a]{Federico Galli,}
\author[a]{Robert C. Myers}
\author[b]{and Jonathan Oppenheim}

\affiliation[a]{Perimeter Institute for Theoretical Physics, Waterloo, ON N2L 2Y5, Canada}
\affiliation[b]{University College of London, Department of Physics \& Astronomy, London, WC1E 6BT}
 
\emailAdd{abernamonti,\,fgalli,\,rmyers\,@perimeterinstitute.ca; j.oppenheim@ucl.ac.uk}

\begin{document} 
\abstract{
Recently, it has been shown that for out-of-equilibrium systems, there are additional constraints on thermodynamical evolution besides the ordinary second law.  These form a new family of second laws of thermodynamics, which are equivalent to the monotonicity of quantum R{\'e}nyi divergences. In black hole thermodynamics, the usual second law is manifest as the area increase theorem. Hence one may ask if these additional laws imply new restrictions for gravitational dynamics, such as for out-of-equilibrium black holes? Inspired by this question, we study these constraints within the AdS/CFT correspondence. First, we show that the R{\'e}nyi divergence can be computed via a Euclidean path integral for a certain class of excited CFT states. Applying this construction to the boundary CFT,  the R\'enyi divergence is evaluated as the renormalized action for a particular bulk solution of a minimally coupled gravity-scalar system. Further, within this framework, we show that there exist transitions which are allowed by the traditional second law, but forbidden by the additional thermodynamical constraints. We speculate on the implications of our findings. 
}

\maketitle
\flushbottom
 
\section{Introduction}

The conventional second law of thermodynamics tells us that the entropy of a closed system increases. If the system can be in one of many possible microstates, we can describe its state in terms of a density matrix $\rho$. We can then think of the second law as placing a restriction on which density matrices $\rho(t)$ are thermodynamically accessible from some initial $\rho(0)$. 
The second law is a necessary condition which any state transformation must satisfy, regardless of the underlying physical laws. Because of this, the second law has broad applicability, allowing us to understand macroscopic properties of common materials, and finding application in cosmology, accelerator physics, astrophysical systems, and fields as diverse as computer science and gravity. 

In the latter case, Bekenstein and Hawking famously showed that the event horizon of a black hole carries entropy given by $S_{BH}=A/4G_N$, where $A$ is the surface area of the black hole's horizon and $G_N$ is Newton's constant \cite{JB1,Bekenstein:1973ur,Hawking:1974rv,Hawking:1974sw}.
The second law of thermodynamics then demands that the area of the event horizon must always increase, as can be geometrically proven for any classical processes \cite{area,Hawking:1973uf}. This deep connection between thermodynamics and black hole physics provides one of the most important clues we have to reconciling quantum theory with gravity.  Hence a better understanding of the second law of thermodynamics not only sheds light on emergent phenomena in many areas of physics, but it may also provide insight into the fundamental theory underlying the unification of quantum mechanics and general relativity. 

Recently it was found that in addition to the standard second law, there are additional constraints on how a thermodynamical system can evolve~\cite{ruch1978mixing,janzing2000thermodynamic,uniqueinfo,dahlsten2011inadequacy, del2011thermodynamic,horodecki2013fundamental,aaberg2013truly,brandao2013second,faist2015minimal,
egloff2015measure,cwiklinski2015limitations,lostaglio2015description,lostaglio2014quantum}. These are akin to having a family of second laws of thermodynamics \cite{brandao2013second} which apply to 
out-of-equilibrium systems. One might therefore wonder whether these additional second laws also place constraints on how black holes can evolve.   It seems plausible that general black hole dynamics obeys a new family of second-law-like constraints. Further, these new constraints may then supply us with additional clues as to what form a consistent theory of quantum gravity should take.

As we will  discuss in section \ref{intro:thermo}, the additional second laws are related to the distance between the state of the system and a thermal state, as measured by the quantum R\'{e}nyi divergences \cite{petz1986quasi,HiaiMPB2010-f-divergences,Muller-LennertDSFT2013-Renyi,WildeWY2013-strong-converse}. The latter are an important distance measure in information theory, and here we show how they can be computed for some class of excited states in a conformal field theory (CFT). Specifically, in section~\ref{PIapproach}, we will show how for this class of states, the R\'{e}nyi divergences can be expressed as a particular partition function obtained from a Euclidean path integral. Because these divergences place additional second-law-like constraints on the evolution, in principle, these new techniques provide us with  information about extra restrictions that quantum thermodynamics places on state transitions in the CFT. 

However, this path integral approach also allows us to evaluate these quantities in the setting of the AdS/CFT correspondence \cite{Aharony:1999ti,Kiritsis,Ammon} and to explore the implications  of quantum thermodynamics in a holographic setting.\footnote{Recently, it was argued \cite{Engelhardt:2017aux} that in a holographic framework, the second law of thermodynamics in the boundary theory should be associated with the area increase of the so-called trapping or dynamical horizon \cite{Hayward:1993wb,Ashtekar:2002ag,Ashtekar:2003hk,Gourgoulhon:2005ng, Bousso:2015mqa,Bousso:2015qqa,Sanches:2016pga}.} As noted above, the new second laws will constrain equilibration processes in the boundary CFT. In the holographic context, we can ask what these additional second laws correspond to in the bulk gravitational description. In particular, do they constrain how out-of-equilibrium black holes can evolve in the bulk. In fact, we will be able to demonstrate that there are transitions, both in the boundary CFT and in the dual gravitational system, that are possible if only the ordinary second law holds, but are ruled out by the additional constraints. In contrast to the second law of black hole thermodynamics which applies broadly to black holes in any setting and is derived via thermodynamics, our present discussion relies on the holographic framework and appeals to the microscopic quantum derivations of the new second laws holding in the boundary field theory.  
Nonetheless, we expect that the simple constraints that we find for the evolution of our holographic black holes should have broader applicability, and point towards the existence of a new class of second-law-like  constraints on how black holes can evolve quite generally. 
We summarize the main findings of our holographic approach and the content of the rest of the paper in section~\ref{ss:summary}.

\subsection{New constraints from R\'{e}nyi divergences}
\label{intro:thermo}

The traditional second law has a number of different formulations and
interpretations, from Carnot and Clausius, to Boltzmann and Gibbs, and to
more modern versions such as the Jarzynski equality~\cite{jarzynski1997nonequilibrium,crooks1999entropy}, and the eigenstate thermalization hypothesis \cite{srednicki1994chaos,deutsch1991quantum}. 
The traditional second law places a single constraint on the evolution of a system, for example that the entropy of a closed system can only increase. For out-of-equilibrium systems,  there are many ways to increase their entropy, however, it turns out that not all of them are allowed. The additional restrictions on how entropy can increase constitute additional second laws, because like the conventional second law they place constraints on what a state may evolve into. We shall in particular, focus on the constraints introduced in~\cite{brandao2013second} which are related to the so-called {\it thermo-majorization criteria}~\cite{ruch1978mixing,uniqueinfo,horodecki2013fundamental}.
Their mathematical structure is similar to the traditional second law of thermodynamics, and one that is found in many {\it quantum resource theories}~\cite{horodecki_are_2002,horodecki2013quantumness,brandao2015reversible}.

For an open system in contact with a heat bath at temperature $T$, the second law is equivalent to the statement that the free energy
\begin{align}
F(\rho) = \<E(\rho)\> -TS(\rho),
\label{eq:helmfreeenergy}
\end{align}
must decrease in any cyclic process, with $S(\rho)$ the entropy of the system and $\<E(\rho)\>=\tr (\rho H)$  its average energy. If we consider the equilibration of a closed system so that the energy is conserved,\footnote{Or we might consider the simple case where the Hamiltonian $H$ is trivial, \ie $H=0$.} the decrease in the free energy is equivalent to an increase in the entropy.
This version of the second law holds not only for the thermodynamical entropy, but also the statistical mechanical entropy $S_B=\log N$, where $N$ is the number of microstates, as well as the von Neumann entropy $S(\rho)=-\tr\rho\log\rho$ (see for example the discussion in \cite{Neumann}), which generalises the statistical mechanical entropy to the case where the system is quantum, and where the probability of being in any microstate is not necessarily equal. Here, we wish to consider quantum systems out-of-equilibrium, where typically the probabilities of being in a microstate are not uniform, hence the von Neumann entropy is the appropriate one to consider, and our operations can include coarse graining or tracing out information about the system.

When such a system relaxes to its equilibrium state, there are many
intermediate states, or {\it paths} it might take. The traditional
second law does not rule any of them out, as long as the free energy
decreases. However, there are additional restrictions which apply,
and  have a similar form. Both the usual second law 
and these restrictions can be thought of in terms of the 
distance of the initial state of the system to its equilibrium state, and this distance can only decrease. We can re-write the traditional second law in terms of the relative entropy distance to the thermal state, where the relative entropy is defined as
\begin{align}
D_1(\rho\Vert\sigma) \equiv \tr \rho\log\rho-\tr\rho\log\sigma \,.
\label{relX} 
\end{align}
Then, noting that 
\begin{align}
D_1(\rho\Vert\rho_\beta)=\b \(F(\rho) - F(\rho_\beta)\)
\label{eq:tradsecond}
\end{align}
with $\rho_\beta={e^{-\beta H}}/{Z_\beta}$ the thermal state, we see that in case the Hamiltonian does not change (which is the case we consider in this paper), 
the traditional second law for out-of-equilibrium systems is equivalent to the statement that the relative entropy distance to the thermal state has to
decrease. 

This is actually true for any distance measure. Namely, a distance $D$, which provides a measure of how distinguishable two states are, should have the property that it decreases
under the action of some arbitrary dynamics $\rho(t)=\Lambda_t(\rho)$, so
that 
\begin{align}\label{contract}
D\big(\rho(t)\Vert\s(t)\big)\leq D\big(\rho(0)\Vert\sigma(0)\big)\,.
\end{align}
We say the measure is {\it contractive under completely positive  trace preserving (CPTP) maps}. This is because an experimenter who is trying to distinguish whether a system is in one of two states could apply the map $\Lambda_t$ to the system, thus a measure of distinguishability should only decrease under her actions. Next, we use the fact that by definition, equilibrium states satisfy the property that for almost all times\footnote{A more precise statement appears in \cite{MuellerOppenheim}, where we also consider the role of approximation, \ie $\Lambda_t(\rho_\beta)\approx\rho_\beta$. We also discuss
the role of approximation further in Section \ref{sec:closed}. } $\Lambda_t(\rho_\beta)=\rho_\beta$ and we thus have
\begin{align}
D\big(\rho(t)\Vert\rho_\beta\big)\leq D\big(\rho(0)\Vert\rho_\beta\big)\,.
\label{eq:monotonicity}
\end{align}

Now the standard discussion \cite{brandao2013second} applies the above inequality with $\rho_\beta$ being the equilibrium state into which $\rho(t)$ will evolve (at large $t$). 
Note that eq.~\eqref{eq:monotonicity} holds even if the map is non-linear, provided that eq.~\eqref{contract} still holds. 
Furthermore, an interesting extension \cite{MuellerOppenheim} comes from realizing that, in fact, eq.~\reef{eq:monotonicity} holds for any equilibrium state $\rhoR$ of the system, \ie for any states which remain invariant under the evolution of interest. For example, for closed system dynamics where $\rho$ represents the full degrees of freedom of the system, all thermal states will be preserved and so we may apply eq.~\reef{eq:monotonicity} with $\rho_\beta$ replaced by $\rhoR$, a thermal state with an arbitrary temperature $1/\br$. Therefore any contractive distance measure to any equilibrium state gives a restriction on what is possible for the evolution in such thermodynamical systems.  In contrast, for closed system dynamics where $\rho$ represents a coarse-grained description of the system, or for an arbitrary pure state in the support of $\rho$, one typically has that only $\rho_\beta$ is preserved. Furthermore, the monotonicity property \reef{eq:monotonicity} holds even though the dynamics is still defined by the Hamiltonian evolution of the microscopic degrees of freedom (rather than an effective coarse-grained dynamics). Moreover, we note that there are many different versions of the second law, some of which are contingent on the particular dynamics, or which only hold for most times or only on average.  Here, we are able to use any version which holds that thermal states are preserved by the dynamics under appropriate conditions.\footnote{These conditions might include time averaging, or averaging over the initial micro-state, or including the caveat ``for almost all times''.} Hence, with an appropriate choice of distance measure, one finds an entire family of constraints indexed by $\br$, the inverse temperature of the reference state. To the best of our knowledge, this new family of thermodynamical constraints has not been studied previously, and we return to this idea in sections \ref{sec:Xeno} and \ref{sec:discussion}, as well as with a more detailed examination in \cite{MuellerOppenheim}.

At this point, we have not made precise the distance measure in eq.~\reef{eq:monotonicity}. One might consider a number of distance measures which are contractive, 
and hence provide thermodynamical constraints. An important example would be the quantum R{\'e}nyi divergences of  \cite{petz1986quasi,HiaiMPB2010-f-divergences,Muller-LennertDSFT2013-Renyi,WildeWY2013-strong-converse,JaksicOPP2012-entropy} (see also \cite{frank2013monotonicity,beigi2013sandwiched}).
We shall study in particular those of \cite{petz1986quasi}
\begin{align} \label{eq:RD}
D_\alpha(\rho\Vert\gibbs)\equiv\frac{{\rm sgn}(\alpha)}{\alpha-1} \log\tr \(  \rho^\alpha \gibbs^{1-\alpha}\) \,,
\end{align}
where ${\rm sgn}(\alpha)$ is defined by
\begin{align}
{\rm sgn}(\alpha)\equiv \left \{\begin{array}{cc}
1 &\quad {\rm for}\ \ \alpha\geq 0\,, \\
-1 & \quad {\rm for}\ \ \alpha < 0\,.
\end{array}\right. 
\label{hfunction}
\end{align}
The relative entropy \reef{relX} is then defined using eq.~\eqref{eq:RD} via the limit: $\lim_{\a\to 1}D_\a(\rho\Vert\gibbs)$. 
When $[\rho,\gibbs]=0$, the R{\'e}nyi divergences with $\alpha\geq 0$ give necessary and sufficient conditions for transitions to be possible \cite{brandao2013second}. In the case where $[\rho,\gibbs]\neq 0$, the R{\'e}nyi divergences 
with $0\leq \alpha\leq 2$ provide necessary conditions. Also, there are other quantum versions of the R{\'e}nyi divergence which
are equivalent to those in eq.~\eqref{eq:RD} in the commuting case, and some of them, as the sandwiched R\'enyi divergence \cite{Muller-LennertDSFT2013-Renyi,WildeWY2013-strong-converse}
\be
\tilde D_\alpha(\rho\Vert\gibbs)\equiv \frac{1}{\alpha-1} \log \tr \( \rho_\b^{\frac{1- \a}{2\a}} \rho\, \rho_\b^{\frac{1- \a}{2\a}} \)^\a
\label{eq:sandwhiched-RD}
\ee
 have properties~\cite{frank2013monotonicity,beigi2013sandwiched} that also allow them to provide constraints on thermodynamical transitions. 
 We will not consider them further here, except to note that calculating them is an interesting open question that could be conceivably addressed by extending our path integral approach. 
 
In the thermodynamic limit when correlations and interactions are not long range, all the $D_\alpha(\rho\Vert\gibbs)\approx D_1(\rho\Vert\gibbs)$ and thus these additional constraints are all just equivalent to the traditional second law \cite{horodecki2013fundamental,brandao2013second}. 
However, these additional second laws may still play a role for a single out-of-equilibrium system when there are long-range correlations. This is the case that we will consider here, where we perturb the thermal state of a 2d CFT with a single, correlated deformation. 

To give a sense of what these additional constraints correspond to, let us consider a simpler situation, which more closely mirrors our intuition about entropy.  In particular, we will consider the simple situation with trivial Hamiltonian where, as we explain below, the new constraints are expressed in terms of R\'enyi entropies.
First, let us recall some properties of R\'enyi entropies: Consider the eigenvalues $p_i$ of a density matrix $\rho$ corresponding to microstate $i$ and the R{\'e}nyi entropies defined for $\alpha\in {\mathbb R} \setminus\{0,1\}$  
\be
S_\alpha(\rho) \equiv \frac{{\rm sgn}(\alpha)}{1-\alpha} \log \sum_{i=1}^n p_i^\alpha\,.
\label{eq:renyi-entropy}
\ee
For $\alpha \in \{-\infty,0,1,\infty\}$, we define $S_\alpha  (\rho)$ by taking limits of the above expression, \ie  
\bea
&& S_0(\rho)=\log \rank(\rho )\,,\quad S_1(\rho)=-\sum_{i=1}^n p_i \log  p_i\,,\nonumber\\
 && S_{\infty}(\rho)= -\log p_{\max}\,,\quad
S_{-\infty}(\rho)= \log p_{\min}\,,
\eea 
where $\rank(\rho)$ is the number of nonzero elements of $\rho$, and  $p_{\max}$ and $p_{\min}$ are the maximal and minimal 
elements of $\rho$, respectively.  Of course, $S_1$ corresponds to the usual von Neumann entropy $S$.  

Now as we suggested above, let us consider the simple situation where the Hamiltonian is trivial, \ie $H=0$. 
Then we have $\rho_\beta=\mathbb{1}/d$,  where $d$ is the dimension of the Hilbert space, and thus for positive $\alpha$: $D_\alpha(\rho\Vert\rho_\beta)=\log{d}-S_\alpha(\rho)$. Further, if the dimension $d$ does not change, the decreasing of the R\'{e}nyi divergence corresponds to increasing the R\'{e}nyi entropy and we can think of these  additional second laws as just stating that all these entropies must increase. 
For systems in equilibrium for which all microstates are equiprobable, all the  R{\'e}nyi entropies are approximately equal, and in particular,
equal to the ordinary von Neumann entropy. Thus, these additional second laws tell us nothing new for equilibrium systems. However, for out-of-equilibrium systems,
where the probabilities for being in a particular microstate can be different, these additional second laws place additional constraints on how a system can evolve. 
For example, it is conceivable for a system to increase its Shannon entropy while, at the same time, increasing
its largest eigenvalue (\ie decreasing $S_{\infty}(p)$), or decreasing its rank (\ie decreasing $S_0(p)$). 
However, these two last possibilities are expressly forbidden by these additional second laws. 

\subsection{Summary}\label{ss:summary}

In section~\ref{PIapproach}, we show that the R\'enyi divergence can be obtained in terms of a Euclidean path integral for a specific class of excited CFT  states.  In particular, our discussion there focuses on the simple example where the excited state is prepared by turning on a relevant deformation on the thermal cylinder. However, we expect that our path integral approach should extend to a much broader family of excited states, as we discuss in section \ref{sec:discussion}. With this example, the trace function   $\tr\!\( \rho^\alpha \gibbs^{1-\alpha}\)$  that computes the  R\'enyi divergences  for $\a \in [0,1]$  can be obtained as  the CFT partition function $Z_{\rm CFT}$  with the deformation  turned on along a portion $\a \beta$  of the thermal circle. 

The remainder of section \ref{sec:quench} is devoted to applying the above path integral construction in the context of the AdS/CFT correspondence,\footnote{For a review of the AdS/CFT correspondence see for instance \cite{Aharony:1999ti} and the textbooks \cite{Kiritsis,Ammon}.} and explicitly evaluating the R\'enyi divergence  with a holographic computation. In the holographic bulk dual, our excited state corresponds to a Euclidean black brane geometry in presence of a massive scalar field with non-trivial Dirichlet boundary conditions at the AdS boundary. Following the standard holographic dictionary, the trace function is given by the bulk partition function
\be
Z_{\rm bulk} \approx e^{-S_{\rm ren}}
\ee
evaluated in terms of the renormalized Euclidean on-shell  bulk action. 

We perform this computation perturbatively in the amplitude   of the scalar field (or equivalently in the coupling of the CFT deformation), around the thermal black brane background.
At leading non-trivial order, we have
\be
S_{\rm ren}= - \frac{1}{16 \pi G_N} \int d^2 x \,\left[ M + \frac{\D-1}{4}\,  \vp_{(0)} \vp_{(\D)} \right]  +O(\l^3) \,.
\ee
where $M =(2\pi/\b)^2$ is energy or mass density of the AdS black brane.  $\vp_{(0)}$ and  $\vp_{(\D)}$ denote respectively the non-normalizable  and normalizable mode of the bulk scalar field, and  are holographically related to the source $\l$ and the expectation value $\langle \odd \rangle$ of the operator deforming the CFT thermal state.  An analogous computation can be performed directly in the dual two-dimensional CFT in conformal perturbation theory.  

Our Euclidean path integral construction leads us to identify 
\begin{align}
\log \tr\!  \(\rho^{\alpha} \rho_\beta^{1-\alpha}\) \approx - S_{\rm ren}\,.
\end{align}
In section~\ref{sec:RenyiDiv}, with the above expression in hand, we explicitly evaluate   the  holographic R\'enyi divergences for $\a \in [0,1]$
\be
D_\a (\rho  \|  \rho_\b) = \frac{1}{\alpha-1}\log \frac{\tr\(\rho^{\alpha} \rho_\beta^{1-\alpha}\) }{\( \tr \rho\)^\a \(\tr \rho_\b\)^{1-\a}}\,.
\ee
Here the two traces in the denominator are included to ensure the proper normalization, since our path integral approach yields
$ \tr \rho\ne 1\ne \tr \rho_\b$. 

The R\'enyi divergences depend parametrically on the index $\a$, as well as on the precise states we consider, through the operator conformal dimension $\D$ and amplitude of the source $\l$. 
Our construction of  $D_\a$ satisfies the expected properties of positivity, monotonicity and continuity in $\a$, as well as concavity of $(1-\a)D_\a$ \cite{Erven}. However, for the class of states which we construct, $D_\a$ has various UV divergences in general whose precise structure is parametrized by the conformal dimension $\D$. In the end, we focus much of our discussion on states in the range $0<\D<1$ for which no such divergences appear. 

\begin{figure}[ht]
\centering 
\includegraphics[width=.6\textwidth]{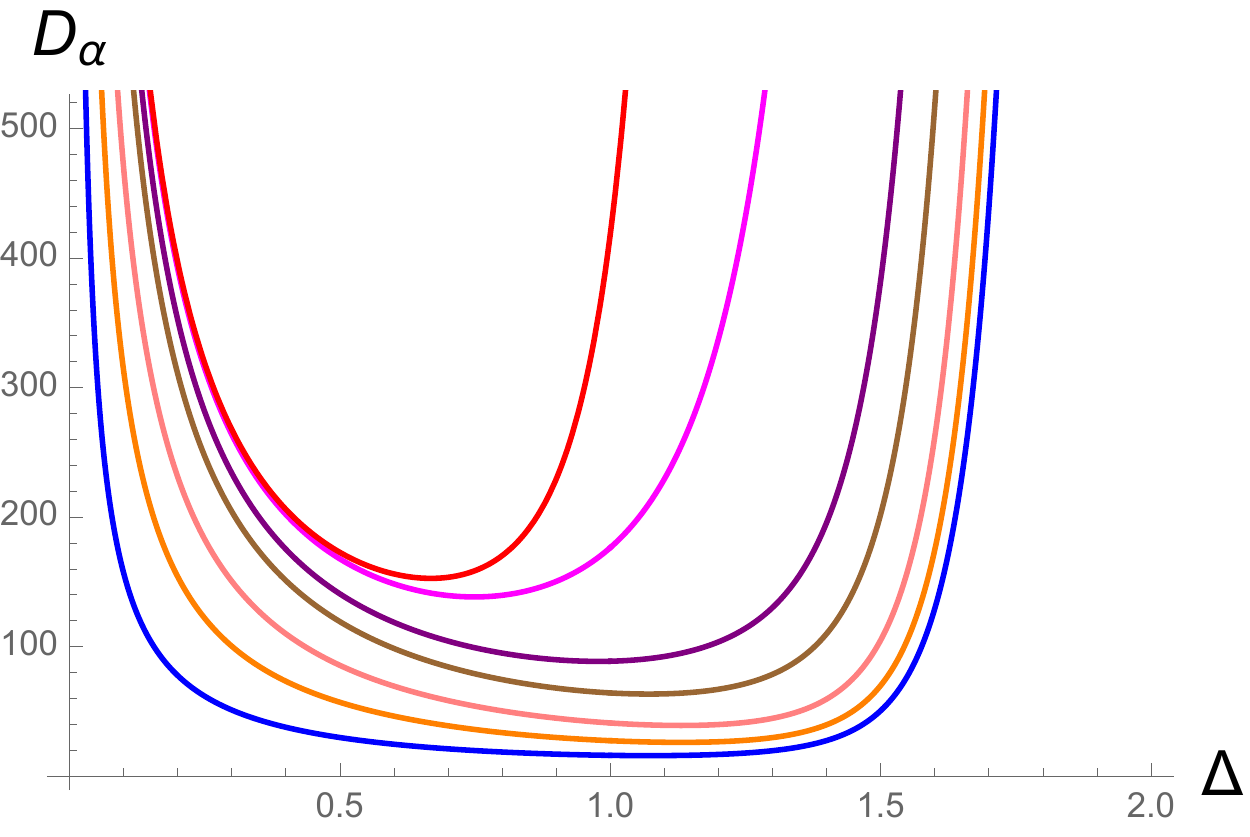} 
\caption{Sample plot of the R\'enyi divergences. The different curves correspond to $\a =0.2,0.4,0.6,0.8,0.9,0.99,1$ from the bottom up. } 
\label{fig:Da_comp}
\end{figure}

We find that depending on the specific excited states we are comparing, the monotonicity constraints 
\be
D_\a  (\rho  \|  \rho_\b) \ge D_\a  (\rho' \|  \rho_\b) 
\ee
for a transition from $\rho$ to $\rho'$ are or not all equivalent. 
As an example, we plot in figure~\ref{fig:Da_comp} one such direction of the R\'enyi divergence parameter space.  Here we indeed  see that curves of different $\a$ have different minima, meaning the additional second laws do   forbid some of the transitions that would be classically allowed.  We develop this point in detail in the discussion section.

In section \ref{sec:closed}, we examine the implications of these additional thermodynamical constraints for closed-system dynamics.
In particular, we consider the extension to applying eq.~\reef{eq:monotonicity} with arbitrary reference states to the present holographic calculations. In section~\ref{sec:discussion} we present some detailed calculations applying the new constraints to our holographic model and in particular, we show that there are transitions which are classically allowed, but which are ruled out by the additional second laws of quantum thermodynamics. Further, by scanning different reference states, we are able to recognize that the excited states do not actually thermalize by unitary time evolution alone. In this closing section, we also give a broader perspective of the implications of our results and the outlook of the new second laws in the context of holography and more general gravitational systems.

Finally, appendix~\ref{app:integral} describes some technical details of the holographic computation, while appendix~\ref{app:EVaidya}  presents a different holographic Euclidean construction for a trace function of the type
\be \label{eq:trVaidya}
\tr \(\rho_{\rm out}^{\a_{\rm out}} \rho_{\rm in}^{1-\a_{\rm in}}\) \,,
\ee
where now $\a_{\rm in} +\a_{\rm out}  \neq 1$. This is a new Euclidean shell  solution, which we obtain, in a Vaidya-like fashion, by gluing together two portions of Euclidean black brane spaces of different masses along a particular family of geodesics connecting boundary endpoints. This quantity does not generically satisfy the data processing inequality, but it does in the situation we consider in the geometric construction of app.~\ref{app:EVaidya}. There in fact $\rho_{\rm in}$ is a thermal density matrix and \eqref{eq:trVaidya} can be recast in the form of a R\'enyi divergence with general reference state, as those studied in sec.~\ref{sec:closed}. This trace function also satisfies Lieb's concavity theorem \cite{Lieb} for the range of the parameters $\a_{\rm in}, \a_{\rm out}$ for which we are able to define it.


\section{Euclidean quench in amplitude expansion} \label{sec:quench}

In this section, we set up the computation of the R\'enyi divergence \eqref{eq:RD} for a class of excited states in any conformal field theory. In particular, we focus on excited states which are prepared by a Euclidean path integral where a relevant deformation is turned on. By considering holographic CFTs and applying the usual AdS/CFT dictionary \cite{Aharony:1999ti,Kiritsis,Ammon}, these states are related to gravitational backgrounds where a black hole is surrounded by  scalar field excitations. The free evolution (where the source of the relevant operator is removed) of these states will then simply involve the collapse of the scalar hair into the black hole and eventually, the gravitational system will settle down to a ``hairless" black hole with a slightly higher mass (and temperature). Hence by exploiting the holographic framework, we go on to evaluate the R\'enyi divergences and examine the constraints which the second laws of quantum thermodynamics \eqref{eq:monotonicity} may impose on the evolution of these black holes. 

\subsection{R\'enyi divergences from path integrals } \label{PIapproach}
 
Before exploring the R\'enyi divergences in the holographic context, we  consider a particular Euclidean path integral  construction that  can be identified as computing $\tr \(\rho^\alpha \gibbs^{1-\alpha}\)$ for excited states in  CFT.
While the present discussion focuses on a special class of excited states, we expect that the path integral approach described below can be extended to a much broader family of excited states. For further discussion of this matter, see section \ref{sec:discussion}.

First, the reference thermal state has density matrix, which can be identified with a Euclidean path integral (with appropriate boundary conditions) on a slab of width $\beta$ in the Euclidean time direction, \ie
\be
\rho_\b = e^{-\b H} =\raisebox{-0.5\height}{
\begin{overpic}[scale=0.8]{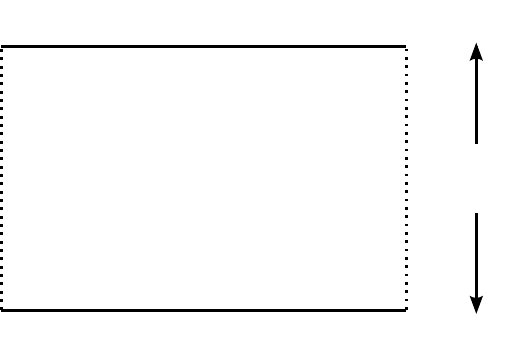}
\put (910,340) {$\b$}
\end{overpic}
} \,. \label{mini}
\ee
Now we extend this well-known construction to a particular class of excited states prepared via an analogous Euclidean path integral in which a relevant deformation is turned on, \ie
\be
\rho= \int {\mathcal D}\phi \ e^{- S_{\mt{CFT}} [\phi] - \int d^dx\, \l \, \odd(x)}= 
\raisebox{-0.5\height}{
\begin{overpic}[scale=0.8]{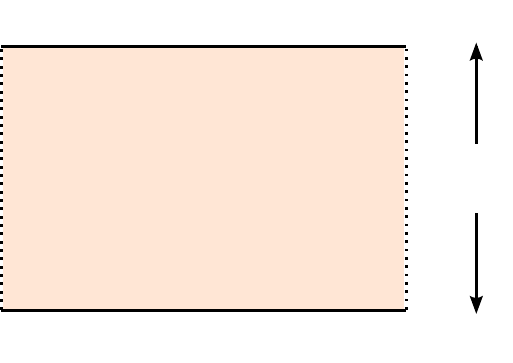}
\put (910,340) {$\b$}
\end{overpic} \label{jumbo}
}  \,,
\ee
where the colored shading represents the presence of the deformation. 
We may alter the state by varying the amplitude of the (constant) 
source $\l$ or by choosing different relevant operators $\odd(x)$. 
We could think of the excited state \reef{jumbo} as the thermal state defined with respect to a new Hamiltonian consisting of the original CFT Hamiltonian deformed by the relevant operator, \ie $H'=H +\lambda\,\odd$. However, we wish to emphasize that we are still thinking of this state as an excited state within the same theory, \ie within the CFT governed by the Hamiltonian $H$.

For this particular choice of excited states, the trace appearing in the R\'enyi divergence can then be computed by sewing together the two path integrals represented in eqs.~\reef{mini} and \reef{jumbo},
\be
Z_{\rm CFT} = \tr \(\rho^\alpha \gibbs^{1-\alpha}\) =
\raisebox{-0.5\height}{
\begin{overpic}[scale=0.8]{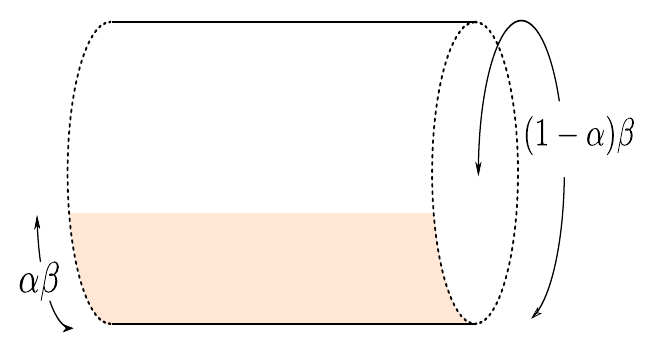}
\end{overpic}
} \,.
\label{jumbo2}
\ee
That is, the desired trace in eq.~\reef{eq:RD} is evaluated as the partition function in the CFT on a thermal cylinder of circumference $\beta$ with the relevant deformation turned on for fraction of the full span, namely, a period of Euclidean time $\a\b$. This path integral construction thus allows to compute $D_\a$ with index $\a \in [0,1]$.  

With some abuse of language we refer to this construction as a {\it Euclidean quantum quench} since we are disturbing the system with a time dependent source, \ie
\be \label{eq:profile}
\l(\tau) = \left[ \theta(\tau) - \theta( \tau- \a \b)\right]   \l \,.
\ee
Hence the relevant deformation is abruptly turned on at $\tau=0$ and just as abruptly turned off again at $\tau=\a\b$. Hence, in some respects,
our construction resembles an instantaneous quench of \eg \cite{card1,card2,card3}, where the initial excited state is prepared by evolving the system with one (time-independent) Hamiltonian, but the latter is instantaneously swapped for another (time-independent) Hamiltonian which then controls the future time evolution.\footnote{Foreshadowing certain technical details of the calculations in section \ref{holorenormX}, we warn the reader that such instantaneous quenches can lead to UV divergences if the conformal dimension of the relevant operator is not sufficiently small, \eg \cite{abrupt1,abrupt2,abrupt3}.}\\ 

Now we turn to the computation of this partition function \reef{jumbo2} in holography, where the thermal state is given by a black brane bulk geometry and the relevant operator ${\mathcal O}$  corresponds to a bulk scalar field $\Phi$ sourced on the boundary by $\l(\tau)$. To obtain analytical or semi-analytical results we restrict to AdS$_3$/CFT$_2$ and work at first non-trivial order in  a perturbative expansion in the amplitude $\l$.  The next two subsections contain the technical details of the holographic computation. For the final result, the reader can move directly to section~\ref{sec:Eaction}, where we also comment on how this bulk computation is directly equivalent to performing the conformal perturbation theory expansion of $Z_{\rm CFT}$ on the thermal cylinder. 

\subsection{Bulk setup} \label{sec:setup}

Following the discussion above, we consider Einstein gravity in 2+1 dimensions  minimally coupled to a massive scalar field, with Euclidean bulk action  
\bea 
S_{E} = -\frac{1}{16 \pi G_N} \int_{\mathcal{M}} d^3x\,\sqrt{g} \[R - 2\Lambda - \frac{1}{2} g^{\mu\nu}\partial_{\mu}\Phi\partial_{\nu}\Phi - \frac{1}{2}m^{2}\Phi^{2} \] \qquad\qquad \label{eq:action}\\  -\frac{1}{8 \pi G_N} \int_{\del \mathcal{M}}  d^2x \,\sqrt{\gamma}\, K \,,
\nonumber 
\eea
where the cosmological constant $\Lambda = -1$ and as a result, the radius of curvature in AdS geometry is also set to one. The boundary term is the usual Gibbons-Hawking-York term, with  the extrinsic curvature defined as $K_{\mu\nu} = \nabla_{(\mu}\hat n_{\nu)}$ and $\hat n$ being the outward-directed normal vector to the boundary $\del \mathcal{M}$.
The bulk scalar field is dual to an operator of conformal dimension $\Delta$, and its mass squared is $ m^2 = \Delta(\Delta  - 2 )$. We are interested in relevant deformations with $0<\Delta<2$, corresponding to negative values of $m^2$ all the way down to the Breitenlohner-Freedman bound ($m^2 \ge -1$) \cite{BF1,BF2}.  
The equations of motion following from the above action are 
\begin{align}
R_{\mu\nu} - \frac{1}{2}g_{\mu\nu}\( R - 2\Lambda -  \frac{1}{2} (\partial \Phi )^2   - \frac{1}{2} m^{2} \Phi^{2}  \)  +\frac{1}{2} \del_{\mu}\Phi \del_{\nu}\Phi &=0  \,, \label{Ein} \\
\left(-\Box + m^{2}\right)\Phi
&=0 \, . 
\end{align}

Now the reference state of our path integral construction is a thermal state of the CFT on the infinite line at inverse temperature $\b$. The corresponding background solution of the bulk gravity theory is therefore the Euclidean planar black hole geometry\footnote{For example, see \cite{Witten:1998zw,Maldacena:2001kr}.}
\be\label{eq:BHmetric}
ds^2  =  \frac{1}{z^{2}}\left[ (1-M z^{2})d\tau^{2}+\frac{dz^{2}}{1-M z^{2}}+dx^{2}\right]
\ee 
and a vanishing scalar field, \ie $\Phi=0$. 
Here $z \in (0,1/\sqrt M]$,  $\sqrt M  = 2\pi/\b$ and the AdS boundary is at $z \to 0$. 

To compute the R\'enyi divergence \eqref{eq:RD}, we consider the backreaction induced by a spatially homogeneous boundary source $\lambda(\t)$ for the operator dual to the scalar field $\Phi$. Turning on this source amounts to imposing the Dirichlet boundary condition 
\be
\lim_{z\to 0}  z^{\Delta-2} \Phi (z,\t) =  \lambda(\t) \,. \label{eq:scalb}
\ee
To make progress analytically we consider a small amplitude expansion on top of the 3d bulk Euclidean background. 
To leading order in the amplitude $\lambda$ of the deformation there is no backreaction of the scalar on the geometry and we simply solve for a scalar field in the black hole background \eqref{eq:BHmetric}.  The solution satisfying the boundary condition \eqref{eq:scalb} can be written in terms of the Euclidean  bulk-to-boundary propagator
\be
K(z,x -x',\t - \t')=C_{\Delta} \frac{z^\D} { \left[ \frac{2}{M}\left(\cosh(\sqrt M(x-x'))-\sqrt{1-M z^{2}}\cos(\sqrt M(\t-\t'))\right)\right]^\D}\, ,
\ee
with normalization
\be \label{eq:CDelta}
C_{\Delta}=\frac{\D-1}{\pi }\, .
 \ee
This is such that  %
\be
\left(-\Box+m^{2}\right)K(z,x,\tau)=0 \qquad \quad \text{and } \qquad  \lim_{z\rightarrow0}z^{\Delta-2}K(z,x,\t)=\delta(x)\delta(\t)\,,
\ee
so that the solution we are looking for is expressed as 
\be \label{eq:sourced}
\Phi(z, \t) = \int dx'  d\t' ~K(z, x - x',\t -\t') \l(\t') \,.
\ee
At second order in $\l$, the geometry also backreacts, as Einstein's equations are sourced by a non-vanishing energy-momentum tensor at this order. Given the linearity in $\Phi$ of the scalar equation of motion and the absence of first order corrections to the metric, there is however no further correction for the scalar field at second order. The effect of the source at second order is therefore to modify the background metric $g_{\mu\nu}$ by a correction $\delta g_{\mu\nu}$
\be
g_{\mu\nu} \to  g_{\mu\nu} + \dg_{\mu\nu} \,.
\ee

Let's now consider the explicit ingredients which we will need to evaluate the on-shell action.  Using the equations of motion order by order in the amplitude expansion, the general form of the on-shell action at second order can be written as 
\be
S_{E} =  \frac{1}{16 \pi G_N} \left\{  \int_{\mathcal{M}}\!\! d^3 x \sqrt{g} \, 4 -  \int_{\del \mathcal{M}} \!\!d^2 x \sqrt{\gamma} \[  2 K   + \( \gamma^{ij} K - K^{ij} \) \delta \gamma_{ij} - \frac{1}{2}  \Phi \, \hat{n}\!\cdot\! \del \Phi \] \right\}\, .
\label{eq:onshellall}
\ee
The bulk and extrinsic curvature $2K$ terms are simply the zero order contributions, representing the action of the purely gravitational  background solution. The remaining contributions incorporate the $\lambda^2$ corrections to the background value of the action and are evaluated on the boundary $\del \mathcal{M}$.  $\delta \gamma_{ij}$ denotes the metric induced on the boundary by the correction $\delta g_{\mu\nu}$ to the bulk background metric.  The extrinsic curvature and its trace  appearing in \eqref{eq:onshellall}  are all computed in terms of the background metric. 

\subsection{Holographic renormalization} \label{holorenormX}

In this section, we use the standard holographic renormalization techniques \cite{deHaro:2000vlm,Skenderis:2002wp} to evaluate the renormalized on-shell action.     For this, we choose the Fefferman-Graham gauge 
\be
ds^2  = \frac{d\rho^2}{4\rho^2} + \frac{1}{\rho} \bar g_{ij}(\tau,\rho) dx^i dx^j \,  , \label{eq:FG}
\ee
where the coordinates $x^{i}$ indicate the boundary directions $\tau$ and $x$. The conformal boundary  $\del \mathcal{M}$ is the fixed $\rho$ surface at $\rho \to 0$.
Following the discussion above, when solving in a perturbative expansion in the amplitude of the deformation $\lambda$, the metric has an expansion of the form 
\be
  \bar g_{ij}=\gf_{ij} + \dgf_{ij} + O(\lambda^3)\, .
\ee
The background metric  in these coordinates \reef{eq:FG} takes  the form
\be \label{eq:EBTZ}
ds^2 = \frac{d \rho^2}{4 \rho^2} + \frac{(1- M \rho)^2}{4 \rho} d \tau^2 + \frac{(1+ M \rho)^2}{4 \rho} d x^2 \, ,
\ee
where the radial coordinate $\rho$ is related to the $z$ coordinate above through
\be
\sqrt \rho = \frac{z}{1+\sqrt{1-M z^2}}\,, \qquad \qquad z = \frac{2 \sqrt \rho}{1+ M \rho} \,. 
\ee
For the perturbatively backreacted metric, we consider the general ansatz consistent with homogeneity in the spatial $x$-coordinate
 \be
ds^2 = \frac{d \rho^2}{4 \rho^2} + \frac{(1- M \rho)^2}{4 \rho} d \tau^2 + \frac{(1+M \rho)^2}{4 \rho} d x^2
+\frac{1}{\rho}\[ \dgf_{\t\t}(\tau, \rho) d\tau^2 + \dgf_{xx}(\tau, \rho) dx^2\] \, .
\ee
Further, the asymptotically  AdS boundary conditions imply 
 \be \label{eq:AAdS}
 \dgf_{ij}(\t,\rho) \to 0 \qquad \text{for~ } \rho \to0\, .
 \ee
The latter are consistent with the relevant perturbation that sources the backreaction of the metric, as can be checked solving the equations of motion in an asymptotic expansion. 

The boundary metric in Fefferman-Graham coordinates \eqref{eq:FG} is $\bar \gamma_{ij}  = \frac{1}{\rho} \bar g_{ij}$ and it inherits in a natural way the splitting between the background and the leading,  second order in $\l$, correction induced by the scalar source 
\be
\gamma_{ij} = \frac{1}{\rho} \gf_{ij} \, ,\qquad   \delta\gamma_{ij} = \frac{1}{\rho}  \dgf_{ij} \, .
\ee

We introduce a regulator surface in the bulk at $\rho  = \eps$, 
which corresponds to introducing a short-distance cutoff in the boundary CFT. With this cutoff surface in place, one can evaluate the regulated on-shell action and determine the relevant counterterms. Considering first the gravitational part of the action \eqref{eq:onshellall}
\be \label{eq:grav}
S^{(G)}_{\rm reg} =   \frac{1}{16 \pi G_N} \left\{  \int_{\mathcal{M}}\!\! d^3 x \sqrt{g} \, 4 -  \int_{\del \mathcal{M}} \!\!d^2 x \sqrt{\gamma} \[  2 K   + \( \gamma^{ij} K - K^{ij} \) \delta \gamma_{ij} \] \right\}\, ,
\ee
the appropriate counterterm to make the regularized action finite is 
\be  \label{eq:gravct}
S^{(G )}_{\rm ct} =  \frac{1}{16 \pi G_N}\int_{\rho=\epsilon}\!\!d^2 x \sqrt{\bar \gamma} ~ 2 \,.
\ee
This is the standard counterterm which renormalizes the background action
\be
S^{(G)}_{\rm ren}  = \lim_{\eps\to 0}  \( S^{(G )}_{\rm reg}     + S^{(G )}_{\rm ct}  \)  =  -\frac{1}{16 \pi G_N} \int d^2 x  ~  M  +O(\l^2) \,.
\ee
In a perturbative expansion, this counterterm also renormalizes the leading correction to the background metric, \eg see  \cite{Detournay:2014fva}.  
The outgoing normal to the constant $\rho$ boundary surface is $\hat n = - \sqrt{g_{\rho\rho}}\del_{\rho}$ and the extrinsic curvature computed with the background boundary metric is 
\be
K_{ij} = \nabla_{(i} n_{i)} = - \rho \del_\rho \gamma_{ij}
=   \frac{1- M^2 \rho^2}{4 \rho} \delta_{ij}\ \, 
\ee
and
\be
 \g^{ij} K - K^{ij} = \frac{4 \rho }{1-M^2 \rho^2} \delta^{ij} \, .
\ee
Thus the order $\l^2$ contribution to the gravitational part \eqref{eq:grav} of the action  is
\begin{align}
\delta S^{(G)}_{\rm reg}&= -\frac{1}{16 \pi G_N}\int_{\rho =\epsilon} \!\! d^2 x \sqrt{\gamma}    \( \gamma^{ij} K - K^{ij} \) \delta \gamma_{ij} \nonumber  \\
& =  - \frac{1}{16 \pi G_N} \int_{\rho= \eps} d^2 x \frac{1}{\rho} \( \dgf_{\t\t} + \dgf_{xx} \)  
 \label{eq:gravl2contr} \,  
\end{align}
 Expanding \eqref{eq:gravct}  in the source amplitude:
 \bea 
S^{(G)}_{\rm ct} &=&  \frac{1}{16 \pi G_N}\int_{\rho=\epsilon} d^2 x \sqrt{\bar\gamma } ~ 2   =  \frac{1}{16 \pi G_N}\int_{\rho=\epsilon} d^2 x \sqrt{\gamma} ~ \( 2 + \gamma^{ij}\delta\gamma_{ij} \)  + O(\l^3)\,.  
\eea 
The second order contribution in the limit where the regulator is taken to zero gives 
 \be 
\delta S^{(G )}_{\rm ct} =   \frac{1}{16 \pi G_N}\int_{\rho=\epsilon} d^2 x \sqrt{\gamma} ~  \gamma^{ij}\delta\gamma_{ij}  =  \frac{1}{16 \pi G_N}\int_{\rho=\epsilon} d^2 x \(\frac{\delta^{ij}}{\rho} + O(\rho^0)\)\dgf_{ij}
\ee
and therefore  completely cancels \eqref{eq:gravl2contr}, leaving contributions that because of the asymptotycally AdS boundary conditions \eqref{eq:AAdS} go to zero as $\eps \to 0$. 

Therefore up to order $\l^2$ included,  the complete renormalized contribution coming from the gravitational part coincides with the zero-order renormalized result
\be \label{eq:gravreno}
S^{(G)}_{\rm ren}  = \lim_{\eps\to 0}  \( S^{(G )}_{\rm reg}     + S^{(G )}_{\rm ct}  \)  =  -\frac{1}{16 \pi G_N} \int d^2 x  ~  M  +O(\l^3) \, ,
\ee
which is the (negative) on-shell action of an AdS$_{3}$ black brane geometry.  

Next we want to evaluate the scalar part of the action, which is purely second order in the amplitude of the source $\l$ and consists only of boundary terms
\be
\delta S^{(\Phi)}_{\rm reg}= \frac{1}{16 \pi G_N}    \int_{\rho =\epsilon} d^2 x \sqrt{\gamma} \, \frac{1}{2} \Phi \, \hat{n}\cdot\del \Phi \, .
\ee
For  this  we only need to know the asymptotic solution, which in the range $0< \Delta <2 $ is\footnote{
We will not treat in the following the special case $\D=1$, which contains logarithmic terms.
}
\be
\Phi =  \rho^{\frac{2- \D}{2}} \vp_{(0)}(\t)  + \rho^{\frac{\Delta}{2}} \vp_{(\D)}(\t) + \dots   \label{eq:scalasympt}
\ee
with $\dots$ indicating subleading contributions that will not enter in our analysis. 
Notice that depending on whether  the conformal dimension is in the range $0< \Delta <1 $ or  $1< \Delta <2 $ the leading mode will be $\vp_{(\D)}$ or $\vp_{(0)}$ respectively, but according to  \eqref{eq:scalb} we are always identifying $\vp_{(0)}$ with the source  of the boundary deformation. 
As the range of $\D$ affects the structure of divergences, we analyze the two cases separately. 

\subsubsection*{$\bullet~1< \Delta <2 $}
 
In this range of conformal dimensions, the divergences of the scalar action and the associated counterterms  are the standard ones. 
Using the asymptotic form of the solution, the part of the action directly involving the scalar field has the following regularized structure
\bea
\delta S^{(\Phi)}_{\rm reg} &=& \frac{1}{16 \pi G_N}    \int_{\rho =\epsilon} d^2 x \sqrt{\gamma} \, \frac{1}{2} \Phi \,  \hat{n}\cdot\del \Phi \nonumber  \\ 
&=&   -\frac{1}{16 \pi G_N}     \int d^2 x \frac{1}{4} \left[  \eps^{1-\D} \frac{2-\D}{2} \vp_{(0)}^2  + \vp_{(0)} \vp_{(\D)} + \dots\right] 
\eea
up to terms that vanish in the limit $\eps\to 0$. This is renormalized by the counterterm  action    
\bea
S^{(\Phi)}_{{\rm ct}} &=&  \frac{1}{16 \pi G_N}\int_{\rho =\eps} d^2 x \sqrt{\gamma} \left( \frac{2-\Delta}{2} \Phi^2 \right)  \,,
\eea
which, as $\eps \to 0 $, leads to the following scalar renormalized action 
\bea
S^{(\Phi)}_{{\rm ren}} &=&  -\frac{1}{16 \pi G_N} \int d^2 x \, \frac{1}{4}  (\D-1) \vp_{(0)} \vp_{(\D)} \, . 
\eea
Combining this with  \eqref{eq:gravreno}, up to second order in the amplitude of the source $\l(\t)$, we get 
\be
S_{\rm ren}=\lim _{\eps \to 0 }\( S_{\rm reg} + S_{\rm ct}  \)  = -\frac{1}{16 \pi G_N} \int \frac{d^2 x}{4}  ~\left[4 M + (\D-1) \vp_{(0)} \vp_{(\D)} \right]  +O(\l^3)\, . 
\label{waggle}
\ee

\subsubsection*{$\bullet~0< \Delta <1 $}

The regulated scalar action in this case is 
\bea 
\delta S^{(\Phi)}_{\rm reg} &=& \frac{1}{16 \pi G_N} \int_{\rho =\epsilon} d^2 x \sqrt{\gamma} \,   \frac{1}{2}  \Phi \,  \hat{n}\cdot\del \Phi  \nonumber \\
&=&  -\frac{1}{16 \pi G_N} \int d^2 x \frac{1}{4}  \left[   \eps^{\D-1} \frac{\D}{2} \vp_{(\D)}^2  + \vp_{(0)} \vp_{(\D)} + \dots \ \right]
\eea
and together with the corresponding counterterm action
\bea
S^{(\Phi)}_{{\rm ct}} &=&  \frac{1}{16 \pi G_N}\int_{\rho =\eps} d^2 x \sqrt{\gamma} \left( \frac{2-\Delta}{2} \Phi^2 \right)  
\eea
gives as $\eps \to 0 $ the scalar renormalized action 
\bea
S^{(\Phi)}_{{\rm ren}} &=&  \frac{1}{16 \pi G_N} \int d^2 x \, \frac{1}{4}  (\D-1) \vp_{(0)} \vp_{(\D)} \, . 
\eea
However, working in the alternate quantization, in order for the Ward identities to hold, this is not sufficient. One also needs to include  a Legendre term in the scalar action \cite{Compere:2008us,Andrade:2011nh,Andrade:2011dg,Casini:2016rwj} 
\bea \label{eq:legendre}
S_{\rm Legendre} &=&  \frac{1}{16 \pi G_N}\int_{\del M} d^2 x \sqrt{\bar \gamma} \left( \Phi \, \hat n \cdot \del \Phi  - \Delta \Phi^2   \right)  \nonumber  \\
&=&  \frac{1}{16 \pi G_N}\int  d^2 x   \( 2(1-\D) \vp_{(0)} \vp_{(\D)} + \dots \) + O(\l^3)\, .
\eea
Notice that this term is simply $-2\(\delta S^{(\Phi)}+S^{(\Phi)}_{{\rm ct}}\) $, so it is finite and its unique effect on the renormalized on-shell action for the scalar is to flip the overall sign 
\be
S^{(\Phi)}_{{\rm ren}}+S_{\rm Legendre}  = - S^{(\Phi)}_{{\rm ren}}   \,. 
\ee
Therefore, also in the range $0<\D<1$, once the Legendre term  \eqref{eq:legendre} is included, the total renormalized action   gives 
\be
S_{\rm ren}=\lim _{\eps \to 0 }\( S_{\rm reg} + S_{\rm ct} + S_{\rm Legendre}  \)  = -\frac{1}{16 \pi G_N} \int \frac{d^2 x}{4}  ~\left[4 M + (\D-1) \vp_{(0)} \vp_{(\D)} \right]  +O(\l^3) \, . 
\ee
Of course, we observe that this result for the renormalized action takes a form which is identical to that in eq.~\reef{waggle} for $1<\Delta<2$.
 
\subsection{On-shell Euclidean action}\label{sec:Eaction}

The holographically renormalized on-shell action associated to the configuration in which we are interested takes the form
 \be \label{eq:SrenTOT}
S_{\rm ren}  = -\frac{1}{16 \pi G_N} \int dx ~ d\t ~ \frac{1}{4}   ~\left[4 M + (\D-1) \vp_{(0)} \vp_{(\D)} \right]    \, . \\
\ee
at leading non-trivial order in the amplitude of the perturbation.  

To extract explicitly the mode $ \varphi_{(\D)}$ we should expand the bulk profile \eqref{eq:sourced} for $z \to 0$ and read the coefficient of the mode $\sim z^\D$. That is
\begin{align}
\varphi_{(\D)}\sim  \frac{(\D-1) M^{\D}}{2^\D \pi}  \int dx'   d\tau'  \lambda(\t')\left[\cosh \(\sqrt M (x-x')\)-\sqrt{1- \tilde \eps^2}  \cos \(\sqrt M(\tau - \tau')\) \right]^{-\D}  \label{eq:deltamode} \nonumber\\
\end{align}
where we introduced the $z$-coordinate cutoff $\tilde \eps  = 2 \sqrt{ \eps M } $ and, after performing the integral, only keep terms that are finite as $\tilde \eps \to 0$. 
We are however interested in evaluating \eqref{eq:SrenTOT}, which contains an additional integral over $dx ~ d\t$. It turns out that it is easier to first perform both integrations over $x',\t'$ and $x,\t$ for finite $\tilde \eps$ and then extract the relevant contributions as we send $\tilde \eps \to 0$. That is,   we compute 
\bea  
S_{\rm ren}  \simeq && -\frac{L}{16 \pi G_N} \Bigg\{ 2 \pi \sqrt{M}    \, +  \l^2 \frac{(\D-1)^2 M^{\D}}{2^{\D+2}\, \pi}  \times    \label{woah} \\  
&&  \times    \int_{0}^{ \frac{2 \pi \a } {\sqrt{M}}} d\tau  \int_{-\infty}^{\infty} d x' \int_{0}^{ \frac{2 \pi \a } {\sqrt{M}}} d\tau'  \left[\cosh \(\sqrt M x'\)-\sqrt{1- \tilde \eps^2}  \cos \(\sqrt M(\tau - \tau')\) \right]^{-\D} \Bigg\}
\nonumber 
\eea
where we used translational invariance in $x$ and regulated the overall spatial integral by introducing $L$ as the spatial volume (\ie with $d=2$, the length of a fixed time slice). 
Performing a change of coordinates and defining 
\be \label{eq:IaD}
I(\alpha, \D)_{\rm reg} \equiv \int_{0}^{2\pi \alpha} d \tau  \int_{-\infty}^{\infty} d x' \int_{0}^{2\pi \alpha} d \tau'  \left[ \cosh x'- \sqrt{1- \tilde \eps^2} \cos( \t- \t') \right]^{-\D}\,,
\ee
eq.~\reef{woah} can be re-expressed as 
\beq 
 S_{\rm ren}  \simeq     -\frac{L}{16 \pi G_N} \Bigg\{ \frac{(2 \pi)^2}\b    \, +    \lambda^2 \frac{ (\D-1)^2}{2^{\D+2} \pi} \(\frac{2\pi}\b\)^{2\D-3} I(\alpha, \D)_{\rm reg} \Bigg\}  \, ,
\label{woah2}
\eeq
where we have also used $\sqrt M  = 2\pi/\b$.
As we explain in the next section and in Appendix~\ref{app:integral}, in doing so we introduce an additional divergence $\sim \tilde \eps^{2(1-\D)}$. This arises from integrating the non-normalizable mode of the bulk scalar over the boundary, and we simply drop it in the final result. 
 
Notice however that when integrating $\vp_{(\D)}$ with the inhomogeneous source $\l(\t)$ in  \eqref{eq:SrenTOT}, there will be also physical divergences arising. These are associated to the Euclidean path integral construction we are using, and more in particular to the fact that we are sharply localizing the profile of the source $\l$ along the Euclidean time circle. \\

In purely field theoretic terms, the on-shell action reads
 \be \label{eq:actionlo}
S_{\rm ren}  = - \int d x \ d\t   ~\left[\frac{\pi\,c }{6\, \beta^2}+ \frac 1 8 \l(\t)\, \langle \odd(\t) \rangle \right]   \,  \\
\ee
where the expectation value of the dual operator is related to the normalizable mode of the scalar field by
\be \label{eq:modeop}
16 \pi G_N \,  \langle \odd(\t) \rangle =  2 (\D-1)\,   \varphi_{(\D)} (\t) \,  ,
\ee
where $\varphi_{(\D)} (\t)$ is the normalizable mode of the bulk scalar, as given in eq.~\reef{eq:deltamode}.
Further, $\beta = 2 \pi / \sqrt M $ and we used the Brown-Henneaux central charge $c = 3/(2G_N)$. 
Indeed from the boundary point of view, the computation we are performing is the conformal perturbation theory expansion
\begin{align}
Z_{\rm CFT} &=\int {\mathcal D} \phi \, e^{- S_{\rm CFT}(\phi) - \int d^2x \, \lambda (\tau)\, \odd(\tau,x)} \nonumber \\
&= \int {\mathcal D} \phi \, e^{- S_{\rm CFT}(\phi)} \(1-  \int   \lambda (\tau) \odd(\tau,x) + \frac 1 2 \int   \int   \lambda (\tau) \lambda (\tau') \odd(\tau,x) \odd(\tau',x') + \dots \) \nonumber  \\ 
&= Z_{\rm CFT}\Big|_{\l=0} \(1+ \frac 1 2 \int \int \lambda (\tau) \lambda (\tau') \langle\odd(\tau,x)\, \odd(\tau',x')\rangle + \dots \)\,,
\end{align}
where we used $\langle  \odd \rangle_{\l =0} =0$ and on a cylinder
\be
\langle \odd(\tau, x) \odd(\tau', x')\rangle_{\l=0, {\rm cyl }} = \frac{1}{2^\D} \(\frac{2\pi}{\beta}\)^\D \left[ \cosh \frac{2\pi (x-x')}{\beta} - \cos \frac{2\pi (\tau - \tau')}{\beta}\right]^{-\D}\,.
\ee
With the identification \eqref{eq:modeop}, the holographic and conformal perturbation theory results $\d Z_{\rm bulk}$ and $\d Z_{\rm CFT}$ thus only differ by overall multiplicative terms and in that the holographic procedure directly renormalizes the divergences associated to contact points in the two-point function. 
 
\section{Holographic R\'enyi divergences}\label{sec:RenyiDiv} 

For the  R\'enyi divergence of an excited state  $\rho$ prepared by Euclidean path integral turning on a relevant deformation in the thermal state $\rho_\b$, the holographic construction of the previous section leads us to identify
\begin{align} \label{eq:trSren}
\log \tr  \(\rho^{\alpha} \rho_\beta^{1-\alpha}\) \approx - S_{\rm ren}\,,
\end{align}
with $S_{\rm ren}$ given by the expression in  eq.~\eqref{woah2}, together with the integral in eq.~\eqref{eq:IaD}.

As we anticipated,  to explicitly evaluate $S_{\rm ren}$ and the R\'enyi divergence we find it more convenient at the technical level to first compute the related quantity given in eq.~\reef{eq:IaD}
\be\label{in99}
I(\alpha, \D)_{\rm reg} =\int_{0}^{2\pi \alpha} d \tau  \int_{-\infty}^{\infty} d x' \int_{0}^{2\pi \alpha} d \tau'  \left[ \cosh x'- \sqrt{1- \tilde \eps^2} \cos( \t- \t') \right]^{-\D}\,,
\ee
to all orders in $\tilde \eps$, and then to extract from it what will be the relevant contributions to \eqref{eq:trSren} as we take $\tilde \eps \to 0$. 

We evaluate explicitly the integral $I(\alpha,\D)_{\rm reg}$  in appendix~\ref{app:integral}. As we remove the regulator $\tilde \eps$, the integral is finite for all $0< \D < 1$. For $\D >1$, it contains two different types of divergences. 
The first is the same divergence that we discussed in section~\ref{sec:Eaction}, which is of the form  $\sim \tilde \eps^{2(1-\D)}$, and its coefficient is linear in $\a$. This arises from the fact that  the integrand in \eqref{eq:IaD} is the full bulk-to-boundary scalar field propagator, rescaled by a factor $z^{\D}$. As such, it contains also the contribution of the non-normalizable mode of the bulk scalar, which is responsible for the $\sim \tilde \eps^{2(1-\D)}$ divergence.  
We drop this divergent contribution, which is absent in the holographically renormalized $S_{\rm ren}$, in what we define below as the renormalized quantity $I(\alpha, \D)$. This corresponds to a particular choice of contact terms in the boundary theory. The fact that these divergences are physically unimportant is also evident since generally they would not contribute to $D_\alpha  (\rho  \Vert  \rho_\b)$ even if they were not removed at this stage. However, we must add that there remains a residual effect at $\a\to1$ for $\D\ge1$. These details are explained below.  

The second type of divergence has the form $ \sim \tilde \eps^{3 - 2\D}$. This is a physical divergence arising from the specific form of the excited states we are considering in our analysis. In the Euclidean path integral construction, it is associated to the fact that we are working with source $\l(\t)$ that gives  a sharp discontinuity in the Euclidean path integral. However, we also note that this divergence is absent in the limit $\a \to 1$ (see figure \ref{fig:Iaclose1} in appendix \ref{app:integral}), where the path integral becomes smooth. 

At the practical level, we define the renormalized quantity as 
\be \label{eq:Iaren}
I(\alpha, \D) = I(\alpha, \D)_{\rm reg} + \frac{2^{\D+1} \pi^2 \a }{1-\D}\, \tilde \eps^{2(1-\D)}\,,
\ee
by subtracting the contribution  arising from the  non-normalizable mode of the scalar field  (see eq.~\eqref{eq:Iren} in appendix~\ref{app:integral}). For $\a =1$, this can be evaluated analytically and gives
\be \label{eq:I1}
I(1, \D) = \frac{ 2\pi^{3/2} \G\(\frac{1-\D}{2}\) \G\(\frac \D 2\)^2}{\G(\D)\G\(1- \frac \D 2\)} \,. 
\ee
For $\a <1$, we find it convenient to write the regulated expression as 
\begin{align}
I(\alpha, \D)_{\rm reg}&= \frac{2^{2-\D}\sqrt \pi  \G(\D)}{\G\(\D+\frac 1 2\)} \int_{0}^{2\pi \alpha} \!\!\!dp \, (2\pi \a -p ) \, F\!\[\D, \D, \D+ \frac 1 2, \frac{1+ \sqrt{1- \tilde \eps^2} \cos p}{2}\]  \,,
\end{align}
and perform the remaining integration numerically.

The trace function \eqref{eq:trSren} we are interested in is then evaluated in terms of the renormalized quantity $I(\alpha, \D)$ simply as  
\begin{align}
\log \tr & \(\rho^{\alpha} \rho_\beta^{1-\alpha}\) \approx \frac{c}{24 \pi} L  \left\{ \frac{(2\pi)^2}{\beta} +  \lambda^2 \frac{ (\D-1)^2}{2^{\D+2} \pi} \( \frac{2\pi}{\beta} \)^{2\D-3} I(\a, \D) \right\} \,, \label{oneX}
\end{align}
where  $c = 3/(2G_N)$.
The density matrices $\rho$ and $\rho_\b$ computed in this way are not normalized to one, as can be immediately seen taking the limit  $\alpha \to  1$
\begin{align} \label{eq:logrho}
\log \tr &\rho \approx \frac{c}{24 \pi} L  \left\{ \frac{(2\pi)^2}{\beta} +  \lambda^2 \frac{ (\D-1)^2}{2^{\D+2} \pi} \( \frac{2\pi}{\beta} \)^{2\D-3} I(1, \D) \right\} \,,
\end{align}
and $\alpha \to 0 $
\begin{align}
\log \tr & \rho_\beta \approx \frac{c  \, \pi  L }{6 \beta}\label{threeX}
\end{align}
of the expression above.  Hence to account for this normalization in  the R\'enyi divergences, we write the following expression 
\bea
D_\alpha  (\rho  \Vert  \rho_\b) &=& \frac{ 1}{\alpha-1}\log \frac{\tr\(\rho^{\alpha} \rho_\beta^{1-\alpha}\) }{( \tr \rho)^\a ( \tr \rho_\b)^{1-\a}} \nonumber  \\
&\approx& \l^2 \frac{c\,L}{6 \pi\b} \,   \frac{(\D-1)^2}{2^{\D+3}} \( \frac{2\pi}{\beta} \)^{2(\D-2)} \frac{I(\alpha, \D)-\a I(1, \D) }{\alpha-1}   \,  .\label{eq:Dalphareg}
\eea
The second line above gives the leading order result for the holographic R\'enyi divergences, which we see is second order in the amplitude $\lambda$ of the deformation. We should note that since this amplitude is dimensionful, our perturbative expansion is properly described in terms of the dimensionless quantity
$\big(\frac{2\pi}{\b}\big)^{\D-2}\,\l$.\footnote{Let us add that while the condition $(2\pi/\b)^{\D-2}\,\l \ll 1$ is required for the validity of our perturbative expansion, it also ensures that the  excited state \reef{jumbo} will have a (relatively) simple interpretation in terms of the CFT excitations. Otherwise the relevant perturbation will drive the new state in the initial theory far
away from the conformal phase, \ie far from the thermal state \reef{mini}. In the dual gravitational description, the latter means that the dual scalar field grows in the region outside the event horizon to such an extent that its backreaction will significantly deform the black hole geometry (and that any nonlinearities in the scalar potential will become important), \eg see discussion in \cite{Buch1,Buch2}.}  \\ 

\begin{figure}[ht]
\centering 
\includegraphics[width=.6\textwidth]{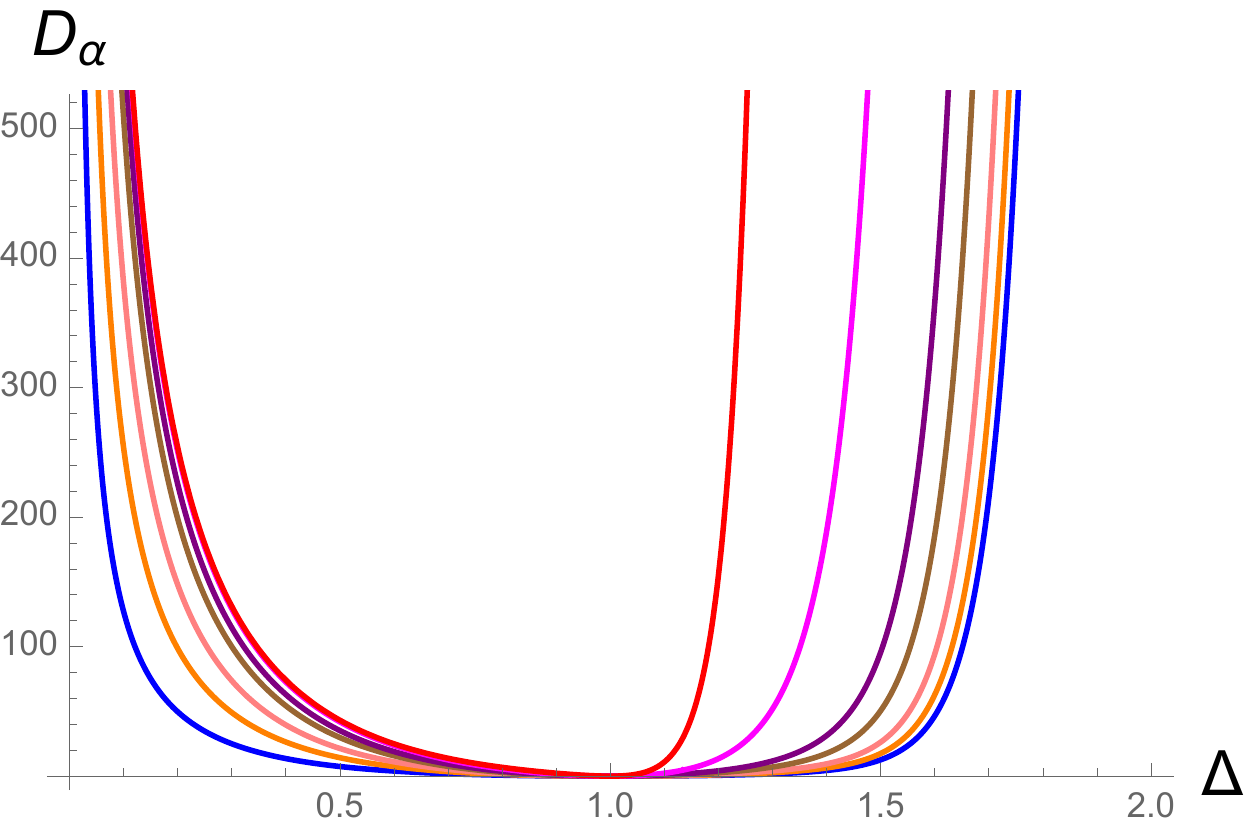}  
\caption{Holographic R\'enyi divergence for a cutoff $\tilde \eps = 0.0001$ and inverse temperature $\b =2\pi$. $\a =0.2,0.4,0.6,0.8,0.9,0.99,1$ from the bottom up. We  rescaled the $\D$-independent prefactor $ \l^2 \frac{c}{ 3 \pi^2 2^5 } L $.}
\label{fig:Da_hol}
\end{figure}

In figure~\ref{fig:Da_hol}, we plot a number of representative curves for the R\'enyi divergences as a function of the conformal dimension $\D$, setting for convenience  $\b = 2\pi$. As noted above, when we take the regulator $\tilde \eps \to 0$, we find a single UV divergence of the from $\sim \tilde \eps^{3-2\D}$ for most values of $\a$ (and $\D\ge3/2$). However, there is also a residual divergence of the form $\sim \tilde \eps^{2(1-\D)}$, which appears at $\a=1$, as we show next. 

In the limiting case $\a =1$, the R\'enyi divergence becomes the relative entropy, which in turn can be written as the difference of ordinary free energies
\begin{align}
D_{1} (\rho \| \rho_\beta) &= \tr \rho \log \rho - \tr \rho \log \rho_\beta 
=  \b \( F(\rho) - F(\rho_\b)\)\,,
\end{align}
and can also be computed explicitly. Namely,
\begin{align}
D_1  (\rho  \|  \rho_\b) &\approx \l^2 \frac{c\,L}{6 \pi\b}  \frac{(\D-1)^2}{2^{\D+3}} \( \frac{2\pi}{\beta} \)^{2(\D-2)} \left\{ \del_\a I(\alpha, \D)\big|_{\a=1}- I(1, \D)\right\}   \nonumber \\  
&= \lambda^2 \frac{\pi\,c\,L}{3\,\b} \frac{(\D-1)^2}{8} \( \frac{2\pi}{\beta} \)^{2(\D-2)} \left\{  \frac{ \G(\frac{1-\D}{2}) \G(\frac \D 2)^2}{2^\D \sqrt \pi \G(\D)\G(1- \frac \D 2)}- \frac{ \tilde  \eps^{2(1-\D)}}{1-\D}   \right\}  \,,\label{eq:Dalpha1/2}
\end{align}
where we used eq.~\eqref{eq:Iprime1}:
\be \label{eq:Iprime1main}
\del_\a I(\a, \D) \Big|_{\a=1}= - \frac{2^{\D+1} \pi^2 }{1-\D}\tilde \eps^{2(1-\D)} + 2  I(1, \D) \,.
\ee

The double zero at $\D=1$ appearing in the numerator of \eqref{eq:Dalphareg} forces all curves $D_\a$ to have the same unique minimum. This prefactor comes from the bulk-to-boundary normalization \eqref{eq:CDelta} and the on-shell action computed in holographic renormalization \eqref{eq:SrenTOT}. In such a case, the monotonicity constraints
\be
D_\a  (\rho  \|  \rho_\b) \ge D_\a  (\rho'  \|  \rho_\b) 
\ee
are equivalent for all $\a \in (0,1]$, as can be seen from figure~\ref{fig:Da_hol}. According to the second laws of quantum thermodynamics \cite{brandao2013second}, a transition between a state prepared via a relevant deformation of conformal dimension $\D$ and one with $\D'$ is therefore possible only if $\D < \D' < 1$ or $1 < \D' < \D$.

\begin{figure}[ht]
\centering 
\includegraphics[width=.6\textwidth]{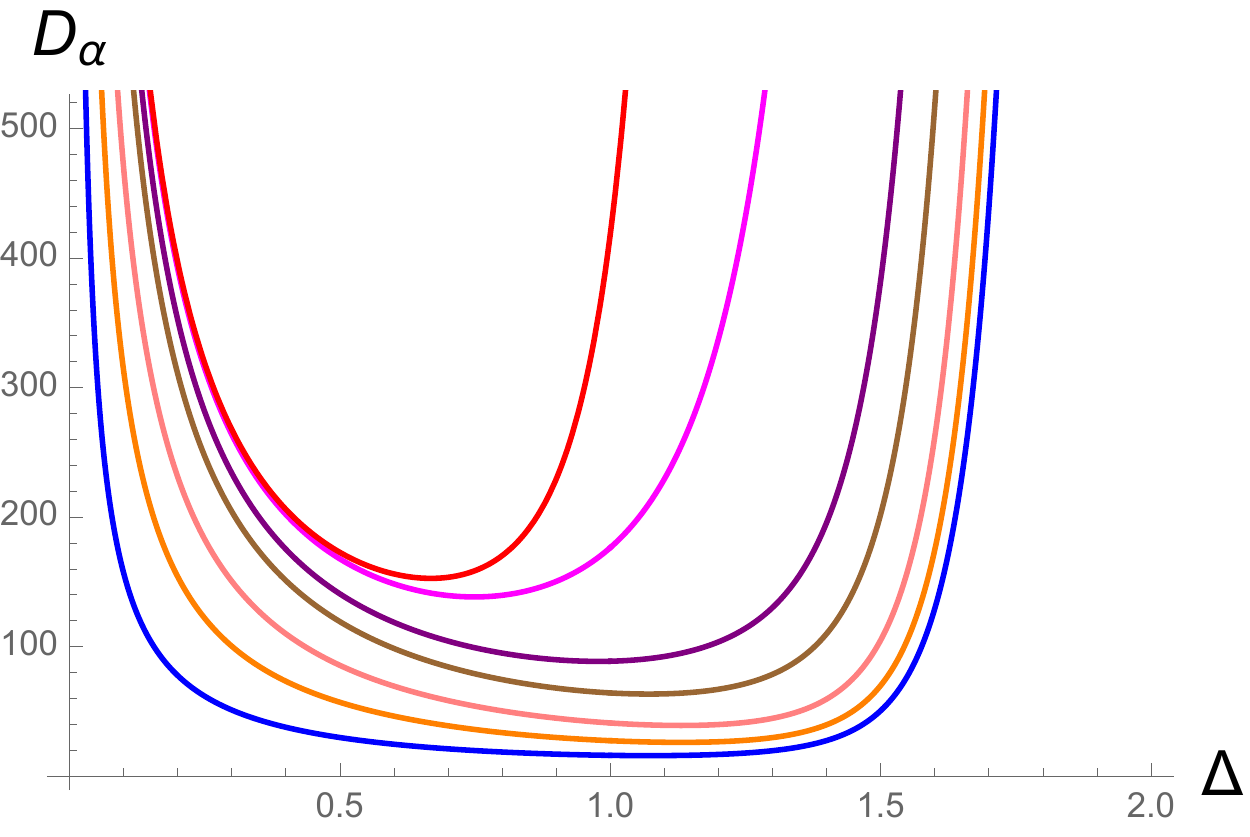} 
\caption{Holographic R\'enyi divergence   with $\lambda = \tilde \lambda  / |\D - 1|$ and $\tilde \lambda (2\pi/\b)^{\D-2}$ fixed, for a cutoff $\tilde \eps = 0.0001$ and inverse temperature $\b =2\pi$. $\a =0.2,0.4,0.6,0.8,0.9,0.99,1$ from the bottom up. We have rescaled the $\D$-independent prefactor $ \tilde \l^2 \frac{c}{ 3 \pi^2 2^5 } L $.} 
\label{fig:Da_CFT}
\end{figure}

However, implicitly we assumed above that the coefficient $\lambda$ (or rather the dimensionless quantity $(2\pi/\b)^{\D-2}\,\l$, as in producing the plot we have set $\b =2\pi$)  was the same for both deformations. It is important to remember that we still have the freedom of varying the amplitude of the source $\l$, and thus of  modifying the quantum $\a$-free energies in a non-trivial way. 
For fixed $\a$, the plot of $D_\a  (\rho  \|  \rho_\b)$ is in fact effectively three-dimensional, as a function of both dimensionless parameters $\D$ and $\l (2\pi/\b)^{\D-2}$. For example, we could for instance have considered the source amplitude of the form $\lambda = \tilde \lambda  / |\D - 1|$ and held $\tilde \lambda (2\pi/\b)^{\D-2}$ fixed. This would  effectively rescale the formula above by a factor $(\D-1)^2$ and give the result plotted in figure~\ref{fig:Da_CFT}. 
As curves of different $\a$ now have distinct minima, in these directions the second laws are not equivalent and would pose non-trivial constraints for a Lorentzian evolution allowing transitions between states associated to different relevant deformations --- see further discussion in section \ref{sec:discussion}.

\begin{figure}[ht]
\centering 
\includegraphics[width=.48\textwidth]{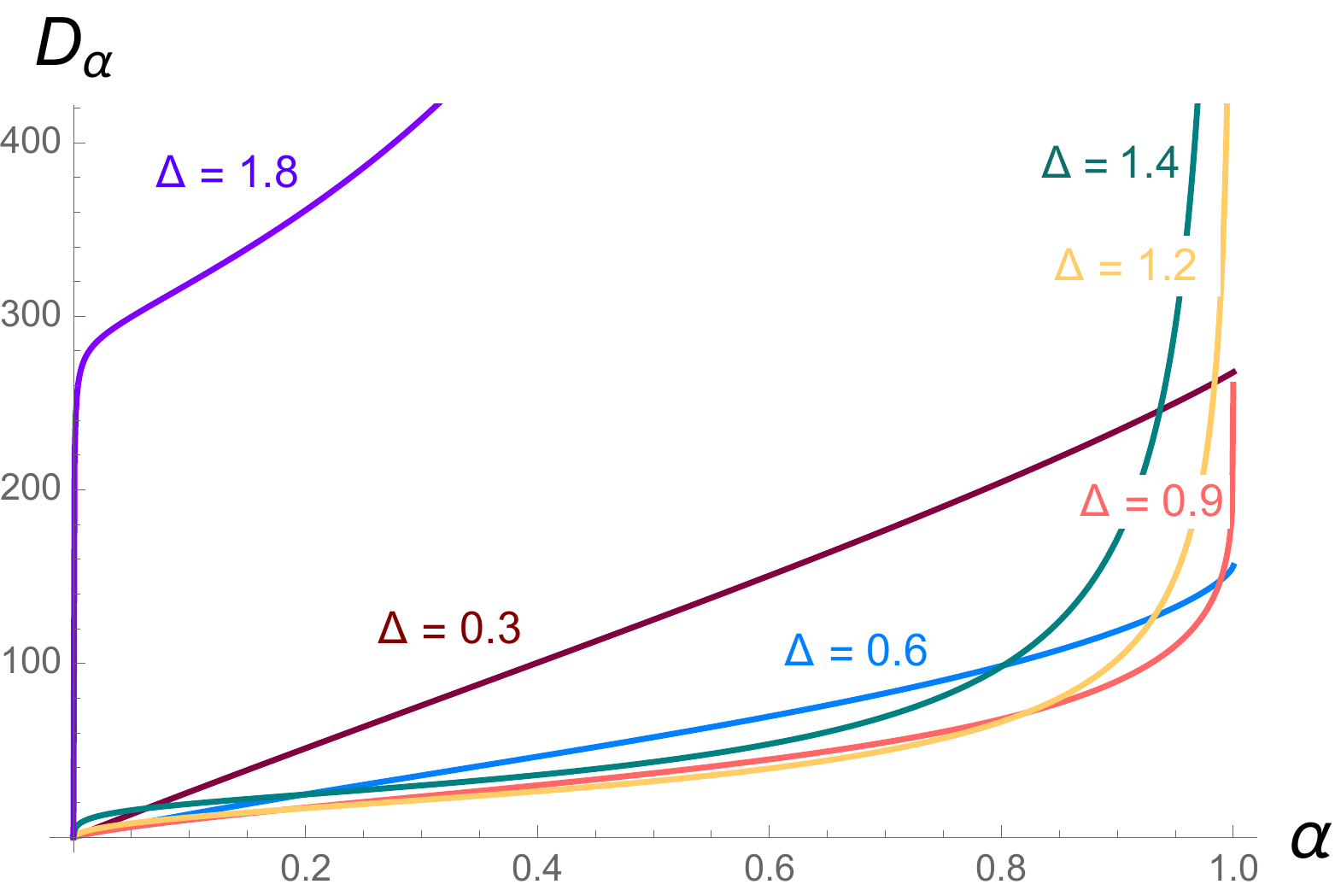}  \hfill \includegraphics[width=.48\textwidth]{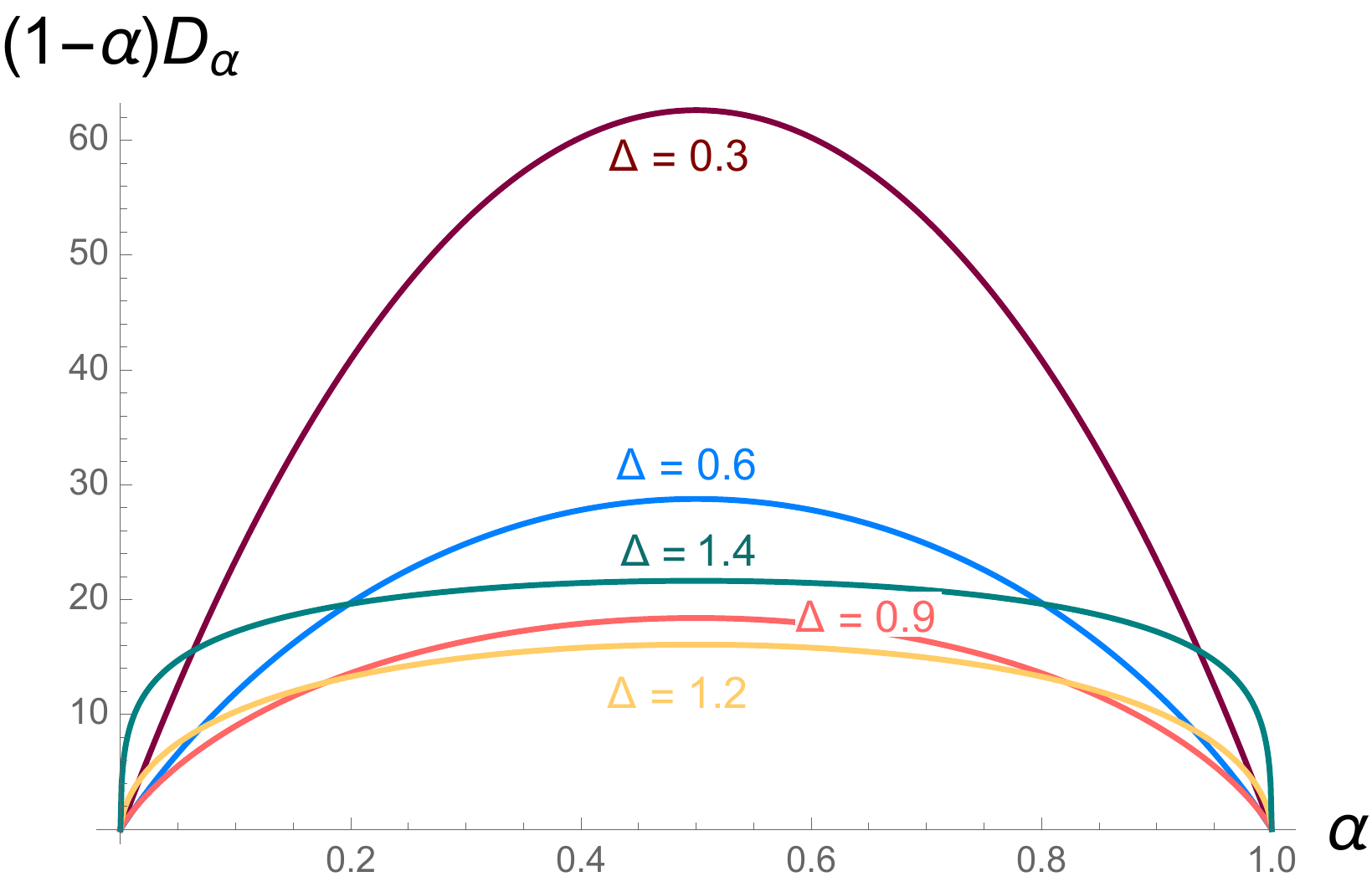} 
\caption{$D_\a$ (left)  and $(1-\a)D_\a$ (right), which is concave in $\a$. Here $\lambda = \tilde \lambda  / |\D - 1|$ and $\tilde \lambda (2\pi/\b)^{\D-2}$ fixed. In both plots $\D =0.3,0.6,0.9,1.2,1.4$ and $\tilde \eps = 0.001$ to be able to fit all curves in the same plot. (In the right panel we excluded the curve $\D =1.8$, which has much bigger magnitude than the others.) The inverse temperature is $\b =2\pi$ and in all curves we rescaled an overall factor $ \tilde \l^2 \frac{c}{ 3 \pi^2 2^5 } L $.} 
\label{fig:Da_adep}
\end{figure}

Also notice that we find that our result consistently satisfies the expected properties of R\'enyi divergences for $0 \le \a \le 1$ \cite{Erven}: 
\begin{itemize}\setlength\itemsep{0em}
\item {\bf Positivity}: $D_\a \ge 0$; 
\item {\bf Monotonicity in $\a$}: $D_\a$ is nondecreasing in $\a$; 
\item  {\bf Continuity in $\a$};
\item {\bf Concavity}: $(1-\a)D_\a$ is concave in $\a$.
\end{itemize}
This can be directly  seen from figure~\ref{fig:Da_adep}, where we plot the $\a$ dependence of $D_\a$ and $(1-\a)D_\a$ for various representative values $\D$. 

Before proceeding, we wish to return to the UV divergences in our results. First recall that the regulated integral \reef{in99} contained a divergence of the form $\tilde \eps^{2(1-\D)}$, which we removed in eq.~\eqref{eq:Iaren}. However, we would first like to note that since the divergence that we removed there is linear in $\a$, it would  have canceled out in the R\'enyi divergence \reef{eq:Dalphareg} even if we worked directly with the regulated integral $I(\alpha, \D)_{\rm reg}$. Again this simply reflects the fact that this divergence is physically unimportant and can be removed with a particular choice of contact terms in the boundary theory. However this description is not complete since, as we see in eq.~\reef{eq:Dalpha1/2}, there is a residual $\tilde \eps^{2(1-\D)}$ divergence at $\alpha=1$.\footnote{This softens to a logarithmic divergence at precisely $\D=1$.} As explained in appendix \ref{app:integral}, the divergence in the regulated integral actually has a step-function-like coefficient, which makes a rapid transition in the vicinity of $\alpha=1$ and so the previous cancellation fails there (and in a narrow band of width $\delta\a\sim\tilde \eps$ about $\a=1$) --- see figure \ref{fig:I1primeBTZ}. As shown in eq.~\reef{eq:Dalpha1/2}, this divergence then appears in the relative entropy $D_1  (\rho  \Vert \rho_\b)$, but further in quantities like the energy and entropy of the excited states with $\D\ge1$ --- see eqs.~\reef{creep4} and \reef{creep5} below. 

Of course, as we already commented, for states prepared with operators of higher conformal dimensions, namely $\D\ge3/2$, there is an even more pervasive divergence proportional to $\tilde \eps^{3-2\D}$ --- again this softens to logarithmic at precisely $\D =3/2$. For these states, the R\'enyi divergence \reef{eq:Dalphareg} contains this divergence for all values of $\a$ except at $\alpha=1$. As described above, these UV divergences can be understood as an effect of the source $\l(\tau)$ that changes instantaneously from zero to some fixed value in our path integral construction \reef{jumbo2}. 

The final conclusion is that any of our results for $\D\ge1$ do not have a physical interpretation unless we imagine that there is a finite UV regulator in place. However, we must also recall that we have formulated our calculations in a perturbative framework in the (dimensionless) expansion parameter $(2\pi/\b)^{\D-2}\l$. Hence even with a finite UV regulator, if the $O(\l^2)$ term is proportional to $1/\tilde{\eps}^a$ (for some positive $a$), then we must limit our calculations to $(2\pi/\b)^{\D-2}\l\ll \tilde{\eps}^a$. That is, there is a tension between our perturbative expansion and these UV divergences.\footnote{Examining figures~\ref{fig:Da_hol} or \ref{fig:Da_CFT}, we also note that the R\'enyi divergences also appear to diverge in the limit $\Delta\to0$. For $\a=1$, this divergence can be explicitly seen as a $1/\D$ pole in eq.~\reef{eq:Dalpha1/2}. These divergences are independent of the UV regulator, but they will also limit our perturbative calculations for very small values of $\D$.} Therefore in our further examination and discussion of the R\'enyi divergences in the next two sections, we will limit our attention to excited states corresponding to conformal dimensions $\D<1$, which do not exhibit any UV divergences and remain well defined in the limit
$\tilde\eps\to0$.


\section{Closed-system thermodynamics and further considerations}
\label{sec:closed}

Before we discuss the results, a few more detailed points about the role of R{\'e}nyi divergences in the context of closed systems are worth making. The first, is that one
is sometimes interested in {\it smoothed} R{\'e}nyi divergences \cite{RR-phd,dahlsten2011inadequacy,horodecki2013fundamental,brandao2013second}. Namely, 
there are cases where we are not just interested in exact transformations of a state into another, but just in approximate ones. For example, if we are trying to extract work from a state transition, we may only be able to extract a small amount if we want to be completely certain that we extracted work, but if we 
are willing to tolerate an $\epsilon$-small probability of failing, then we may be able to extract a lot more. This is also the case if we only care about average work. Likewise, if we are considering state transitions, as in eq.~\reef{contract}, we may not care that we produce the exact state we want, and so may be content with an approximate transformation which still produces a state close to the desired one. The term {\it smoothing} is used to denote the process of minimising the quantities under consideration over initial or final states which are in an $\epsilon$-sized ball of the states of interest.  Here, we restrict to considering exact transitions, and leave the case of approximate transitions to further study.

Perhaps more importantly for the case of closed systems, we may also want to consider versions of the second law which pertain to dynamical processes which only approximately preserves thermal states. Indeed, requiring that a map is linear and exactly preserves thermal states at all temperatures is a severe restriction on the map. In particular, such maps will generically only approximately thermalize an arbitrary state \cite{MuellerOppenheim}.  If we were to extend our discussion to include approximately thermalizing maps, then the quantity of interest is \cite{MuellerOppenheim}
\begin{align}
D^\epsilon_\alpha(\rho\Vert\rho_R):= \inf_{\sigma\in{\cal B^{\epsilon}}(\rho_R)} D_\alpha(\rho \Vert\sigma)
\end{align}
where infimum is taken over an $\epsilon$-ball around $\rho_R$. Computing the  thermodynamical constraints that this quantity imposes is beyond the scope of the present work, but will be discussed further in  \cite{MuellerOppenheim}. However, one should keep in mind that transitions which are forbidden by maps which exactly preserve thermal states might be allowed by maps which are only approximately thermalizing.  

Thirdly, it is worth noting that the derivation of the thermodynamical constraints given by thermo-majorization and the R{\'e}nyi divergence was originally done in the context of a system in contact with a thermal reservoir of arbitrary temperature \cite{brandao2013second}. Here instead we want to consider a closed system, which will equilibrate to a thermal state of the same energy, \ie to a thermal state with inverse temperature $\beta'$ such that $\tr ( \rho H ) =\tr (\rho_{\beta'} H)$. Nonetheless, eq.~\eqref{eq:monotonicity}, the monotonicity of a distance measure to a reference thermal state, can hold for any temperature of the thermal state. This should be clear from its derivation, and holds provided the dynamics are such that any thermal state is a fixed point.
This is typically the case when $\rho$ represents all the degrees of freedom of the system. This means that for such closed system dynamics, varying both $\alpha$ and $\beta$ provides a new two-parameter family of constraints. On the other hand, when $\rho$ represents a coarse grained description of the system, one expects that only $\rho_\beta$ is preserved by the dynamics. We examine both of these possibilities more closely below for our holographic model in sections \ref{sec:Xeno} and \ref{sec:discussion}.

\subsection{Work function}

For closed systems, another quantity of interest is
\begin{align}
\work(\rho)\equiv T \left( D_\alpha(\rho\Vert\rho_\beta)-D_\alpha(\rho_{\beta'}\Vert\rho_\beta)\right) \, .
\label{eq:w-alpha}
\end{align} 
For $\alpha=1$, we have $W_{1,\beta}=T \delta S$, where $\delta S$ is the change in entropy of the state as it equilibrates to $\rho_{\beta'}$. It is the work which could be extracted from its increase in entropy, were we to put it in contact with a bath of temperature $T$.
Indeed, the ordinary free energy constrains how a state evolves during a thermodynamical process, and determines how much work can be extracted from a state transformation (the latter is in fact a special case of the former).

Likewise, the R{\'e}nyi divergences constrain state transformations and $\work(\rho)$ tells us how valuable a resource a particular state is.
For example, the quantity $\inf \work$ is the {\it work distance} \cite{brandao2013second}, which gives the deterministic work which could be extracted as the system equilibrates were we to couple it to an ancilla at temperature $T$. We can however, say more.
If we have two states, such that $\work(\rho_1)\geq \work(\rho_2)$ for all $\alpha,\beta$, then we can conclude that $\rho_1$ is a better thermodynamical resource during its equilibration. To see this, consider a third ancillary system in state $\rho_a$ which we want to force to make a transition to $\rho_a'$. Then if
the transition $\rho_a\otimes\rho_2\rightarrow\rho_a'\otimes\rho_{\beta_2'}$ is
possible, \ie 
\begin{align}
D_\alpha(\rho_a\Vert\rho_\beta)+D_\alpha(\rho_2\Vert \rho_\beta)
\geq D_\alpha(\rho_a'\Vert\rho_\beta)+D_\alpha(\rho_{\beta_2'}\Vert\rho_\beta) \qquad \forall~\alpha,\beta \, , \label{boot88} 
\end{align}
 then the transition $\rho_a\otimes\rho_1\rightarrow\rho_a'\otimes\rho_{\beta_1'}$ is less constrained. 
We may trivially re-express eq.~\reef{boot88} as follows:
\be
D_\alpha(\rho_2\Vert\rho_\beta)-D_\alpha(\rho_{\beta_2'}\Vert\rho_\beta)
\geq D_\alpha(\rho_a'\Vert\rho_\beta)-D_\alpha(\rho_a\Vert\rho_\beta)\,,
\label{boot99}
\ee
and we see that $\beta \work(\rho_2)$ is on the left hand side and determines
how useful $\rho_2$ is as a thermodynamical resource to induce transitions in
an ancilla in the sense of imposing more or less constraints. The larger the $\work(\rho_2)$, the more freedom we have to induce  a transition $\rho_a\rightarrow\rho_a'$.

Although $\work$ provides constraints for any reference state $\rho_\beta$, in the case where it is the equilibrium state, we have
$\beta'\,W_{\alpha,\beta'}(\rho)=D_\alpha(\rho\Vert\rho_{\beta'})$. Thus the R\'enyi divergence has a more direct physical interpretation in terms of the work function when the reference state is $\rho_{\beta'}$. 
With the perturbative expansion in which we are working, we have  $ \beta\, \work(\rho)\approx D_\alpha(\rho\Vert\rho_\beta)$, because the extra term in eq.~\reef{eq:w-alpha}, \ie $D_\alpha(\rho_{\beta'}\Vert\rho_\beta)$, is higher order than $D_\alpha(\rho\Vert\rho_\beta)$. In particular, we will show below (see eq.~\reef{creep6})
that the final equilibrium temperature $1/\beta'$ differs from $1/\b$ by an $O(\l^2)$ correction, \ie
\beq
\frac{1}{{\beta'}} =  \frac{1}{\beta}\left\{1 +  \kappa \( \frac{2\pi}{\beta}\)^{2(\D-2)} \l^2\right\} \,, \label{gamble2}
\eeq
where $\kappa$ is some numerical factor. Now substituting this expression into
\begin{align}
D_\a \(\rho_{\b'} \Vert \rho_\b\) &=  \frac{1}{\alpha-1}\log \frac{\tr\(\rho_{\b'}^{\alpha} \rho_\beta^{1-\alpha}\) }{( \tr \rho_{\b'})^\a ( \tr \rho_\b)^{1-\a}} \nonumber \\
&= \frac{\pi c L}{6} \frac{1}{\a-1} \left\{ \frac{1}{\a \b' + (1-\a)\b} - \frac{\a}{\b'} - \frac{1-\a}{\b} \right\}
\nonumber\\
&\simeq\frac{\pi\,c\,L}{6\,\b}\, \a\,
\kappa^2 \( \frac{2\pi}{\beta}\)^{4(\D-2)} \l^4
\label{gamble}
\end{align}
That is, $D_\a \(\rho_{\b'} \Vert \rho_\b\) \sim O(\l^4)$, and thus it is beyond the order to which we are evaluating the R\'enyi divergence in our perturbative expansion. 
 
Notice also that if $\work(\rho_1)\geq \work(\rho_2)$ for a range of $\alpha$ (but not necessarily all $\alpha$), then the ordering of $\work$ still gives us physically relevant information about the relative usefulness of $\rho_1$ and $\rho_2$. In particular, it tells us that for
a family of constraints, $\rho_1$ is a better resource than $\rho_2$. 
In the case where $[\rho_1,\rho_\beta]=[\rho_2,\rho_\beta]=0$, and when the reference state is the equilibrium state, we could in fact find an ancilla with
$\work(\rho_2)\leq \work(\rho_a)\leq \work(\rho_1)$
for $0\leq\alpha\leq 1$ and $\work(\rho_a)=0$ elsewhere,
and because the R\'enyi divergences are necessary and sufficient conditions in the commuting case \cite{brandao2013second}, we would be able to induce a transition in the ancilla using $\rho_1$ but not $\rho_2$.

However, in general, the positivity of $\work(\rho_a)-\work(\rho_i)$ only gives necessary conditions that the transition of the ancilla needs to satisfy. There may also be additional constraints coming from other quantum R\'enyi divergences (\eg the sandwiched R\'enyi divergences of eq.~\eqref{eq:sandwhiched-RD}, or the decohered divergences of \cite{brandao2013second}). 
In fact, for the set of states $\rho$ that we are considering, we can compute  $\work(\rho)$ not only for different $\alpha$ but also for different values of reference state inverse-temperature $\beta$ --- see below. 

This rich set of constraints means we should be careful to only compare the relative usefulness of two states in terms of the strength of the constraints they impose. So, while it is physically meaningful to compare the $\work(\rho)$ of various states in a particular range of $\alpha,\beta$ in terms of the strength of some second laws, these are necessary conditions and not sufficient ones. We return to this in discussing our holographic R\'enyi divergences in the following section.

\subsection{General reference states}\label{sec:Xeno}

Here, we return to the idea introduced in section \ref{intro:thermo} that the R\'enyi divergence must decrease in physical processes but that we may use any equilibrium state as the reference  state --- see discussion around eq.~\reef{eq:monotonicity}. We will show below that it is straightforward to extend our holographic calculations in sections \ref{sec:quench} and \ref{sec:RenyiDiv} to incorporate this generalization. We are then able to use these new results in section \ref{sec:discussion} to explore how varying the reference state modifies the constraints imposed on our holographic model by demanding the monotonic reduction of the R\'enyi entropies.

As described in section \ref{PIapproach}, our example focuses on a special family of excited states \reef{jumbo}, which are defined by a path integral on an interval in Euclidean time, \ie we can think of these states as thermal states defined with a modified Hamiltonian
$H'$. Now in the partition function \reef{jumbo2}, both the thermal reference state and the excited state are defined with the inverse temperature $\beta$. However, even if the reference thermal state was chosen with $\br$ (which is unrelated to $\beta$), then the partition function takes essentially the same form of a path integral on a thermal circle (with the deformation turned on for some fraction of the full circumference) and in principle then, it remains straightforward to evaluate the R\'enyi divergences for this general situation. 

In the case with a new reference state $\rhoR$ with inverse temperature $\br$, eq.~\reef{jumbo2} is replaced by
\be
Z'_{\rm CFT} = \tr \(\rho^\alpha \rhoR^{1-\alpha}\) =
\raisebox{-0.5\height}{
\begin{overpic}[scale=0.8]{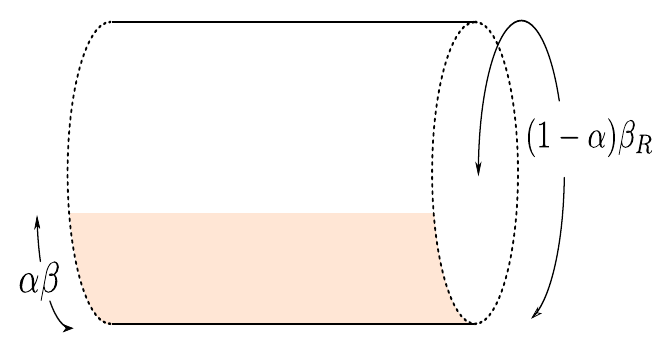}
\end{overpic}
} \,.
\label{jumbo4}
\ee
Here, we see the total circumference of the thermal circle is given by
\be
C=(1-\a)\,\br+\a\,\beta\ \,.
\label{CCx}
\ee
The interval over which the deformation is present is still $\ell =\alpha\,\beta$ and hence the fraction of the total thermal circle in which $H'$ acts is 
\be
f=\frac{\ell }{C}=\frac{\a\,\b}{(1-\a)\,\br+\a\,\beta}\,.
\label{ffx}
\ee
Now, let us set aside our perturbative calculations for the moment, and imagine that the partition function in eq.~\reef{jumbo2} can be evaluated and takes the form $Z_{\rm CFT} = P(\b,\a)$. Then if the strength and conformal weight of the deformation are chosen with the same values in eq.~\reef{jumbo4}, we will find
 $Z'_{\rm CFT} = P(C,f)$. Hence for this class of excited states, if we succeed in the initial R\'enyi divergence $D_\a(\rho\Vert\rho_\b)$, then evaluating the generalized quantities $D_\a(\rho\Vert\rhoR)$ is straightforward.

Let us illustrate the latter observation using our explicit perturbative calculations in sections \ref{sec:quench} and \ref{sec:RenyiDiv}. In particular,  beginning with eq.~\reef{oneX}, the above prescription yields for our generalized construction \reef{jumbo4}, 
\be
\log \tr \! \(\rho^{\alpha} \rhoR^{1-\alpha}\) \approx \frac{c}{24 \pi} L  \left\{ \frac{(2\pi)^2}{C} +  \lambda^2 \frac{ (\D-1)^2}{2^{\D+2} \pi} \( \frac{2\pi}{C} \)^{2\D-3} I(f, \D) \right\} \,, \label{oneY}
\ee
where $C$ and $f$ are given by eqs.~\reef{CCx} and \reef{ffx}, respectively. Of course here, the calculation is still perturbative in the source amplitude $\l$. Further, eq.~\reef{eq:logrho} is unchanged since we are using the same excited state, but eq.~\reef{threeX} is simply replaced by
\be
\log \tr \rhoR = \frac{\pi\,c \, L }{6 \br}   \label{threeY}
\ee
for the new reference state. Combining these ingredients then yields
\bea
D_\alpha  (\rho  \Vert  \rhoR) &=& \frac{ 1}{\alpha-1}\log \frac{\tr\!\(\rho^{\alpha} \rhoR^{1-\alpha}\) }{( \tr \rho)^\a ( \tr \rhoR)^{1-\a}} \label{New1}  \\
&\approx& D_\alpha(\rho_\b\Vert\rhoR)+
\frac{\l^2}{\alpha-1} \frac{c\,L}{6\pi\b}    \frac{(\D-1)^2}{2^{\D+3}}  \(\frac{2\pi}{\beta}\)^{2(\D-2)} \left[  \frac{ I\!\(\frac{\a}{(1-\a) x+\a}, \D\)}{((1-\a) x+\a)^{2\D-3}}-\a\, I(1, \D)  \right] \,  ,
\nonumber \eea
where we have introduced $x\equiv\br/\b$ and
\beqa
D_\a\! \(\rho_{\b} \Vert \rhoR\) 
&=& \frac{\pi\,c \,  L }{6}  \frac{1}{\a-1} \left\{ \frac{1}{(1-\a)\,\br+\a\,\beta} - \frac{\a}{\b} - \frac{1-\a}{\br} \right\}
\label{gamble5}\\
&=& \frac{\pi\,c \,  L }{6\,\br} \,   \frac{\a\,(1-x)^2}{(1-\a)\,x+\a}  \,.  
\nonumber
\eeqa
Of course, it is straightforward to see that with $x=1$ (\ie $\br=\b$), the above expression vanishes and eq.~\reef{New1} reduces to the R\'enyi divergence given in eq.~\reef{eq:Dalphareg}.

Now as argued below eq.~\reef{eq:monotonicity}, in principle, we have a two-parameter family of new constraints based on the decrease of $D_\a(\rho\Vert\rhoR)$, \ie we demand that this quantity decreases for all values of $\a$ and $x$. However, it is important to keep in mind that implicitly this argument relies on the fact that we are considering the evolution in a closed system, and in particular, in which any thermal state remains unchanged. That is, the system cannot be in contact with an external heat bath since then a general reference state $\rhoR$ would not be a fixed point of the dynamics.\footnote{Of course, one could pick an external thermal bath for which the inverse temperature matches some particular $\br$, and then the $D_\a(\rho\Vert\rhoR)$ with that precise $\br$ would provide constraints on the evolution, but not for any other value of $\br$.}  For such closed-system dynamics, the usual conservation of energy becomes an important constraint to consider before examining $D_\a(\rho\Vert\rhoR)$. In particular, we see above that the new R\'enyi divergences \reef{gamble5} include a non-vanishing contribution at $O(\l^0)$, which is equivalent to the R\'enyi divergence comparing two purely thermal states, as in eq.~\reef{gamble}. Now let us consider a particular excited state $\rho_1$ evolving towards its equilibrium, and we wish to ask if a second state $\rho_2$ can appear in its evolution.\footnote{Of course, we are considering $\rho_1$ and $\rho_2$ within the class of excited states constructed in eq.~\reef{jumbo}.} If the corresponding temperatures, $\beta_1$ and $\beta_2$, are not equal, then the difference between the R\'enyi divergences appears to be dominated by the $O(\l^0)$ contributions noted above. But if $\beta_1\ne\beta_2$, we already know that the energies of the corresponding thermal states is different, and so we can immediately rule out the transition from $\rho_1$ to $\rho_2$ using energy conservation.

\begin{figure}[ht]
\centering 
\includegraphics[width=.48\textwidth]{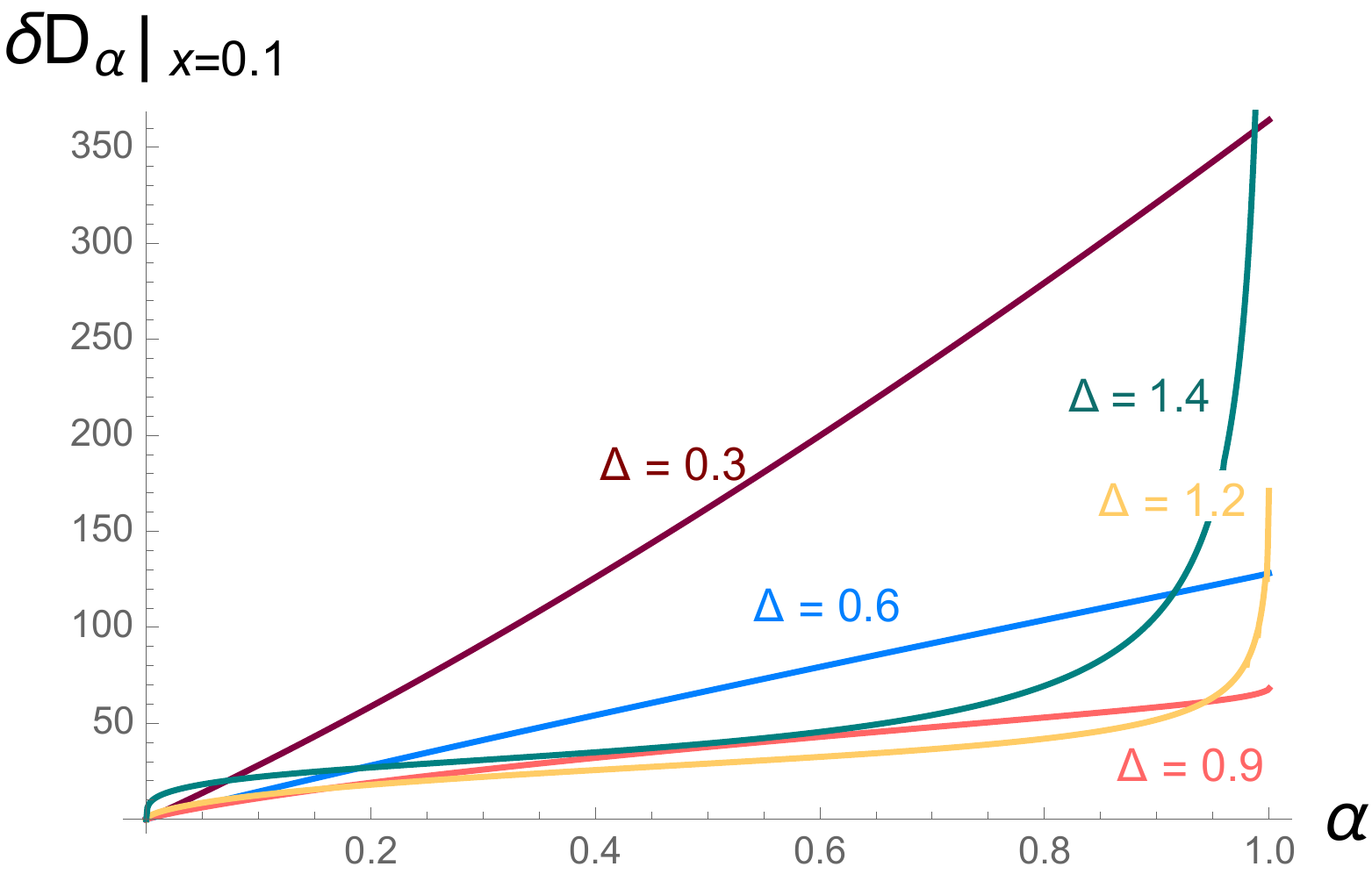}  \hfill \includegraphics[width=.48\textwidth]{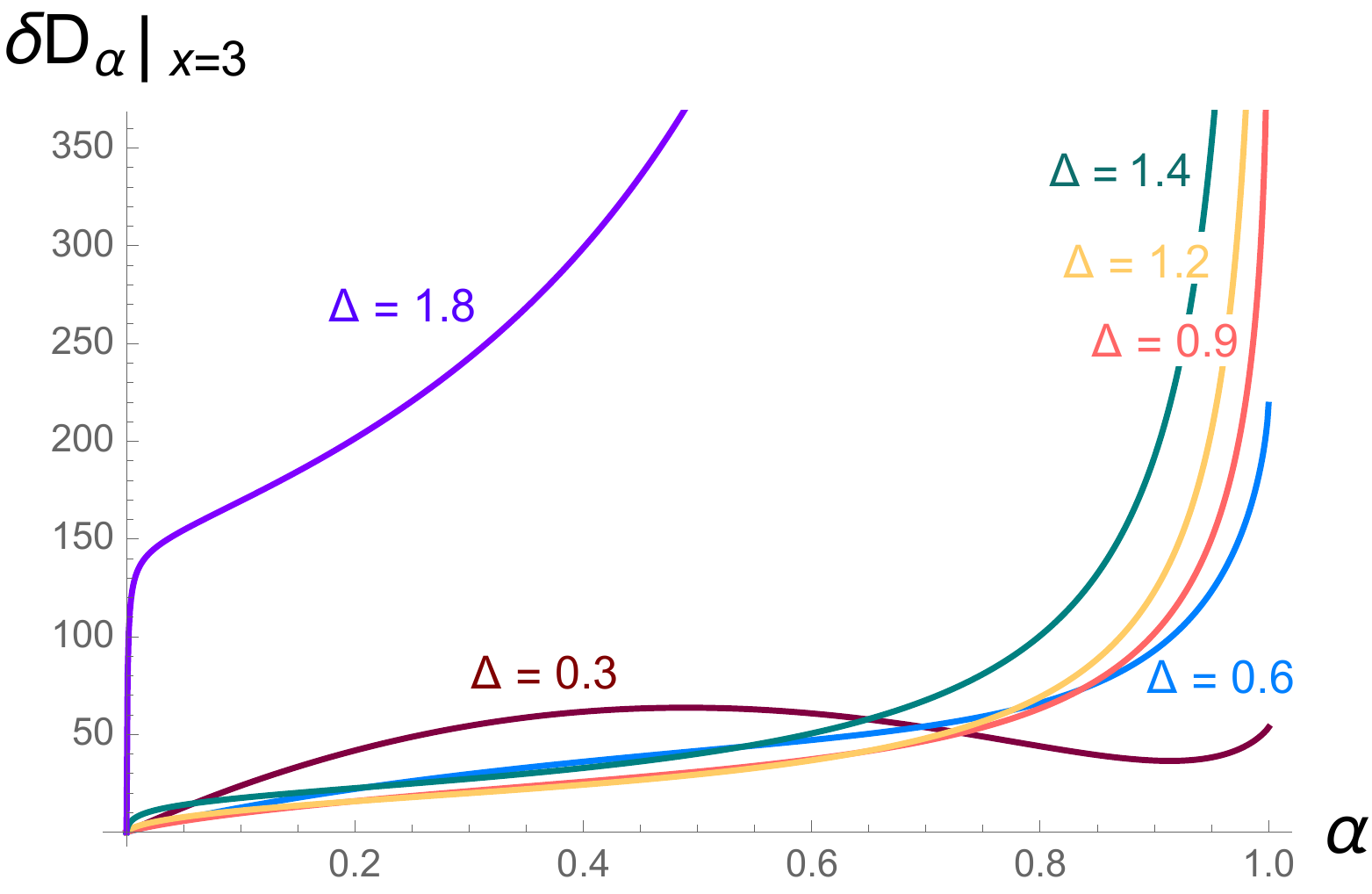} 
\caption{$\d D_\a (\rho  \Vert  \rhoR)= D_\a\! \(\rho \Vert \rhoR\) - D_\a(\rho_\b\Vert \rhoR)$ for $x =0.1$ (left)  and $x=3$ (right). Here $\lambda = \tilde \lambda  / |\D - 1|$ and $\tilde \lambda (2\pi/\b)^{\D-2}$ fixed. In both plots $\D =0.3,0.6,0.9,1.2,1.4$, $\b=2\pi$ and $\tilde \eps = 0.001$. (In the left panel we excluded the curve $\D =1.8$, which has much bigger magnitude than the others.) In all curves we rescaled the prefactor $ \tilde \l^2 \frac{c}{ 3 \pi^2 2^5 } L $.} 
\label{fig:dDax}
\end{figure}

Therefore, for closed-system dynamics, the 
new broader family of constraints provided by $D_\a(\rho\Vert\rhoR)$ can only provide nontrivial constraints on the evolution from $\rho_1$ to $\rho_2$ in the setting of our holographic model when examining excited states with equal or nearly equal temperatures, \ie $\b_2-\b_1\simeq O(\l^2)$, since only in these cases can we match the energies of the two excited states --- see further discussion below. In this case, the difference of the corresponding R\'enyi divergences will be of order $\l^2$ (irrespective of the choice of $\br$), and in comparing two states in our holographic model (with nearly equal temperatures), we can consider the constraints for all values of $\a$ and also for all values of $x=\br/\b$. As an example, in figure~\ref{fig:dDax}, we plot 
\be
\d D_\a (\rho  \Vert  \rhoR)  \equiv D_\a\! \(\rho \Vert \rhoR\) - D_\a(\rho_\b|| \rho_R)\,,
\ee
\ie the $O(\l^2)$ correction in the R\'enyi divergence \eqref{New1}, for excited states all with fixed $\beta$ and fixed $(2\pi/\b)^{\D-2}\l$. Again, this is the contribution that would be relevant in comparing the R\'enyi divergences of excited states with the same $\beta$. However, the two panels show the results for two different reference temperatures, \ie $x=0.1$ (left) and $x=3$ (right). These plots can be compared to the left panel of figure~\ref{fig:Da_adep}, which corresponds to the $x=1$ case. Both of the new graphs show curves for different $\D$ which now cross whereas they did not in figure~\ref{fig:Da_adep} and hence we should expect that with general reference states, the R\'enyi divergences should constrain the dynamics more strongly than if when we only consider $x=1$.

As we noted above, energy conservation plays an essential role in constraining the evolution of excited states when considering closed-system dynamics. Further, the standard second law dictates that the (coarse-grained) entropy must increase. Hence in exploring the generalized $D_\a(\rho\Vert\rhoR)$ constraints (see section \ref{sec:discussion}), we should first consider whether or not these two classical constraints are satisfied. Therefore we discuss here how these two quantities can be extracted from eq.~\reef{New1} for the excited states in our holographic model.

Taking the limit $\a\to1$ of $D_\a(\rho\Vert\rhoR)$ yields the relative entropy \reef{relX}, and as in eq.~\reef{eq:tradsecond} with a thermal reference state with temperature $1/\br$, this becomes
\beqa
D_1 (\rho\Vert  \rho_R) &=& \br \(F(\rho) - F(\rho_R)\)\nonumber\\
&=& \(\br E(\rho) - S(\rho) \)- \(\br E(\rhoR) - S(\rhoR)\)\,.
\label{creep1}
\eeqa
Now the thermal free energies are easily identified using the energy and entropy of the BTZ black brane, \ie
\bea
E(\rho_\b) = \frac{\pi\, c\, L}{6\,\b^2} \,,\qquad
S(\rho_\b) = \frac{A}{4 G_N} = \frac{\pi\, c\, L}{3\, \b}\,. 
\eea
For example, using these expressions, we can evaluate
\bea
D_1(\rho_\b\Vert \rhoR) &=& \(\br E(\rho_\b) - S(\rho_\b) \)- \(\br E(\rhoR) - S(\rhoR)\)\nonumber\\ 
&=&\frac{\pi\, c\, L}{6} \(\frac{\br}{\b^2}-\frac{2}\b + \frac{1}{\br}\) \,\label{creep2}
\eea
and verify that this indeed matches the $\a\to1$ limit of the expression in eq.~\reef{gamble5}. More generally then, we can extract the energy and entropy of our excited states $\rho$ by taking the $\a\to 1$ limit in eq.~\reef{New1}, which yields
\beq
D_1(\rho \Vert \rho_R) = D_1(\rho_\b\Vert \rho_R)  + \l^2 \frac{c\,L}{3 \pi\b} \frac{(\D-1)^3}{2^{\D+3}} \(\frac{2\pi}{\beta}\)^{2(\D-2)}\left\{ \frac{2\D-1}{2(\D-1)}\,x-1\right\} I(1, \D)
\label{creep3} 
\eeq
where we used eq.~\eqref{eq:Iprime1main}, and $D_1(\rho_\b\Vert \rhoR)$ is given above in eq.~\reef{creep2}.\footnote{As discussed at the end of section \ref{sec:RenyiDiv}, we are implicitly assuming that $\Delta<1$ here. Otherwise, a UV divergent term proportional to $x\,\tilde \eps^{2(1-\D)}$ appears in eq.~\reef{creep3}, indicating that the energy of the states with $\Delta\ge1$ diverges in the limit $\tilde\eps\to0$. \label{footyABC}} Above, we identified the term proportional to $1/\br$ in $D_1(\rho_\b\Vert \rhoR)$ as the free energy contribution of the reference state. Now the energy is given by collecting the terms proportional to $\br$ in eq.~\reef{creep2} and to $x=\br/\b$ in eq.~\reef{creep3}, which yields
\beq
E(\rho) = \frac{\pi\, c\, L}{6\,\b^2}\left\{1 +  \l^2\,\frac{2\D-1}{ \pi^2} \frac{(\D-1)^2}{2^{\D+3}} \(\frac{2\pi}{\beta}\)^{2(\D-2)} I(1, \D)\right\}\,.
\label{creep4}
\eeq
Similarly, collecting the terms independent of $\br$ and $x$ gives the entropy,
\beq
S(\rho) = \frac{\pi\, c\, L}{3\,\b}\left\{1 + \frac{\l^2}{ \pi^2} \frac{(\D-1)^3}{2^{\D+3}} \(\frac{2\pi}{\beta}\)^{2(\D-2)} I(1, \D)
\right\}\,.
\label{creep5}
\eeq
We will make use of these expressions in section \ref{sec:discussion} when we explore the generalized constraints provided by $D_\a(\rho\Vert\rhoR)$. In particular, in considering a potential transition from $\rho_1\to\rho_2$, the first step will be to ensure that energy conservation and the traditional second law are satisfied, \ie $E(\rho_2)= 
E(\rho_1)$ and $S(\rho_2)\ge S(\rho_1)$.

We can also extend the discussion introduced at the beginning of this section of using the R\'enyi divergence to examine the utility of different states as a thermodynamical resource. That is, we can evaluate the work function in eq.~\reef{eq:w-alpha} but now with a new general reference state $\rhoR$. Recall that our physical interpretation of $W_{\alpha,\beta_R}$ is that it provides a contraint on how the equilibration of $\rho$ can be used to induce a transition on an ancilla $\rho_a$. This constraint holds not only if $\rho_a$ is itself in contact with another heat bath at inverse temperature $\beta_R$, but also for all values of $\beta_R$ if the ancilla is not in contact with a heat bath.  

In this case, the additional ingredient needed in eq.~\reef{eq:w-alpha} is $D_\alpha  (\rho_{\b'}  \Vert  \rhoR)$, where $\rho_{\b'}$ is the final equilibrium state reached by our excited state. As this R\'enyi divergence is again comparing two thermal states, it takes the form appearing in eq.~\reef{gamble5}. Given eq.~\reef{creep4} for the energy of the excited state, we can determine the final temperature by equating
$E(\rho)=E(\rho_{\b'})=\pi cL/(6\b'^2)$, which yields
\beq
\frac1{\b'} = \frac1\b\left\{1 +  \frac{2\D-1}{2 \pi^2} \frac{(\D-1)^2}{2^{\D+3}} \(\frac{2\pi}{\beta}\)^{2(\D-2)}\l^2\,  I(1, \D)\right\}\,.
\label{creep6}
\eeq
Using this equilibrium temperature and eq.~\reef{gamble5}, we find
\bea
\br\,\Work &=&\frac{\l^2}{\alpha-1}\frac{c\,L}{6 \pi\b}  \frac{(\D-1)^2}{ 2^{\D+3}} \(\frac{2\pi}{\beta}\)^{2(\D-2)} \label{relaX}\\
&&\quad\times\ \left\{         \frac{ I\!\(\frac{\a}{(1-\a) x+\a}, \D\)}{((1-\a) x+\a)^{2\D-3}}
-     \a \,  I(1, \D)  \( \D+\frac12 -\frac{\D-\frac12}{((1-\a)x+\a)^2} \) \right\}\,.
\nonumber
\eea
Recall that $x=\br/\b$. In this case, the $O(\l^0)$ term in eq.~\reef{New1} has been canceled by the same term which appears in $D_\alpha  (\rho_{\b'}  \Vert  \rhoR)$, and as we see above, the resulting work function is $O(\l^2)$ irrespective of the choice of the reference state $\rhoR$. Hence in comparing different excited states, $\rho_1$ and $\rho_2$, for their usefulness as a thermodynamic resource, it seems that we can make interesting comparisons even when $\beta_1\ne\beta_2$. In this case, we can interpret $D_\a(\rho_{\beta'}\Vert\rho_R)$ as accounting for how useful a resource the equilibrium state $\rho_{\beta'}$ is. The fact that we subtract it off in the expression for $\Work$ reflects the fact that we are only inducing the transition in the ancilla during the equilibriation process, and once the state has reached equilibrium, we no longer use it as a resource.

\section{Discussion} \label{sec:discussion}

\subsection*{Path integrals and R\'enyi divergences}

With the path integral approach for evaluating R\'enyi divergences introduced in section \ref{PIapproach}, we have taken the first step towards studying quantum thermodynamics in quantum field theory. 
Our construction considers a special class of excited states \reef{jumbo} in a CFT, which are prepared with Euclidean path integral by turning on a  coupling  $\lambda$ for a relevant operator $\O$ of conformal dimension $\Delta$. In many respects, the resulting partition function \reef{jumbo2} resembles a global quantum quench to a CFT, where, however, we are working in Euclidean signature. In physical processes in which the system achieves equilibrium, the R\'enyi divergences \eqref{eq:RD} provide an ordering of these states \cite{brandao2013second}. That is, given an initial state settling into the equilibrium Gibbs state, we can use the R\'enyi divergence to decide whether or not a third state may participate in this process, \ie whether the system can pass through this third state as it evolves towards its final equilibrium.
As described in section \ref{intro:thermo}, this ordering  provides an extension of the standard thermodynamics rule which demands only that the free energy of the system must decrease as it evolves towards thermal equilibrium. In section \ref{sec:closed}, we also discussed the interpretation of another quantity $\work(\rho)$, given in eq.~\eqref{eq:w-alpha}, as indicating how valuable a state can be as a thermodynamical resource. Further, in the context of our present perturbative calculations, we showed that $\beta \, \work(\rho)\approx D_\alpha(\rho\Vert \rho_\beta)$, \ie from eqs.~\reef{gamble2} and \reef{gamble}, we deduced that the difference is $O(\l^4)$.

As described above, our approach pertains to a very specialized class of excited CFT states, and one future direction would be to generalize this construction. One simple extension would be to consider sources $\lambda(\vec x)$ with a nontrivial spatial profile.
Certainly by introducing a much more complicated (but local) Hamiltonian (including both spatial and time dependence) on part of the thermal circle, we can produce a path integral representation of much more general states. However, identifying the correct Hamiltonian to produce a desired $\rho^\alpha$ would be very challenging. 

In the preceding, we were considering preparing a state (or a power of the density matrix) by Euclidean evolution with conventional local Hamiltonians. More generally, if we are given a particular state $\rho$, we might consider the entanglement Hamiltonian $H'=-\log\rho$, which is expected to be nonlocal for most states of interest. Further, we should expect that identifying $H'$ is another very challenging problem. However, given the entanglement Hamiltonian, we are tempted to formally write the following
\be
\rho^\alpha= e^{-\alpha H'}=\int{\mathcal D}\phi \ e^{- S'_{\rm E} [\phi]}\,.
\label{woggle}
\ee
That is, we would like to express the `Euclidean evolution' by $H'$ in terms of a Euclidean path integral (with appropriate boundary conditions) weighted by a corresponding `entanglement action.'
Of course, working with a conventional local Hamiltonian, the construction of the Euclidean path integral is straightforward. However, as noted above, the entanglement Hamiltonian will typically not be a local operator and so we should certainly not expect the corresponding $S'_{\rm E} [\phi]$ to take a conventional form. But further, beyond the special cases where $H'$ is local (as in our example in eq.~\reef{jumbo}), it is not immediately clear that the path integral in eq.~\reef{woggle} has a meaningful definition. 

In any event, our ultimate goal here should be to find a broader approach and/or new approaches which allow us to efficiently evaluate R\'enyi divergences for more general classes of states and in more general quantum field theories, as well as for a range of $\a$ extending beyond $0\le\a\le1$. One path we plan to explore further is to generalize the replica method for relative entropy \cite{Lashkari:2014yva,Lashkari:2015dia,Ruggiero:2016khg} to the case of R\'enyi divergence. Perhaps, also the techniques developed in \cite{Nozaki:2014hna,Asplund:2014coa,Caputa:2014eta,Marolf:2017kvq} may be of some use in this program. 

\subsection*{Holography and R\'enyi Divergences}
 
Our primary interest was to apply the new techniques in section \ref{PIapproach} in the context of the AdS/CFT correspondence, which relates our computation of the R\'enyi divergences \eqref{eq:RD} in the boundary CFT to a gravitational calculation in the dual AdS space. As usual, the equilibrium Gibbs state in the $d$-dimensional boundary CFT is equivalent to a (static) black hole in a ($d$+1)-dimensional AdS spacetime, as described in section~\ref{sec:quench}. 
Further, according to the holographic dictionary, turning on the source in the boundary theory excites the dual scalar field in the bulk gravitational theory. Thus we can think of the excited CFT states as being dual to a black hole surrounded by a cloud of scalar hair. The natural (Lorentzian) evolution of the latter gravitational configuration will be that the cloud of scalar field collapses and is absorbed by the event horizon when we remove the source (\ie modify the boundary conditions for the scalar) at $t=0$. After a long time then, we expect that the system will settle down to a (static) black hole with a slightly higher mass (and temperature).

While we did not explicitly study this evolution of the gravitational system, we know that the R\'enyi divergences constrain the equilibration process in the boundary theory \cite{brandao2013second}. Hence one is lead to ask a number questions: What do the constraints which quantum thermodynamics imposes in the boundary CFT correspond to in the bulk gravity theory? In particular, do these additional second laws correspond to additional macroscopic laws for black hole evolution? The Bekenstein-Hawking (BH) formula $S_\mt{BH}=A/4G_N$ \cite{JB1,Bekenstein:1973ur,Hawking:1974rv,Hawking:1974sw} provides a translation of the conventional second law, which tells us that entropy always increases, to Hawking's area increase theorem \cite{area,Hawking:1973uf}, which says that in any classical processes the area of the event horizon must always increase. Therefore one might expect that the increase of the R\'enyi entropies in an equilibration process (at fixed energy) 
will have a translation as new macroscopic laws of black hole evolution. 

However, in considering these questions, we must recognize that the BH formula implies that each black hole configuration corresponds to approximately $\exp[S_\mt{BH}]$ microstates. That is, in the full theory of quantum gravity, there are $\exp[S_\mt{BH}]$ microstates which produce essentially the same macroscopic black hole geometry. We certainly expect that the new R\'enyi divergence constraints will limit the evolution of black holes at the level of these microstates. But we must emphasize that the question at hand is whether such constraints will also translate to new second laws restricting the evolution of macroscopic observables which characterize the gravitational solutions. That is, whether in their evolution, features of the spacetime geometry and matter fields are subject to new constraints, which can be traced back to the R\'enyi divergences, in the same way that the area increase theorem can be connected to the second law of thermodynamics for the black holes. 

In fact, interpreting our results presented in section \ref{sec:RenyiDiv} through the AdS/CFT correspondence goes some way towards an affirmative answer to this question, as we now discuss. The only reservation is that if we begin with one of our excited states as the initial data for the bulk equations of motion, we do really not expect the other excited states which we are able to construct to be representative of subsequent configurations appearing in the evolution of the gravitational system. However, we can still examine the constraints implied by the R\'enyi divergences calculated in section \ref{sec:RenyiDiv}.

At this point, let us also remind the reader that while our construction introduces an equilibrium state with inverse temperature $\beta$, the excited state will in fact settle down to a final equilibrium with a slightly different inverse temperature $\beta'$, as given in eq.~\reef{creep6}. In the dual gravitational description, this change arises because absorbing the cloud of scalar field excitations increases the mass and temperature of the black hole. However, a few points are worth mentioning here: First, as established with eq.~\reef{eq:monotonicity}, when $\rho$ represents the state of the entire system the monotonicity of the R\'enyi divergence can hold with any equilibrium state, \ie we do not need to consider the  reference state to be precisely the final state emerging from the equilibration of our system.\footnote{We return to examining this possibility in more detail below.} Notwithstanding this observation, the R\'enyi divergences have a more natural interpretation in terms of the physical equilibration process when the reference state is chosen to be the final equilibrium state, since then, $D_\alpha(\rho\Vert \rho_{\beta'})=\beta'\,W_{\alpha,\beta'}$. The monotonicity of this R\'enyi divergence also holds when $\rho$ represents a coarse-grained description of the system, provided the dynamics is contractive. Further, for the particular example which we are considering here, it is straightforward to show that the change in temperature only effects the final R\'enyi divergence \reef{eq:Dalphareg} at higher orders in our perturbative expansion, \ie at $O(\l^4)$. This result relies on the fact that the change in the inverse temperature \reef{gamble2} is $O(\l^2)$ and appropriately rescaling $\beta$ and $\alpha$ in eqs.~(\ref{oneX})-(\ref{threeX}) for the various ingredients which go into the holographic R\'enyi divergence \reef{eq:Dalphareg}. 

If we consider a situation where there is a single scalar field in the bulk, we can only consider states with different values of $\lambda$, since $\Delta$ is fixed by the mass of the scalar field, as described in section \ref{sec:setup}. In this case, the R\'enyi constraints do not provide any insights on the evolution of the black hole which could not be deduced from the ordinary second law.\footnote{Recall that we refer to the decrease of the free energy and the second law in thermodynamic processes interchangeably. In a process where energy is conserved, the decreasing of the free energy corresponds to increasing the entropy, \ie the second law of thermodynamics for closed systems. Given an initial and final state, the second law places a constraint on what other states our system might evolve into at intermediate times. The R{\'e}nyi divergences have a similar interpretation. On the other hand, the difference in free energy between an initial and final state also determines how valuable a state is, in terms of how much work can be extracted during equilibration. Likewise in section \ref{sec:closed}, a similar interpretation exists in terms of the R{\'e}nyi divergences and how valuable a state is as a resource for driving other processes --- see the discussion in the next paragraph.  \label{footy} }  Recall that our holographic calculations were perturbative in the amplitude $\lambda$ of the scalar field. The only constraint, as can be seen from eq.~\eqref{eq:Dalphareg}, is that $\lambda$ can only decrease in the evolution of the gravitational system. That is, if we begin with the state where the amplitude is $\lambda_0$, then the only states (from our class of excited states) which may appear in the subsequent evolution of the system are those with $\lambda\le\lambda_0$. Of course, this constraint is entirely intuitive, \ie we expect that the amplitude of the scalar hair around the black hole can not increase during the collapse of the scalar cloud into the black hole. Further, since $D_\alpha(\rho\Vert\gibbs)\propto\lambda^2$ in our perturbative calculations, comparing the free energies, \ie the R\'enyi divergences with $\alpha=1$, of the states with different values of $\lambda$ would be sufficient to arrive at this conclusion. Therefore having the full family of R\'enyi divergences provides no additional insight into the evolution of the gravitational system. 

The situation changes if we consider a gravitational theory with more than one scalar field. As example, let us illustrate this situation by considering the case where there are two relevant operators in the boundary theory with $\D_1 =0.9$ and $\D_2=0.6$. These would be dual to two scalars, $\Phi_1$ and $\Phi_2$, in the bulk gravitational theory with masses $m_1^2=-0.99$ and $m_2^2=-0.84$, respectively. If the two scalars are free fields as in eq.~\reef{eq:action}, then they would evolve independently in the collapse of the cloud of scalar hair, \ie excitations in one field will not evolve to generate new excitations in the other. Therefore let us consider an interacting theory where the bulk action \reef{eq:action} is supplemented with the following scalar potential
\beq
U(\Phi_1,\Phi_2)=\frac{g}2\left( \Phi_1\Phi_2^2+\Phi_2\Phi_1^2\right)\,.
\label{pot3}
\eeq
To leading order in our perturbative expansion, this potential will not change the equations of motion for the individual scalars, however, it will produce interactions in the subsequent evolution of the excited states. 
That is, a state in which only $\Phi_1$ is originally excited will evolve to a state where both scalars are excited or possibly where only $\Phi_2$ is excited --- or vice versa. 

\begin{figure}[th]
\centering 
\includegraphics[width=.48\textwidth]{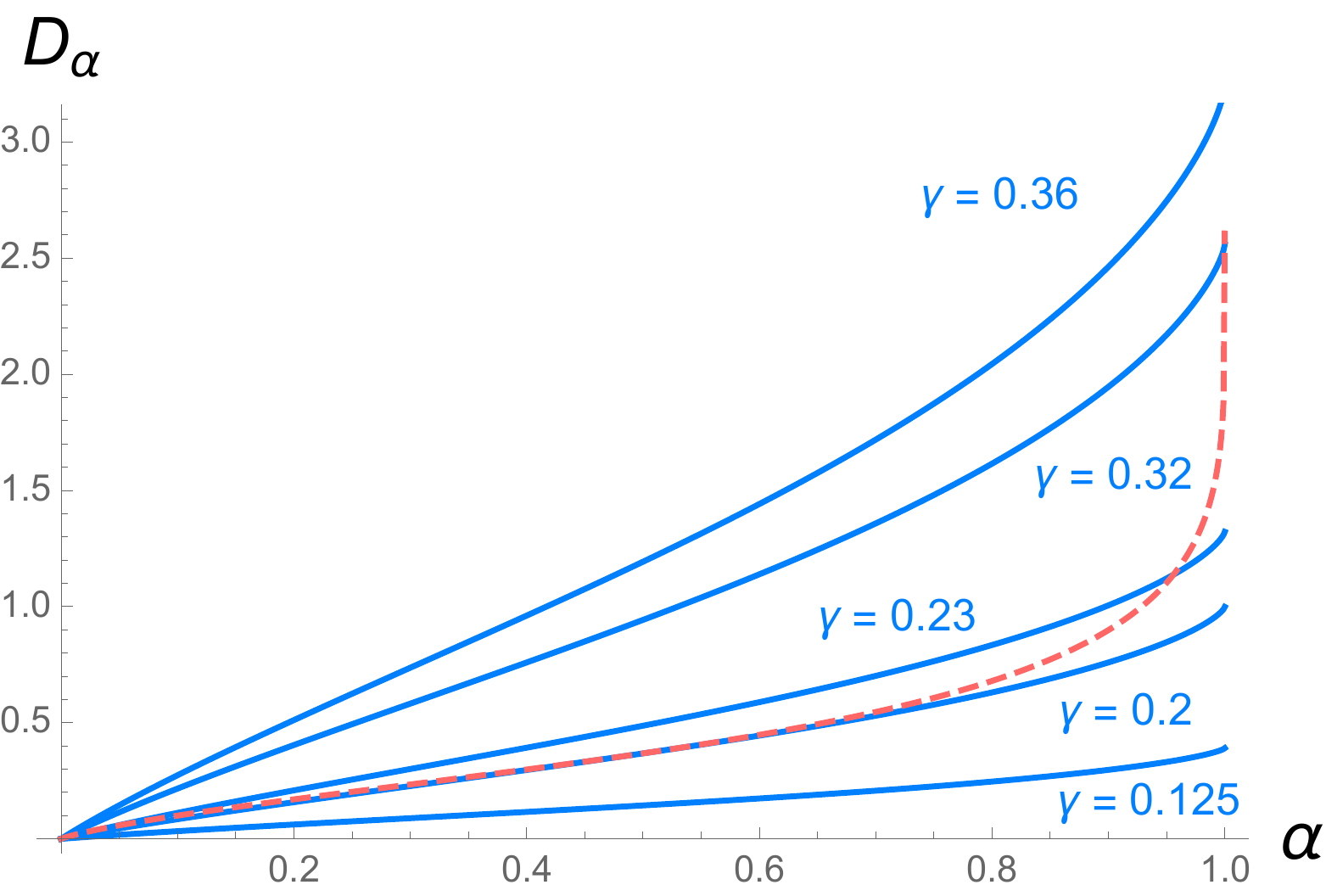} \hfill \includegraphics[width=.48\textwidth]{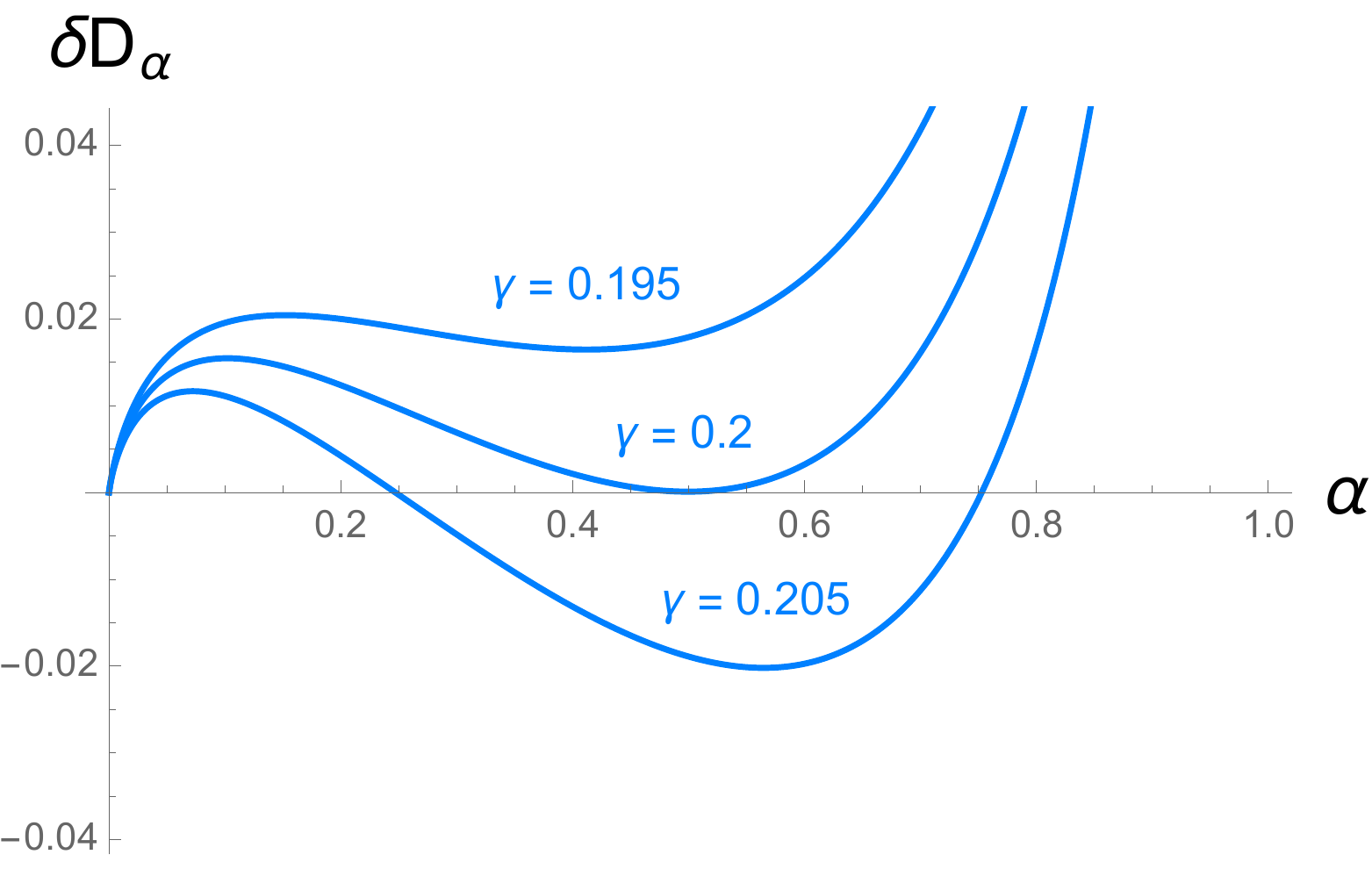} 
\caption{(Left) $D_\a$ from perturbative holographic calculations for $\D_1 =0.9$ (dashed curve) and $\D_2 =0.6$ (solid curves). The different $\D_2=0.6$ curves correspond to different values of the ratio $\gamma$ defined in eq.~\reef{ratio}. In particular, we show  $\gamma= 0.125, 0.2, 0.23, 0.32, 0.36$ from the lowest to the highest. (Right) $\delta D_\alpha \equiv D_\alpha(\rho_1 \Vert\gibbs) - D_\alpha(\rho_2 \Vert\gibbs)$ for values of $\gamma$ around $\g=0.2$. Since the difference is positive for $\g\lesssim0.2$, the transition $\rho_1\to \rho_2$ may be allowed in this regime. In both plots, we rescaled the vertical axis by a factor $\l_1^2 \frac{c L}{ 48 \pi  } \(\frac{2\pi}{\b}\)^{2(\D_1-2)} $.
} 
\label{fig:new}
\end{figure}

Thus we might ask whether a particular state $\rho_2$ with only the $\Delta_2=0.6$ excitation can arise in the evolution of an initial state
$\rho_1$ where the $\Delta_1=0.9$ scalar is excited.\footnote{Let us note here that within our perturbative framework, we can also consider states in which both scalars are excited and eq.~\reef{eq:trSren} would simply extend to 
$\log \tr  \big(\rho^{\alpha} \rho_\beta^{1-\alpha}\big) \approx - S_{\rm ren}(\Phi_1)- S_{\rm ren}(\Phi_2)$. 
We can then examine the R\'enyi divergences of such mixed states to determine whether or not they are allowed to appear in the evolution from the initial state.} Let us begin by comparing the standard free energies (\ie the R\'enyi divergences at $\a=1$). Whether or not this transition is possible will depend on the relative amplitude of the scalars --- see figure \ref{fig:new}. In particular, we should consider the ratio of the (dimensionless) expansion parameter
$({2\pi}/{\b})^{\D-2}\,\lambda$ for the two states in question, \ie
\beq
\gamma\equiv \frac{\big(\frac{2\pi}{\b}\big)^{\D_2-2}\,\lambda_2}{\big(\frac{2\pi}{\b}\big)^{\D_1-2}\,\lambda_1}\,.
\label{ratio}
\eeq 
Then focusing on $\alpha=1$, we see in figure \ref{fig:new} that $D_1(\rho_1\Vert\gibbs)\ge D_1(\rho_2\Vert\gibbs)$ for $\gamma\lesssim 0.32$. Hence the standard second law suggests that the transition $\rho_1\to\rho_2$ is ruled out for $\gamma> 0.32$ but appears possible for smaller values of $\gamma$. However, if we examine the constraints imposed by $D_\alpha(\rho\Vert\gibbs)$ from the full range of $\alpha$, we see that the R\'enyi divergences provide stronger constraints. In particular,  with $\gamma=0.23$ which is well within the allowed regime above, we see in the figure that  $D_\alpha(\rho_2\Vert\gibbs)\ge D_\alpha(\rho_1\Vert\gibbs)$ for 
$0.04 \lesssim \alpha\lesssim 0.96$ and therefore such a transition is actually ruled out. That is, if we consider $\gamma=0.23$, then the excited state $\rho_2$ with the scalar $\Phi_2$ excited can not appear as the initial state $\rho_1$ where $\Phi_1$ was excited evolves towards the final equilibrium black hole.  Hence these additional second laws provide tighter constraints on which equilibration processes will be ruled out. 
In particular, the figure shows that we must have $\gamma\lesssim 0.2$ in order to ensure that $D_\alpha(\rho_1 \Vert\gibbs)\ge D_\alpha(\rho_2 \Vert\gibbs)$ for all values of $\alpha$ across the full range from 0 to 1, and so the transition could be allowed in this case.\footnote{Recall that we must also have
$D_\alpha(\rho_1\Vert\gibbs)\ge D_\alpha(\rho_2\Vert\gibbs)$ for $\alpha\geq 1$, however, our path integral calculations do not allow us to access these higher values of $\alpha$. There may also be further constrains in the case where $[\rho,\rho_\beta]\neq 0$.} 

In passing, let us also consider the inverse transition  $\rho_2\to\rho_1$. In this case, we see in figure~\ref{fig:new} that the classical free energy, \ie $D_1(\rho\Vert\rho_\b)$, provides the most stringent constraint. In particular, such a transition is ruled out for all $\g \lesssim 0.32$. Therefore, for these transitions, the additional R\'enyi divergence constraints appear to be redundant. \\

At this point, we wish to discuss the evolution process that is governing how $\rho_1$ settles into the final equilibrium state in more detail.
As noted above, the usual derivation \cite{brandao2013second} of the constraints imposed by the R{\'e}nyi divergence \reef{eq:monotonicity} involves a system in contact with a thermal reservoir. Hence, in the present setting, we could imagine that $\rho_1$ is in contact with a external thermal bath with temperature $1/\b'$. Even though the reference state $\rho_\b$ does not quite have the appropriate temperature (\ie $\beta'-\beta\sim O(\l^2)$), eqs.~\reef{gamble2} and \reef{gamble} imply that our R\'enyi divergences are only modified at $O(\l^4)$ if we try to correct for this difference. In fact, this reasoning would allow us to choose any reservoir temperature within $O(\l^2)$ of $1/\b$. However, note that if the reservoir temperature is not $1/\b'$, there would be a net energy transfer from the thermal bath and the CFT, which need to be accounted for in the holographic model. In any event, we can conclude that our previous result (\ie transitions are ruled out for $\gamma\gtrsim0.2$) holds for such equilibration processes involving an external bath. On the other hand, we could also consider an evolution without any such external bath, as this is a more natural setup for our holographic model. As pointed out in the discussion of closed-system dynamics in \ref{sec:closed}, the first step is to match the energies of the two excited states under consideration. This matching is easily accomplished by making a small $O(\l^2)$ shift in the temperature $1/\b_2$ of the state with $\D=0.6$ excitations --- see discussion around eq.~\reef{shift3} below. However, $D_\a(\rho_2\Vert\rho_\b)$ is again only modified at $O(\l^4)$ with this shift and hence our constraints also apply when $\rho_1$ thermalizes as a closed system. However, as discussed in section \ref{sec:Xeno},  the R\'enyi divergences with general reference states $\rhoR$ may provide even stronger constraints on the evolution for closed-system dynamics and we examine this possibility in detail below.

In any event, with the simple example above, we see that the R\'enyi divergences can in principle impose new constraints on the collapse of the scalar fields and the evolution of the black hole,\footnote{One may worry that with our perturbative calculations, we are really only constraining the evolution of the matter fields in the bulk. However, the AdS/CFT correspondence puts excitations of the gravitational (\ie metric) degrees of freedom and the matter fields on essentially the same footing. Hence we can easily imagine extending the present class of excited states to states where the metric is also deformed.} which would not be seen if we only considered standard thermodynamic properties, \ie the free energies, of the corresponding gravitational configurations. 

\subsection*{Holography and thermodynamical resources}

Here we turn briefly to our discussion in section \ref{sec:closed} of applying the R\'enyi divergences to compare different states and determine which is a better thermodynamical resource. Here, we might consider a holographic theory with two different scalars as above, or we might consider two different holographic theories with a single scalar field in the bulk where the scalar masses are distinct.\footnote{Hence, the corresponding boundary operators have different conformal dimensions.}
In either case, we would compare two states, where each contains excitations of either scalar, and examine the strength of the constraints which the work functions \reef{eq:w-alpha} place on the ability of these states to drive transitions of an ancillary system. As described above, we have $\beta \, \work(\rho) \approx D_\alpha(\rho\Vert \rho_\beta)$ with our perturbative calculations. Hence going back to the simple example above  with $\D_1 =0.9$ and $\D_2=0.6$, we have 
\be
\b \, \work(\rho_1) \approx D_\alpha(\rho_1 \Vert\gibbs)\ge \b \, \work(\rho_2) \approx D_\alpha(\rho_2 \Vert\gibbs)
\label{toggle}
\ee
for the entire range $0\leq \alpha\leq 1$ when $\gamma\lesssim 0.2$. Therefore we can conclude that in this range (\ie $\gamma\lesssim 0.2$), there are transitions of an ancilla that the state $\rho_1$ might be able to induce  but which $\rho_2$ cannot. Once again, we are able again to draw physical conclusions for this question which can not be drawn from the ordinary free energy, and so having the full family of R\'enyi divergences also provides new insights in terms of how the equilibration of the holographic states could be used to drive other processes.

\subsection*{General reference states, again}

Recall that we argued $D_\alpha  (\rho  \Vert  \rhoR)$ must decrease in physical processes for any reference state $\rhoR$ that is a fixed point of the dynamics, \eg we could choose the reference state to be a thermal state with temperature $1/\br$, which is unrelated to $1/\b'$, the temperature of the final equilibrium state into which $\rho$ will settle --- see also \cite{MuellerOppenheim} for a detailed discussion. However, this argument alone does not indicate whether or not these additional constraints are actually useful in constraining the evolution of the system, \ie that these constraints can impose tighter constraints than those found by considering $D_\alpha  (\rho  \Vert  \rho_{\b'})$
alone. However, using our results from section \ref{sec:Xeno}, we will show below that the extended family of constraints does control the evolution more tightly in our holographic model than the original R\'enyi constraints. This is natural, since the additional family of second laws provided by varying the reference state is appropriate when we do not consider coarse-graining, and thus equilibration will be more constrained.

However, before proceeding with explicit calculations, we would like to make some general observations about the new constraints. In particular, since the new family of constraints is indexed by $\br$, it is insightful to consider a number of simple limits. The first limit to consider is taking $\br\to0$, \ie $T_\mt{R}\to \infty$, for which the spectrum of the reference state should become flat. That is, at infinite temperature, all microstates are equally probable, and so $\rhoR=\mathbb{1}/d$. Hence we find $D_\alpha(\rho\Vert\rhoR)\to\log{d}-S_\alpha(\rho)$   and the new constraints become second laws for the R\'enyi entropies, \ie all R\'enyi entropies must increase. Of course, this result mimics the simple model with a trivial Hamilitonian discussed at the end of section \ref{intro:thermo}, however, here we have produced second laws for the R\'enyi entropies in any system, \eg with an arbitrarily strongly coupled Hamiltonian. Further, one can easily see this result in our path integral construction. In particular if $\br$ shrinks to zero in eq.~\reef{jumbo4}, we are only left with the path integral involving the deformed Hamiltonian $H'$ and it produces powers of the corresponding excited state $\rho$. We should add that in a situation with only short-range correlations, when we take the thermodynamic limit, all the R\'enyi entropies are approximately equal to the von Neumann entropy and we would recover the standard second law, \ie these additional second laws yield no new constraints for the evolution. However, we will see below the long-range correlations in our holographic model allow us to evade this conclusion here.

The second simple limit which we consider is to take $\br\to \infty$, \ie $T_\mt{R} \to0$. In this case, the reference state becomes a projection operator onto the ground state. This is easily seen in eq.~\reef{jumbo4} again. In this case, the thermal circle stretches to infinite size, and so focusing in on the small portion of length $\alpha\,\beta$ which prepares the power of excited state $\rho^\alpha$, we lose sight of the fact that the path integral is on a circle. Instead, we can imagine that we have an infinite Euclidean path integral with $H$ from $\tau = -\infty$ to $-\alpha\beta/2$, which then prepares the ground state of the CFT on this final time slice. Similarly, we would have another from $\tau=\alpha\beta/2$ to $\infty$, which is preparing the ground state on the (Euclidean) time slice $\tau=\alpha\beta/2$. Hence the powers of the excited state are then sandwiched between these two ground states, \ie $D_\alpha(\rho\Vert\rhoR)= -\log \langle 0|\rho^\alpha|0\rangle/(1-\alpha)$. Hence in this limit, the new constraint indicates that the probability of $\rho$ being in the ground state can only increase. 

To conclude these general comments, let us add that this new extension of the R\'enyi divergence constraints is reminiscent of the entropy-energy diagrams of closed system dynamics discussed in \cite{alicki_entanglement_2013,sparaciari2017resource}. There, one may also consider the constraints corresponding to the relative entropy distance to the maximally mixed state or to the ground state, which are analogous to the $\br\rightarrow0$ and $\infty$ limits considered above.
The corresponding monotones give both energy and entropy as important thermodynamical resources for closed systems.

Recall that in section \ref{sec:Xeno}, we showed how our path integral approach is easily extended to evaluate $D_\alpha  (\rho  \Vert  \rhoR)$ and then applied this procedure in our holographic model. Here we would like to begin to explore how varying the reference state modifies the R\'enyi divergence constraints imposed in this holographic setting.

As a simple illustration, we return to the example used above in discussing the constraints imposed by $D_\a(\rho\Vert\gibbs)$. In particular, we consider a bulk theory with two scalar fields,  $\Phi_1$ and $\Phi_2$, dual to relevant operators with $\D_1 =0.9$ and $\D_2=0.6$ in the boundary theory. In principle, an additional scalar potential \reef{pot3} will allow for transitions between the two scalar fields and so we wish to examine whether a particular state $\rho_2$ with only $\Phi_2$ excited can arise when an initial state $\rho_1$ with only $\Phi_1$ excited evolves towards equilibrium. Now as discussed in section \ref{sec:Xeno}, to apply the generalized $D_\alpha  (\rho  \Vert  \rhoR)$ constraints, we must be considering closed-system dynamics in which the evolution respects energy conservation. Hence we must first ask if the energy of the two states is the same, \ie we require $E(\rho_1)=E(\rho_2)$ where the energy is determined by the expression in eq.~\reef{creep4},
\beq
E(\rho) = \frac{\pi\, c\, L}{6\,\b^2}\left\{1 +  \l^2\,\frac{2\D-1}{ \pi^2} \frac{(\D-1)^2}{2^{\D+3}} \(\frac{2\pi}{\beta}\)^{2(\D-2)} I(1, \D)\right\}\,.
\label{creep4a}
\eeq
As commented above, a simple way to achieve this equality is to shift the temperature of the second state slightly, \ie we take $1/\beta_1=1/\beta$ and
\beq
\frac{1}{\beta_2}=\frac1{\beta}\left\{1- \l_2^2\,\frac{2\D_2-1}{2 \pi^2} \frac{(\D_2-1)^2}{2^{\D_2+3}}\(\frac{2\pi}{\beta}\)^{2(\D_2-2)}  I(1, \D_2)\,\left( 1- \frac{\sigma}{\gamma^2} \right)
\right\}\,,
\label{shift3}
\eeq
where $\gamma$ is the ratio given in eq.~\reef{ratio} and
\beq
\sigma={2^{\D_2-\D_1}}\,\frac{2\D_1-1}{2\D_2-1}\, \frac{(\D_1-1)^2}{(\D_2-1)^2}\, 
\frac{I(1, \D_1)}{I(1, \D_2)}\,.
\label{shift3a}
\eeq

We note that eq.~\reef{gamble} implies that an $O(\l^2)$ shift in the temperature like this will only modify $D_\a(\rho_2\Vert\gibbs)$ at $O(\l^4)$. Hence the previous analysis of the R\'enyi divergence constraints in this example is not effected by imposing $E(\rho_1)=E(\rho_2)$. However, when a reference state with temperature $1/\br$ is introduced, the $O(\l^2)$ contribution to $D_\alpha  (\rho_2  \Vert  \rhoR)$ is modified by this shift. In particular, combining eq.~\reef{shift3} with eqs.~\reef{New1} and \reef{gamble5} yields
\bea
D_\alpha  (\rho_2  \Vert  \rhoR)  
&\simeq& D_\alpha(\rho_\b\Vert\rhoR)+
\frac{\l_2^2}{\alpha-1} \frac{c\,L}{6\pi\b}  \frac{(\D_2-1)^2}{2^{\D_2+3}}  \(\frac{2\pi}{\beta}\)^{2(\D_2-2)} \left\{  \frac{ I\!\(\frac{\a}{(1-\a) x+\a}, \D_2\)}{((1-\a) x+\a)^{2\D_2-3}}  \right. \nonumber\\
&&\quad\left.+\,
\a\, I(1, \D_2) \left[\frac{2\D_2-1}{2} \left( 1- \frac{\sigma}{\gamma^2} \right)\left(1-\frac{1}{((1-\a)x+\a)^2}\right)-1\right]\right\}
\, .
\label{New2}  
\eea
Recall that $\gamma$ is defined in eq.~\reef{ratio}. Further, $D_\alpha  (\rho_1 \Vert  \rhoR)$ is given by eq.~\reef{New1} with the simple substitution $\Delta\to\Delta_1$. 

\begin{figure}[ht]
\centering 
\includegraphics[width=.48\textwidth]{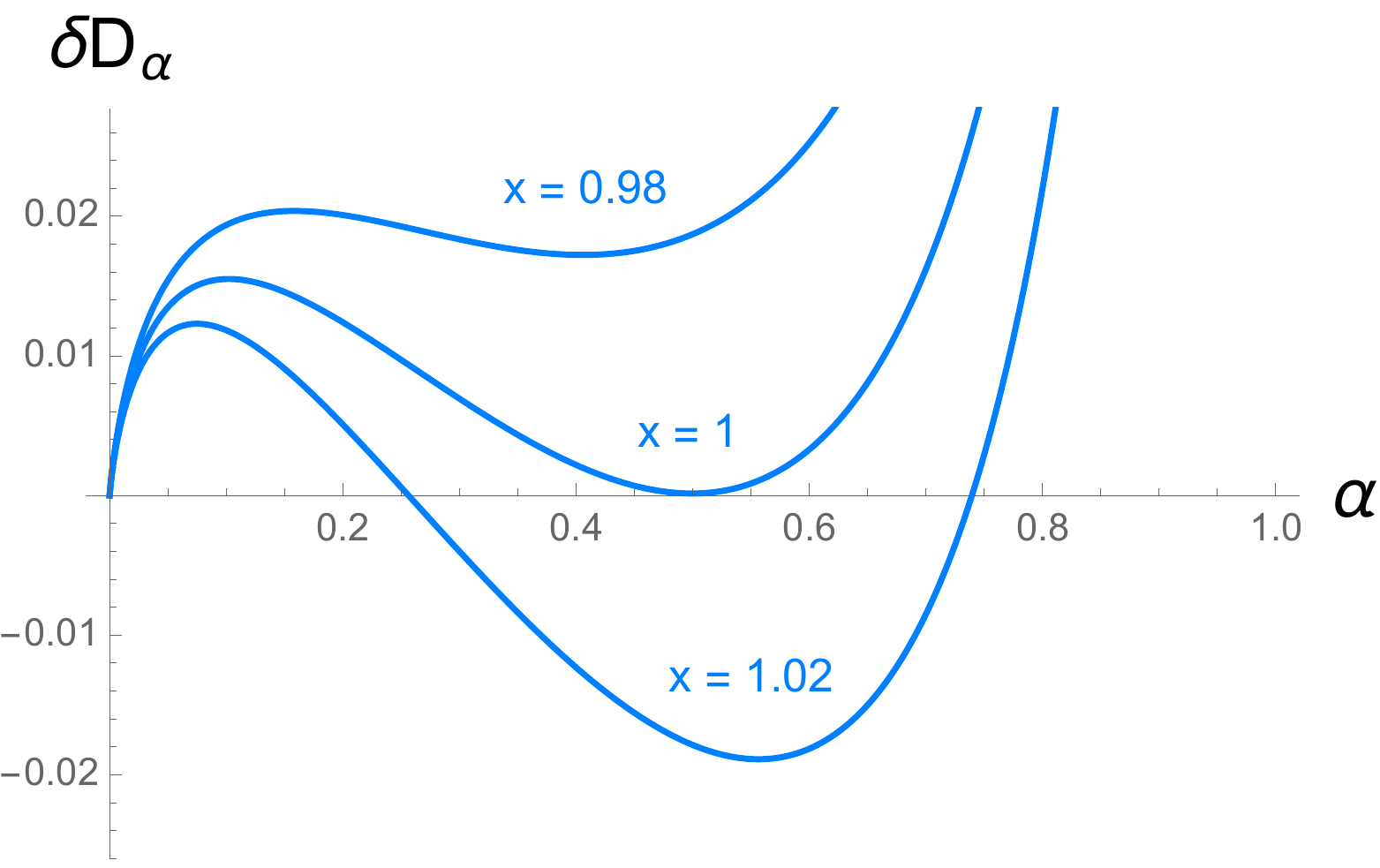} \hfill \includegraphics[width=.48\textwidth]{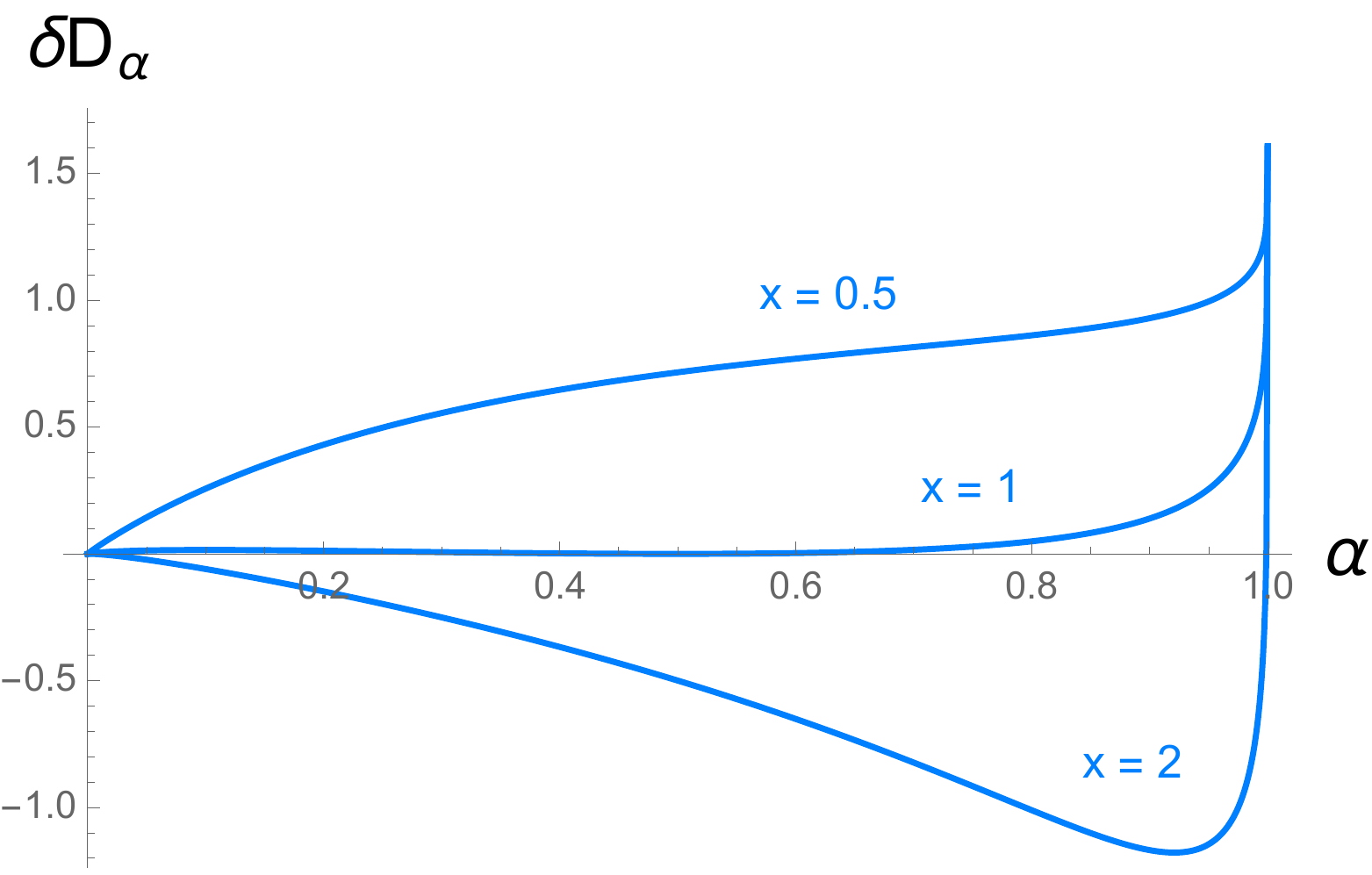} 
\caption{ $\delta D_\a = D_\alpha  (\rho_1 \Vert  \rhoR) - D_\alpha  (\rho_2 \Vert  \rhoR)$ for $\D_1=0.9$ and $\D_2 =0.6$ with varying $x=0.98,1,1.02$ (left) and $x=0.5,1,2$ (right). Here $\g =0.2$ and we rescaled the curves by a factor $\l_1^2  \frac{c L}{ 48 \pi  } \(\frac{2\pi}{\b}\)^{2(\D_1-2)}$.}
\label{fig:new2}
\end{figure}

To leading order, we have $D_\alpha  (\rho_1 \Vert  \rhoR)=D_\alpha  (\rho_\b \Vert  \rhoR)=D_\alpha  (\rho_2 \Vert  \rhoR)$, and hence comparing these generalized R\'enyi divergences, we should focus on the differences in the $O(\l^2)$ contributions. In figure \ref{fig:new2}, we show some examples of $\d D_\a \equiv D_\alpha  (\rho_1 \Vert  \rhoR)-D_\alpha  (\rho_2 \Vert  \rhoR)$ with different values of $x$. 
In particular, we have chosen $\gamma=0.2$ for all of these curves, which was the limiting value for which the transition $\rho_1\to\rho_2$ could be ruled out with the standard R\'enyi divergences in figure \ref{fig:new}, \ie with $\rhoR=\rho_\b$. The new figure shows that with $x< 1$, the constraints are not as strong, however, with $x>1$, we produce tighter constraints than before. In passing, we note that all of the curves in right panel of figure \ref{fig:new2} coincide at $\a\to1$. This agreement can be understood from the discussion in section \ref{sec:Xeno}. In particular, since we matched the energies $E(\rho_2)=E(\rho_1)$, the difference of the R\'enyi divergences at $\alpha=1$ reduces to $S(\rho_2)-S(\rho_1)$ irrespective of the value of $\br$.

\begin{figure}[ht]
\centering 
\includegraphics[width=.8\textwidth]{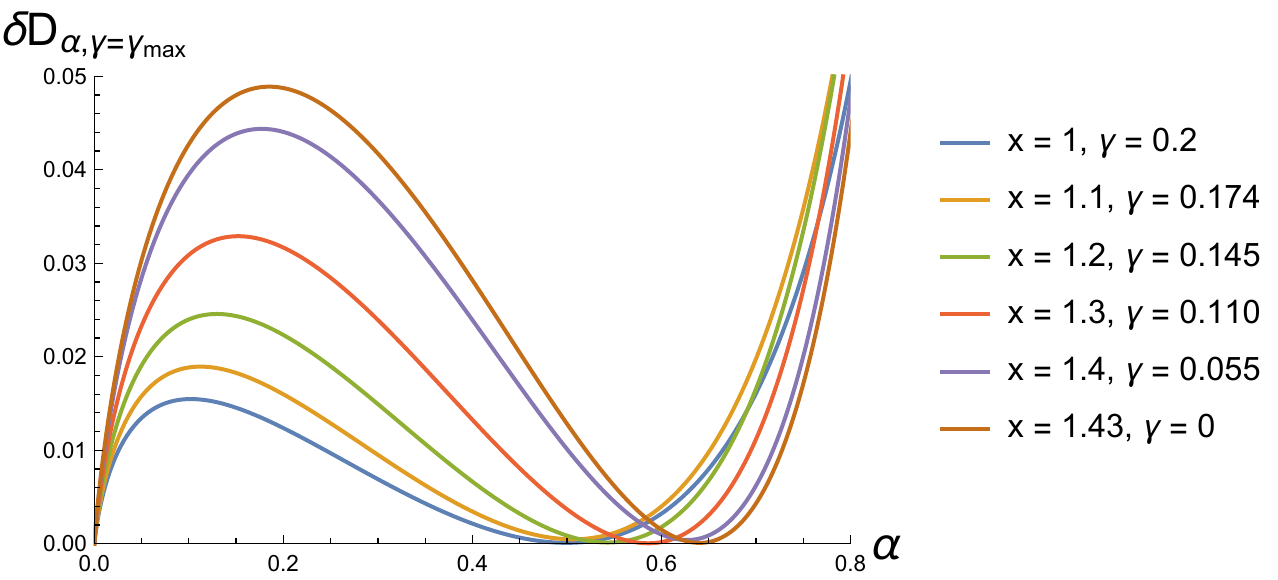}
\caption{$\delta D_\alpha=D_\alpha  (\rho_1 \Vert  \rhoR)-D_\alpha  (\rho_2 \Vert  \rhoR)$ for $\D_1=0.9$ and $\D_2 =0.6$, with varying $x$ and $\g=\gm$. For $x \gtrsim1.43$ there is no $\g$ such that $\delta D_\alpha$ is non-negative in the whole range $0\le\a\le1$, and thus the transition is forbidden. As in the previous plots, we rescaled the curves by a factor $\l_1^2  \frac{c L}{ 48 \pi  } \(\frac{2\pi}{\b}\)^{2(\D_1-2)}$.
} 
\label{fig:new3}
\end{figure}

As noted above, the constraints become tighter when the reference state is chosen with $x>1$, \ie the low reference temperature is decreased, $1/\br<1/\b$. We illustrate this point in figure \ref{fig:new3} where for various values of $x$, we plot the difference $\delta D_\alpha=D_\alpha  (\rho_1 \Vert  \rhoR)-D_\alpha  (\rho_2 \Vert  \rhoR)$ for $\g=\gm$, the maximum value for which $\delta D_\a\ge0$ for all $0\le\a\le1$. In fact, the figure shows the surprising result that we reach $\gm=0$ with $x\simeq1.43\,$! For larger values of $x$, we find that $\delta D_\a$ always dips below zero for some values of $\a<1$ with $\g=0$ (or any other value of $\g$).

Now if we set $\g=0$, we are turning off the source amplitude $\lambda_2$ and $\rho_2$ reduces to a thermal state. Further, since 
we matched $E(\rho_2)=E(\rho_1)$ by shifting the temperature in eq.~\reef{shift3},\footnote{It is straightforward to check that eq.~\reef{shift3} matches eq.~\reef{creep6} in the limit $\g\to0$ (and $\lambda_2\to0)$.} in fact we have $\rho_2=\rho_{\b'}$. That is, our excited state reduces to the thermal state into which we expected $\rho_1$ would evolve. Hence our R\'enyi constraints with $x\gtrsim 1.43$ are indicating that $\rho_1$ will not actually thermalize! We might understand this apparent `paradox' as follows: As we described above, the gravitational description of the evolution certainly involves the system settling down to a (static) black hole with no scalar excitations, which we would naturally interpret in terms of thermalization. However, we are considering the CFT to be a closed system in the current discussion, \ie there is no external thermal bath, and so at a microscopic level, we understand that the system is simply evolving unitarily. Therefore the excited state is not actually thermalizing. By construction, the excited state is a mixed state, \ie a thermal state of the perturbed Hamiltonian  $H'$, and the gravitational description indicates that after unitary evolution for a long time, this microscopic state exhibits properties which are very similar to that of a thermal ensemble, \eg correlation functions would approach thermal correlation functions. From this perspective, our conclusion is that the R\'enyi divergences $D_\alpha(\rho\Vert\rhoR)$ provides a probe which is powerful enough to distinguish the excited state $\rho_1$ from the thermal state $\rho_{\b'}$. An open question would be whether this distinction is accomplished because $D_\alpha(\rho\Vert\rhoR)$ accesses microscopic information about $\rho_1$ or if it can be still be phrased in terms of macroscopic gravitational variables. For example, the scalar field will only vanish after an infinite time and so in principle, measurements with sufficiently high resolution will still detect the decaying scalar field at any finite time. Therefore such very fine measurements of macroscopic observables will distinguish the excited state from the thermal state even at late times. 

It seems that the strongest constraints for these transitions will come from $\br\to\infty$, \ie $T_\mt{R} \to0$. As discussed above, in this limit, the R\'enyi divergences involve a projection of (powers of the excited state) onto the ground state of the system, \ie $D_\alpha(\rho\Vert\rhoR)= -\log \langle 0|\rho^\alpha|0\rangle/(1-\alpha)$. In our holographic model, the ground state is described by the AdS vacuum geometry in the bulk theory. It would be interesting to examine if one can understand this projection and its implications for the bulk dynamics more directly, \ie from a gravitational perspective. In particular, this may shed light on the previous question of whether the generalized R\'enyi constraints and their consequences can be phrased in terms of macroscopic gravitational observables. 

Our results here where the thermalization of certain holographic states is ruled out also bring to mind the possibility of smoothing, discussed at the beginning of section \ref{sec:closed}. In particular, rather than just considering unitary evolution of the excited states, it would be interesting to include some definition of approximate thermalization and examine the effect on the R\'enyi constraints in our holographic model. We hope to pursue these questions in \cite{MuellerOppenheim}.\\

\begin{figure}[ht]
\centering 
\includegraphics[width=.48\textwidth]{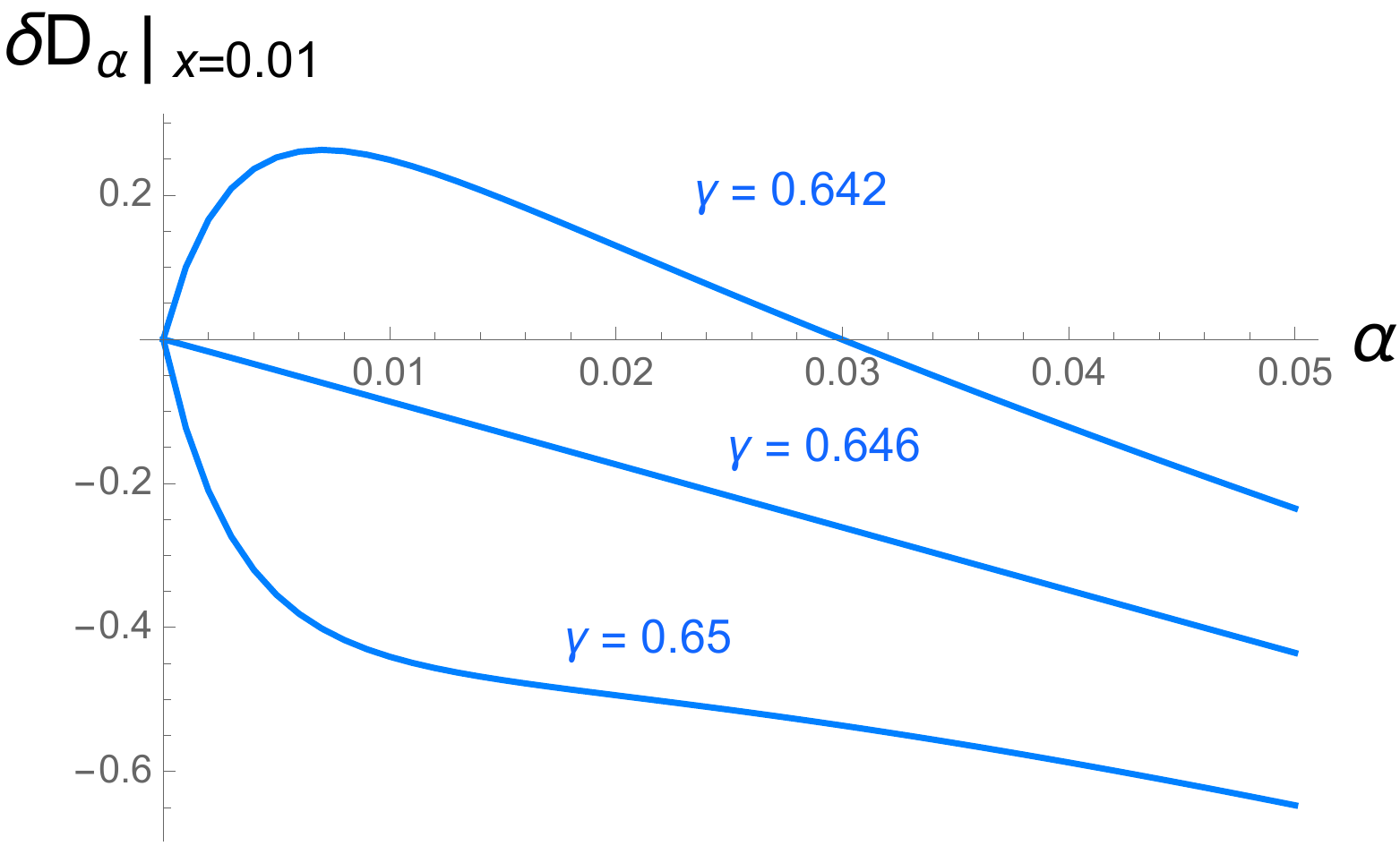} \hfill \includegraphics[width=.48\textwidth]{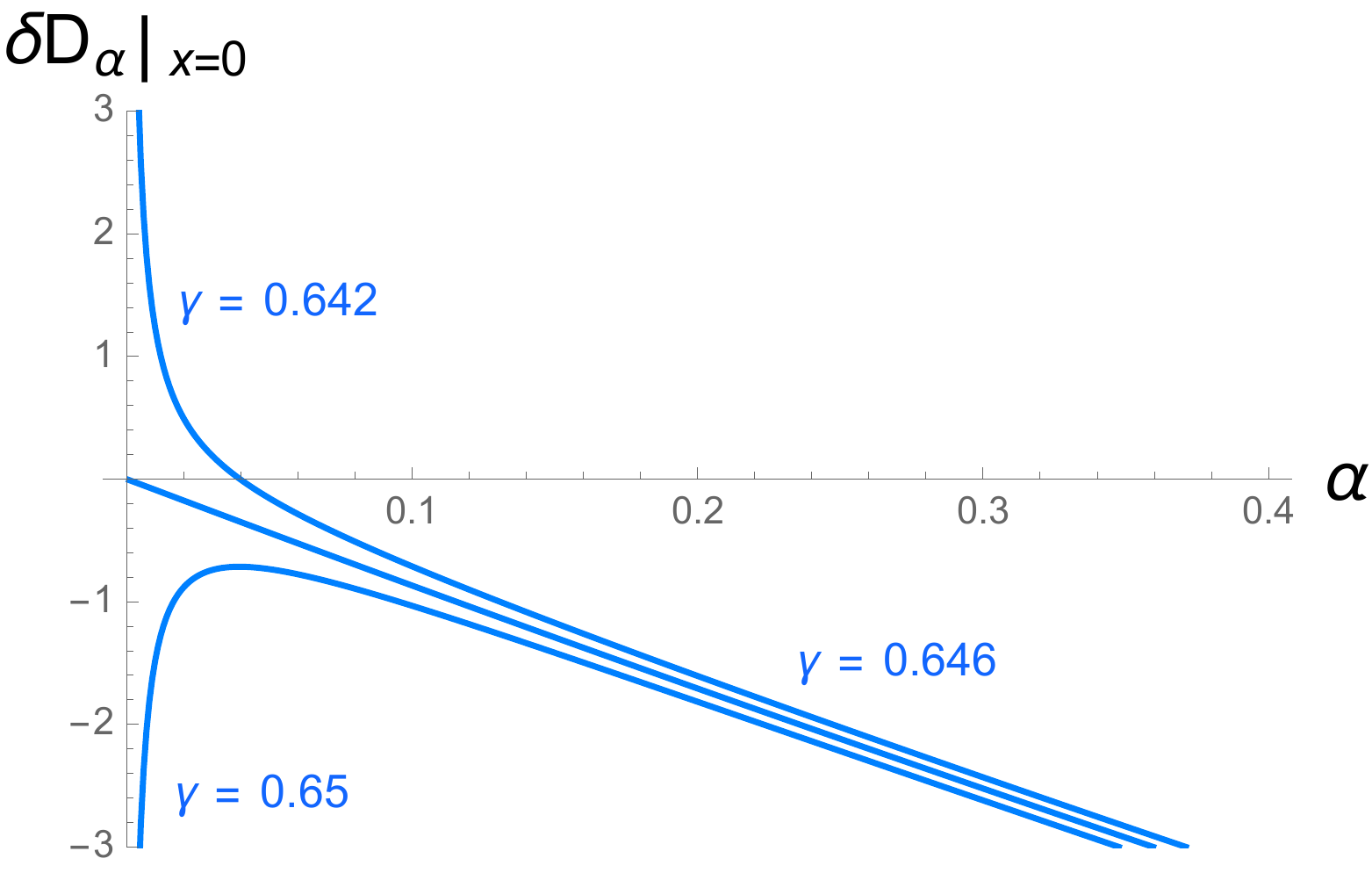}\
\caption{The converse transition $\rho_2\to\rho_1$, here again with $\D_1 =0.9$ and $\D_2=0.6$, is forbidden whenever $\delta D_\a = D_\alpha  (\rho_1 \Vert  \rhoR) - D_\alpha  (\rho_2 \Vert  \rhoR)$ is positive for some $\a$. (Left) The constraints are stronger for $x <1$ and $\a \to 0$. (Right) The limiting case $x =0$, which corresponds to R\'enyi entropies. We see R\'enyi's with $\a <1$ are more constraining than the von Neumann entropy.  All of the curves here are rescaled by $\l_1^2  \frac{c L}{ 48 \pi  } \(\frac{2\pi}{\b}\)^{2(\D_1-2)}$.} 
\label{fig:new4}
\end{figure}

We might also consider the inverse transitions from $\rho_2\to\rho_1$, \ie from a state prepared with the $\D=0.6$ operator to those prepared with $\D=0.9$. In this case with $x=1$, we found that the classical (\ie $\a=1$) constraints were the strongest and that the transition would only be possible for $\g \gtrsim 0.32$, as shown in figure~\ref{fig:new}.  For $x \neq 1$, the transition can now be allowed only if $\delta D_\a = D_\alpha  (\rho_1 \Vert  \rhoR) - D_\alpha  (\rho_2 \Vert  \rhoR)$ is everywhere non-positive, \ie we consider the same difference of R\'enyi divergences as above but ask of the opposite sign. From figure~\ref{fig:new2}, we see that the constraints are more stringent for $x<1$. Further, it turns out the most interesting regime is in the vicinity of $\a=0$. In particular, all of these curves are anchored at $\delta D_\a=0$ for $\a=0$, and the slope at that point is given by
\begin{align}\label{done}
\left.\frac{\del\, \delta D_\a}{\del\a}\right|_{\a=0}
&\ \underset{x \to 0 }{\sim}    \frac{c\,L}{48\pi\b}  \(\frac{2\pi}{\beta}\)^{2(\D_1-2)} \l_1^2 \, \frac{(1- \D_2)^2\,   (2 \D_2-1)\, I(1, \D_2)}{2^{\D_2+1}} \ \frac{  \s-\gamma^2  }{ x^2}\,.
\end{align}
In fact, it is then the sign of this slope which determines whether the transition is allowed or not, \ie the transition is ruled out if the slope is positive. We note that this sign is independent of the precise value of $x$ but rather this sign is controlled by the factor ($\s-\gamma^2$).\footnote{Recall that $\sigma$ is a fixed constant determined by the conformal dimensions, $\D_1$ and $\D_2$, with the expression given in eq.~\reef{shift3a}. In the present case with $\D_1=0.9$ and $\D_2=0.6$, we have $\sigma\simeq 0.417$.} We illustrate this point in the left panel of figure~\ref{fig:new4} with $x =0.01$, where we see that by taking $\g$ large enough the transition can be allowed. However, the constraint here is more stringent than found before, \ie the transition would only be possible with $\g \ge \sqrt \sigma \approx 0.646$. 

As discussed above at $x = 0$, the R\'enyi constraints reduce to constraints on the R\'enyi entropies of the excited states. From the right panel of figure~\ref{fig:new4}, we see the behaviour is more singular at precisely $x=0$. In particular, there is now a pole at $\a\to 0$, with residue proportional to $\s-\gamma^2$. Hence, this same factor controls whether or not the transition is possible or not.
This illustrates the point that at least for certain transitions (\eg $\rho_2\to\rho_1$), the R\'enyi entropies can provide stronger constraints than the usual second law formulated in terms of von Neumann entropy. However, we must add that strictly speaking on our example the pole observed above invalidates our perturbative expansion in that region (\ie we need to confine our considerations to $\alpha$ somewhat larger than zero). Understanding the limit $\a\to0$ and $x\to0$ is therefore  beyond the scope of the perturbative approximation we are considering.  

\subsection*{Lessons for general black holes?}

As discussed above, our holographic model provides explicit examples where the R\'enyi divergences impose new constraints on the evolution of the bulk gravitational system, which would not be seen if we only considered the usual thermodynamic properties of the corresponding bulk configurations. At present, our understanding of the new constraints on the gravitational evolution is rather indirect. In our calculations using the path integral approach, we were able to give a geometric or gravitational description of the R\'enyi divergence using the AdS/CFT correspondence. Let us re-iterate that our present approach only applies for special states in the boundary CFT, and hence for special configurations in the bulk gravity theory. However, the R\'enyi divergences provide a definite ordering of these gravitational configurations, which in principle constrains the evolution towards a final (static) black hole. What we have not done is to articulate these new constraints in a concise manner similar to that in which the area increase theorem encapsulates the standard second law. Therefore we can pose two related questions:\footnote{At the beginning of the discussion, we have already asked the complementary question of how to go beyond the present special states to evaluate the R\'enyi divergences for a general state in the boundary CFT. Certainly advances on this question will provide valuable new tools towards addressing the two gravitational questions above.} The first is how do we frame our calculation of the R\'enyi divergences directly as a computation in the gravitational theory, and in particular whether and how these computations can be extended to general black hole spacetimes, \eg spacetimes which are asymptotically flat. The second question is what is the geometric or gravitational manifestation of the full family of constraints imposed by the decrease of the R\'enyi divergences, \ie what is the analog of the area increase theorem, which is the macroscopic expression of the usual second law of thermodynamics. 

In considering the first of these questions, we might look to holographic entanglement entropy \cite{RT0,RT1} and R\'enyi entropies \cite{Hung1,juan} in the context of the AdS/CFT correspondence for some insight. The Ryu-Takayanagi prescription for holographic entanglement entropy is simply to evaluate the Bekenstein-Hawking formula on an extremal surface with the appropriate asymptotic boundary conditions. Hence one is only measuring certain features of the geometry in the background spacetime and in particular, one does not need to consider modifying the bulk solution. The derivation of this prescription came from a careful translation of the replica trick in the boundary theory to the bulk gravity theory \cite{juan,Dong:2016hjy}. In particular, one begins by evaluating the corresponding R\'enyi entropies (or at least $\tr \rho^\alpha$) but in principle, this requires finding a new gravitational solution for each value of $\alpha$. These new solutions are generated by introducing a codimension-two brane with the appropriate boundary conditions and with a tension $T_\alpha = (\a -1 ) / (4 \a G_N)$ \cite{Dong:2016fnf}. The latter is dual to the twist operator inserted in the boundary theory evaluation of $\tr \rho^\alpha$. Therefore the holographic R\'enyi entropy is not measuring geometric features of the given gravity solution, but rather is measuring the response of the spacetime to the introduction of a new gravitational source. Hence the entanglement entropy plays a distinguished role in that the deformation of the spacetime is eliminated in the limit $\alpha\to1$ in which the R\'enyi entropy reproduces the entanglement entropy. 

Above, we saw that the R\'enyi entropies can constrain certain transitions more strongly that the usual second law. Therefore it is worthwhile to consider that these holographic calculations can be extended to black holes. In particular, the horizon entropy of an eternal black hole in AdS space can be viewed as the entanglement entropy between the two copies of the boundary CFT in the thermofield double state dual to the black hole. Similarly, evaluating the R\'enyi entropy introduces a tensionful brane on the bifurcation surface and so the spacetime geometry is deformed \cite{Hung1,rcm}. At present, we only really understand the latter calculation in very symmetric configurations, \eg spherically symmetric static black holes, where the defect generated by the brane can be eliminated by extending the boundary geometry. It remains a challenge to understand the R\'enyi entropy in more general circumstances. 

More generally, we might ask what insights are provided from the previous comments for the question of evaluating the R\'enyi divergences for out-of-equilibrium black hole configurations. Given the close connection of the R\'enyi divergences and the R\'enyi entropies, we may expect that the former do not simply measure features of the geometry in a given spacetime. Rather it may be that we should be looking to measure the response of the black hole geometry to a new probe which acts as a source in the gravitational equations of motion. This leaves open the second question of finding a succinct expression of these constraints in terms of the gravitational variables. It may even call into question the possibility of formulating a simple description of the constraints. However, the present paper is only a first step in examining the macroscopic consequences of quantum thermodynamics for gravitational systems and in particular, for black holes. It is only by extending the calculations provided here and acquiring experience with more general situations, that we can expect to produce a proper formulation of the new second laws of black hole thermodynamics anticipated by our work. 

While our discussion focused on ``geometric" constraints on the classical evolution of a gravitational system, we should expect that these are a particular limit of the full (quantum) constraints provided by the R\'enyi divergences. That is, we expect that the full expression of these constraints will have an expansion in terms of the Planck length, \ie $\ell_P^2\simeq G_N\hbar$, of the form $D_\alpha\sim
a_0/\ell_P^2 + a_1 + \ell_P^2 a_2 + \cdots$, and we are focusing on the (classical or geometric) $a_0$ term in the present paper. For example, the area increase theorem \cite{area,Hawking:1973uf} is seen as this classical limit of the generalized second law \cite{JB1,Bekenstein:1973ur,Bekenstein:1974ax}.\footnote{For a review and recent advances towards a proof of the latter, see
\cite{Wall:2009wm,Wall:2010cj,Wall:2011hj,Bianchi:2012br}.} However, in a setting where quantum effects accumulate and become important, \eg black hole evaporation, the quantity which increases takes the form $S_\mt{gen} = A/(4G_N\hbar) + S_\mt{out}+\cdots$ where $S_\mt{out}$ is the entanglement entropy of the matter outside of the horizon. Hence if our R\'enyi constraints can be formulated as a simple expression in terms of gravitational variables, we expect that they would also have an analogous quantum (or at least, semiclassical) extension. Undoubtedly, these extended constraints would provide useful new diagnostics to help understand the black hole information paradox and other puzzles in quantum gravity.

However, we wish to close this discussion with a word of caution. Because of the indirect way in which we examine the additional second laws using holography, it remains an open question whether they actually correspond to new macroscopic laws for black hole evolution. Certainly, the conventional second law translates the macroscopic law that the area of a black hole must always increase (in classical processes). However, one logical possibility would be that the ordering of states provided by the R\'enyi divergences simply limits the evolution of the gravitational system only at the level of the underlying microstates. However, our holographic construction, as well as holographic studies of quantum quenches, {\eg} \cite{abrupt3,Asplund:2014coa,AbajoArrastia:2010yt,Albash:2010mv,Balasubramanian:2010ce,
Balasubramanian:2011ur,Aparicio:2011zy,Balasubramanian:2011at,Allais:2011ys,Galante:2012pv,
Caceres:2012em,Wu:2012rib,Balasubramanian:2012tu,Liu:2013iza,Liu:2013qca,Alishahiha:2014cwa,Fonda:2014ula,
Callebaut:2014tva,David:2015xqa,Keranen:2015mqc,Anous:2016kss,Chesler:2008hg,Heller:2011ju,Basu:2011ft,
Erdmenger:2012xu,Buch1,Hubeny:2013hz,Buch2,Nozaki:2013wia,Hartman:2013qma,
Balasubramanian:2013oga,Balasubramanian:2013rva,Asplund:2013zba,Craps:2013iaa,
Hubeny:2013dea,Buchel:2014gta,Ishii:2015gia,Asplund:2015eha,Ziogas:2015aja,Kundu:2016cgh,
Mezei:2016zxg,Jahn:2017xsg,Myers:2017sxr,Arefeva:2017pho,Bagrov:2017tqn}, demonstrates that the out-of-equilibrium character of certain microstates is manifest at the macroscopic level. Therefore our expectation is  that the new thermodynamic constraints will be describable at a macroscopic level in the gravitational theory. One point of tension, however, is that using a general reference state, the R\'enyi constraints revealed that the excited states appear not to exactly thermalize. While not in contradiction with unitary time evolution at the microscopic level, this result does seem to clash with the gravitational description where the excited configuration evolves towards a stationary black hole (without any scalar field excitations). Here, it will be important to understand better the effect that approximation will have on our calculations. Indeed, we do not expect states to exactly thermalize, but only do so approximately. One should then consider both approximate transitions and evolution laws which are both approximately thermalizing or nonlinear in the sense that they also act on additional microscopic degrees of freedom. Certainly resolving this puzzle is one of many intriguing questions which we hope to return to in the future.

\subsection*{Other future directions}

In the main text, we explored the constraints of the additional second laws for a specific class of excited states by working in a perturbative expansion. As we already mentioned, two important extensions of this approach would be to be able to work with a broader class of states and to go beyond the limitations of our perturbative construction. As a first step in this direction, in appendix~\ref{app:EVaidya}, we introduced a different, fully backreacted,  geometric construction that allows evaluating the R\'enyi divergence for another class of excited states. These can be thought as being obtained through the insertion of a shell of CFT operators acting on a thermal state at inverse temperature $\b_{\rm in}$. In the dual bulk, these insertions support a homogenous shell of non-interacting particles, and produce an effective inverse temperature $\b_{\rm out} \le \b_{\rm in}$. Geometrically, this construction is similar in spirit to the well-known AdS-Vaidya geometry \cite{VaidyaAdS}, as it involves joining together two black brane solutions of different mass along a shell comprised of a pressureless perfect fluid. Contrary to the ordinary Vaidya solutions for null dust though, this new solution is also well defined in Euclidean signature, since it relies on massive fluid, following timelike trajectories in Lorentzian time. 

As we explain in appendix, the partition function of this Euclidean shell solution can be interpreted as computing a trace function of the type
\be
\tr \(\rho_{\rm out}^{\a_{\rm out}} \rho_{\rm in}^{1-\a_{\rm in}}\) \,,
\ee
where $\rho_{\rm out}$ denotes the excited state at apparent temperature $1/\b_{\rm out}$, $\rho_{\rm in}$ is the thermal state with temperature $1/\b_{\rm in}$, and the exponents $\a_{\rm in} +\a_{\rm out}  \neq 1$. While this quantity is not generically a monotone under the dynamics, it directly defines one in the special case under consideration in which $\rho_{\rm in}$ is thermal. In fact, by simple identifications of the parameters of the solution, we can write $\rho_{\rm in}^{1-\a_{\rm in}} =\rho_{R}^{1-\a_{\rm out}}$, in terms of a thermal reference state at temperature $1/\b_R$. This allows an interpretation of this quantity as a R\'enyi divergence from the excited state $\rho_{\rm out}$ to a general reference state $\rho_R$, as those discussed above, and studying the monotonicity constraints for different excited states within this class. As described in the appendix, however, our geometric construction only allows exploring fixed trajectories in the $(\a, \b_R)$ parameter space of R\'enyi divergences. It would be interesting to understand whether it is possible to overcome these limitations and fully explore the implications of the second laws in this setting. 
\\

Finally, another interesting direction for future research would be as follows: In this paper we explored the macroscopic requirements imposed by the second laws of quantum thermodynamics. For this, we developed a method for computing the R\'enyi divergence for global states, \ie of the entire system. One natural question we would like to address is whether additional, possibly local constraints for gravitational dynamics could be derived from the R\'enyi divergence of reduced density matrices. This quantity can be computed by extending our path integral construction to allow for deformations, and equivalently scalar fields in the bulk, that only have support in limited spatial regions.  
For a ball shaped region $A$, the properties of positivity and monotonicity for increasing subsystem size of the relative entropy have been used to derive linearized Einstein's equations \cite{Faulkner:2013ica} and to work out new gravitational positive energy theorems \cite{Lashkari:2016idm}. It would thus be interesting to study whether new constraints also follow from properties of the whole family $D_\a(\rho_A \Vert \rho'_A)$ of R\'enyi divergences for subregions. 


\acknowledgments

This work began with an initial collaboration and an early manuscript of one of the authors, JO, with the late Jacob Bekenstein, who initiated the field of black hole thermodynamics. Although it is with sadness that we are not able to continue working together, his ideas continue to reverberate and inspire us, and we humbly dedicate this paper to his memory. \\[1ex]
We also thank Damian Galante for collaboration in the early stages of this project and Alvaro Alhambra, Xi Dong, Michal Horodecki, Hugo Marrochio, Don Marolf, Markus Mueller, Frans Pretorius, Sandu Popescu and Carlo Sparaciari, for useful discussions. Research at Perimeter Institute is supported by the Government of Canada through the Department of Innovation, Science and Economic Development and by the Province of Ontario through the Ministry of Research, Innovation and Science. JO is supported by the Royal Society, and by an EPSRC Established Career Fellowship. RCM and JO  thank the Kavli Institute for Theoretical Physics for its hospitality at one stage of this project. At the KITP, this research was supported in part by the National Science Foundation under Grant No. NSF PHY17-48958. RCM is supported by funding from the Natural Sciences and Engineering Research Council of Canada.  RCM and JO are also funded in part by the Simons Foundation through the ``It from Qubit'' collaboration.

\appendix
\section{Evaluation of $I(\a, \D)$}\label{app:integral}

In this appendix we study the regulated integral
\be \label{eq:IalpharegAPP}
I(\alpha, \D)_{\rm reg} \equiv  \int_{0}^{2\pi \alpha} d \tau  \int_{-\infty}^{\infty} d  x' \int_{0}^{2\pi \alpha} d  \tau'  \left[ \cosh  x'- \sqrt{1- \tilde \eps^2}  \cos( \t-  \t')\right]^{-\D}\,,
\ee
for $0 \le \a \le 1$, $0 < \D < 2$ (but $\D \neq 1$) and $\tilde \eps \ll 1$. \\

To study the structure of UV divergences arising from contact terms, we first consider the simpler expression 
\be
\tilde I(\alpha, \D)_{\rm reg} \equiv  2^{\D+1}   \int_{0}^{2\pi \alpha} d\tau   \int_{0}^{\infty} d x' \int_{0}^{2\pi \alpha} d\tau'  \left[ x'^2 + (\t-\t')^2 + \tilde \eps^2 \right]^{-\D}\,.
\ee
We use the integral representation of the Gamma function to write 
\be
\tilde I(\alpha, \D)_{\rm reg}  = \frac{2^{\D+1}}{\G(\D)}   \int_{0}^{2\pi \alpha} d\tau   \int_{0}^{\infty} d x' \int_{0}^{2\pi \alpha} d\tau'  \int_{0}^{\infty}  ds \, s^{\D-1} e^{- s (x'^2 + (\t-\t')^2 + \tilde \eps^2 )}
\ee
and, since the integrand is positive, compute the multiple integral by iterated integrals, as well as switch the order of integrations.   
We perform the Gaussian spatial integral and the double Euclidean time integral
\bea
\int_{0}^{2\pi \alpha} d\tau   \int_{0}^{2\pi \alpha} d\tau'  e^{- s (\t-\t')^2 }  &=& 2 \int_{0}^{2\pi \alpha} dp\, (2\pi \a - p) e^{- s p^2} \nonumber \\
&=& - \frac {1 - e^{-(2\pi \a)^2 s} }{s}+ \frac{2 \pi^{3/2} \a }{\sqrt s} {\rm erf} (2 \pi \a \sqrt s)\,
\eea
to obtain 
\be
\tilde I(\a, \D)_{\rm reg} = \frac{2^{\D} \sqrt \pi}{\G(\D)} \int_{0}^\infty ds \left\{ - s^{\D - \frac 5 2} \( 1- e^{- (2\pi \a)^2 s}\) + 2\pi^{3/2} \a s^{\D-2}  {\rm erf} (2 \pi \a \sqrt s)  \right\}e^{- s \tilde \eps^2}  \,.\label{eq:IAdSintermediate}
\ee
Using again the integral representation of a Gamma function, the first contribution gives 
\be
- \int_{0}^\infty ds \, s^{\D - \frac 5 2} \( 1- e^{- (2\pi \a)^2 s}\) e^{- s \tilde \eps^2} = - \G\(\D- \frac 3 2\) \left\{ \tilde \eps^{3-2\D} - (2\pi \a)^{3-2\D}\right\}
\ee
which holds for $\D > 3/2$ and where we have dropped terms that vanish as $\tilde \eps \to 0$. It can be analytically continued to all values of $0<\D<2$, except $\D = 1/2, 3/2$ for which the gamma function has simple poles. 

For the second term in \eqref{eq:IAdSintermediate} we perform the change of variable $s = z^2$ and use
\be
\int_0^\infty dz \, z^{\g}  {\rm erf} (a z) e^{- b^2 z^2 } = \frac{a}{\sqrt \pi} b^{-\g-2} \G\(\frac \g 2 +1\) F\[\frac 1 2, \frac \g 2+1, \frac 3 2, - \frac{a^2}{b^2}\]\, ,
\ee
which holds for Re$(b^2)>0$ and Re$(\g )> -2$. Thus for $\D> 1/2$ (but $\D \neq 1$) we obtain
\begin{align}
\int_{0}^\infty ds \,  s^{\D-2}  {\rm erf} (2 \pi \a \sqrt s) e^{- s \tilde \eps^2} &= 4 \sqrt \pi \a \tilde \eps^{1-2\D} \G\(\D- \frac 1 2\) F\[ \frac 1 2, \D- \frac 1 2, \frac 3 2, - \(\frac{2\pi \a}{\tilde \eps}\)^2 \] \nonumber \\
&\approx \G(\D-1) \tilde \eps^{2(1-\D)} + (2\pi \a)^{2(1-\D)} \frac{\G(\D-\frac 1 2)}{(1-\D) \sqrt \pi}  \,,
\end{align}
where we dropped terms that vanish as $\tilde \eps \to 0$. 
This can be continued to  $\D< 1/2$  and all together we find
\be
\tilde I(\a, \D)_{\rm reg} = - \frac{2^{\D+1} \pi^2 \a }{1-\D}\tilde \eps^{2(1-\D)}  - 2^{\D}\sqrt \pi \frac{\G(\D-\frac 3 2)}{\G(\D)} \tilde \eps^{3-2\D}  - 2^{\D-1} \sqrt \pi (2\pi \a)^{3-2\D} \frac{\G(\D-\frac 3 2)}{(1-\D) \G(\D)} \, \label{eq:IAdSreg}
\ee
plus terms that vanish as we remove the UV cutoff $\tilde \eps$. The integral has two types of UV divergences: a divergence $\sim \tilde \eps^{2(1-\D)}$, which is linear in $\a$, and a divergence $\sim \tilde \eps^{3-2\D}$, which is independent of $\a$. 
As we discussed in section~\ref{sec:Eaction}, the first type of divergence corresponds to the integral over the boundary of the non-normalizable mode of the bulk scalar field. This is precisely the divergence that is renormalized in the holographic computation of $\varphi_{(\D)}$ and that we will therefore drop from the result. The second divergence instead arises from the step function profile of the source $\l$ along the Euclidean time circle in our path integral construction. \\

We now  go back to the original integral \eqref{eq:IalpharegAPP}. We use the gamma function integral representation to write 
\be
\left[ \cosh x -\sqrt{1- \tilde \eps^2}   \cos(\t-\t') \right]^{-\D} = \frac{1}{\G(\D)} \int_{0}^{\infty}  ds \, s^{\D-1} e^{- s (\cosh x - \sqrt{1- \tilde \eps^2}  \cos(\t-\t') )}
\ee
and perform the spatial integral, to obtain
\be 
I(\alpha, \D)_{\rm reg}  =  \frac{2}{\G(\D)} \int_{0}^{2\pi \alpha} d\tau   \int_{0}^{2\pi \alpha} d\tau' \int_{0}^{\infty} d s~ s^{\D-1}  K_0(s)   e^{s \sqrt{1- \tilde \eps^2} \cos(\t-\t')}\,.  
\ee
For $\a=1$, the Euclidean time integrals can be performed explicitly and for $\D \neq 1$ give
\bea
I(1, \D)_{\rm reg} &=& 2\frac{(2\pi)^2}{\G(\D)} \int_{0}^{\infty} ds s^{\D-1} K_0(s) I_0(s \sqrt{1- \tilde \eps^2} )  \nonumber \\
& =& - \frac{2^{\D+1} \pi^2 }{1-\D} \, \tilde \eps^{2(1-\D)}+ \frac{ 2\pi^{3/2} \G(\frac{1-\D}{2}) \G(\frac \D 2)^2}{\G(\D)\G(1- \frac \D 2)} \label{eq:I1regBTZ}
\eea
up to terms that vanish as $\tilde \eps \to 0$. It contains the same leading divergence $\sim \tilde \eps^{2(1-\D)}$ as in eq.~\eqref{eq:IAdSreg}, but no additional divergences. 

For $\a <1$, we  perform a simple change of variables and write the remaining integrals as 
\begin{align} \label{eq:Iareg} 
I(\alpha, \D)_{\rm reg}& =  \frac{4}{\G(\D)} \int_0^{2\pi \a} dp \, (2\pi \a -p) \, \int_{0}^{\infty} d s~ s^{\D-1}  K_0(s) e^{s  \sqrt{1- \tilde \eps^2} \cos p} \\
&= \frac{2^{2-\D}\sqrt \pi  \G(\D)}{\G\(\D+\frac 1 2\)} \int_{0}^{2\pi \alpha} dp \, (2\pi \a -p ) \, F\!\[\D, \D, \D+ \frac 1 2, \frac{1+ \sqrt{1- \tilde \eps^2} \cos p}{2}\]  \,.  \nonumber 
\end{align}
In both eqs.~\eqref{eq:I1regBTZ} and \eqref{eq:Iareg}, we subtract the same divergence $\sim \tilde \eps^{2(1-\D)}$ as appears in eq.~\eqref{eq:IAdSreg}. Recall that the latter divergence arises from the boundary integral over the non-normalizable mode of the scalar field. We then consider the renormalized quantity 
\be \label{eq:Iren}
I(\alpha, \D) = I(\alpha, \D)_{\rm reg} + \frac{2^{\D+1} \pi^2 \a }{1-\D} \, \tilde \eps^{2(1-\D)} \,.
\ee
Notice that in any case the coefficient of the divergence  $\sim\tilde \eps^{2(1-\D)}$ is linear in $\alpha$ and therefore such a divergence automatically drops out of the R\'enyi divergence result defined in \eqref{eq:Dalphareg} in terms of normalized density matrices.

\begin{figure}[ht]
\centering 
\includegraphics[width=.45\textwidth]{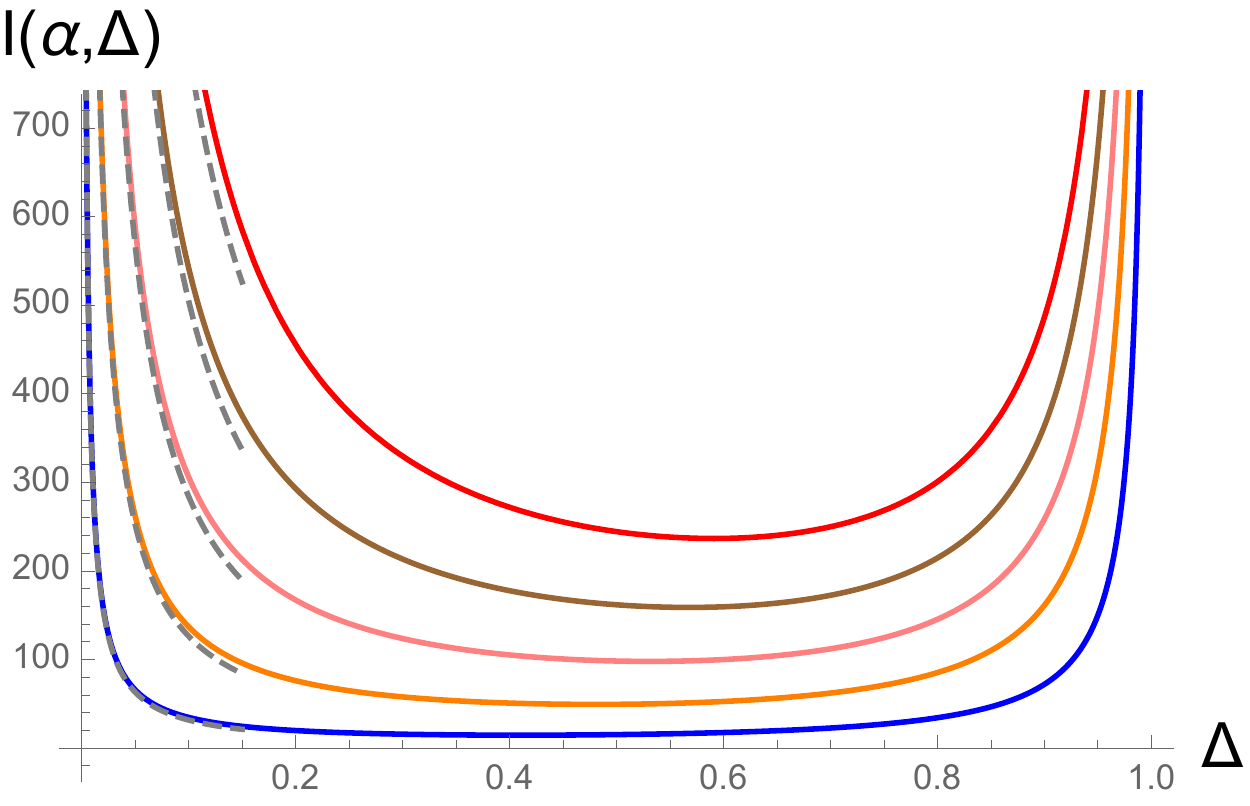}  \hfill \includegraphics[width=.45\textwidth]{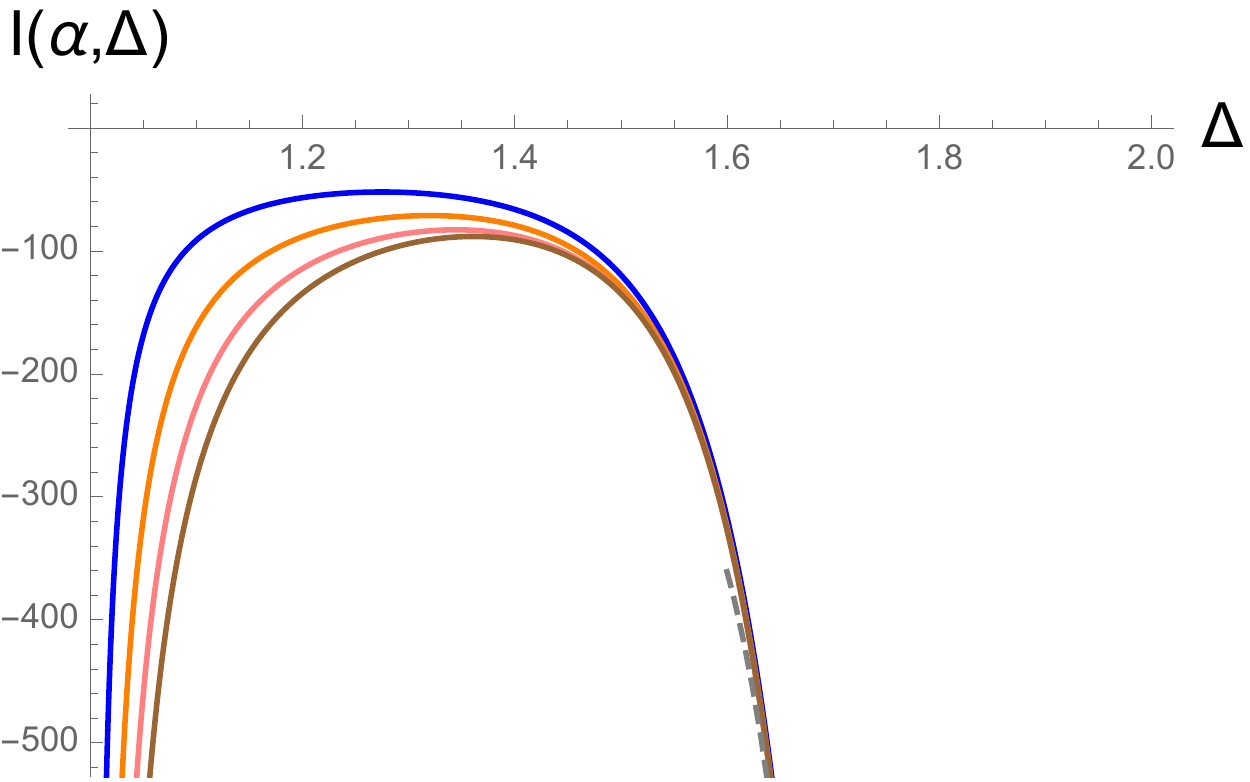}
\caption{(Left) $I(\alpha, \D)$ for $\a =0.2, 0.4, 0.6, 0.8, 1$ from the bottom up between $0 < \D<1$. In gray dashed the known asymptotic $2(2\pi\a)^2/\D$ as $\D \to 0$. (Right) $I(\alpha, \D)$ for $\a =0.2, 0.4, 0.6, 0.8$ from the top down for $1 < \D<2$. In gray dashed the universal divergence $- 2^\D\sqrt{\pi} \G\(\D-\frac 3 2\) \tilde \eps^{3-2\D}/\G(\D)$ as $\tilde \eps \to 0$. In both plots we set the cutoff $\tilde \eps=0.0001$.} 
\label{fig:Ia_BTZ}
\end{figure}

We are unable to  perform the final integral \eqref{eq:Iareg} analytically and we evaluate it numerically. In figure~\ref{fig:Ia_BTZ}  we plot the resulting $I(\alpha, \D) $ as a function of $\D$  for different $\a$ in the range $0\le \a \le 1$. For $0 < \D <1$, $I(\alpha, \D) $  is finite and positive. It increases monotonically in $\a$ and asymptotes to $2(2\pi\a)^2/\D$ as $\D \to 0$. It  has a pole at  $\D = 1$, where it diverges and changes sign, making $I(\alpha, \D)$ negative for $1<\D<2$. 

\begin{figure}[ht]
\centering 
\includegraphics[width=.45\textwidth]{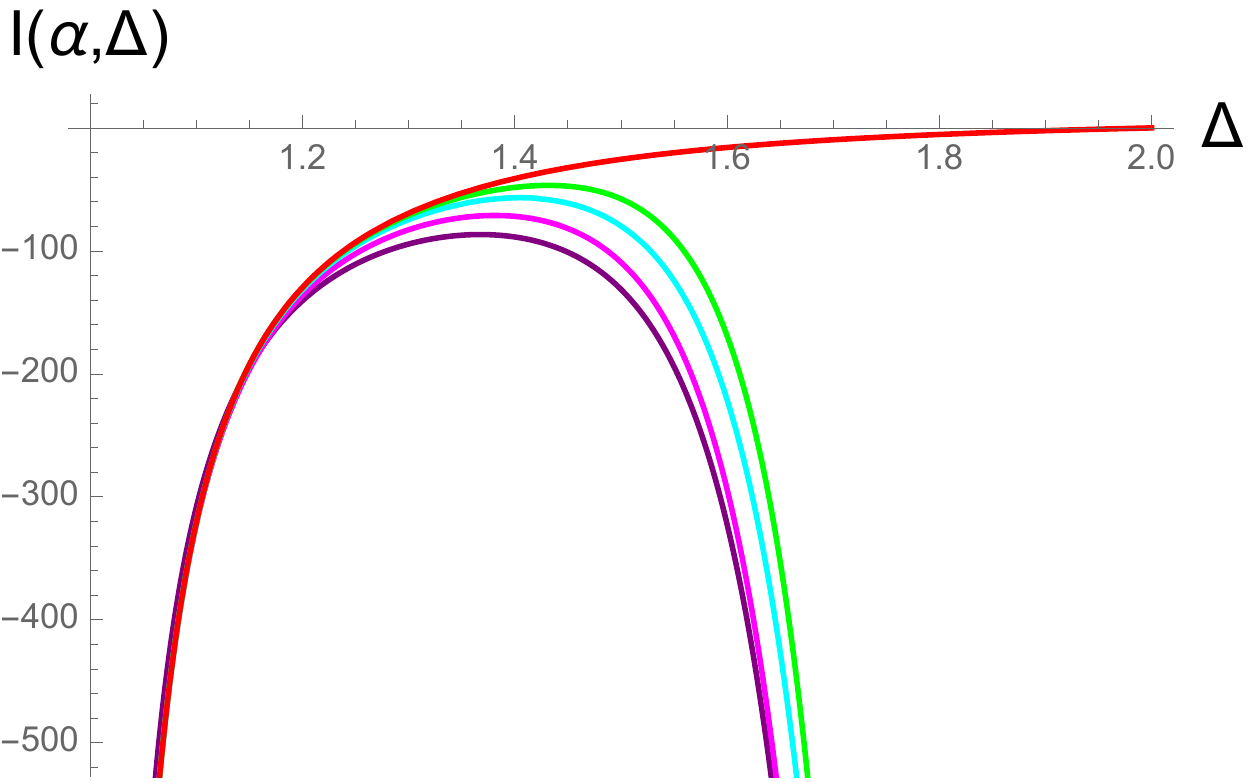}  \hfill \includegraphics[width=.45\textwidth]{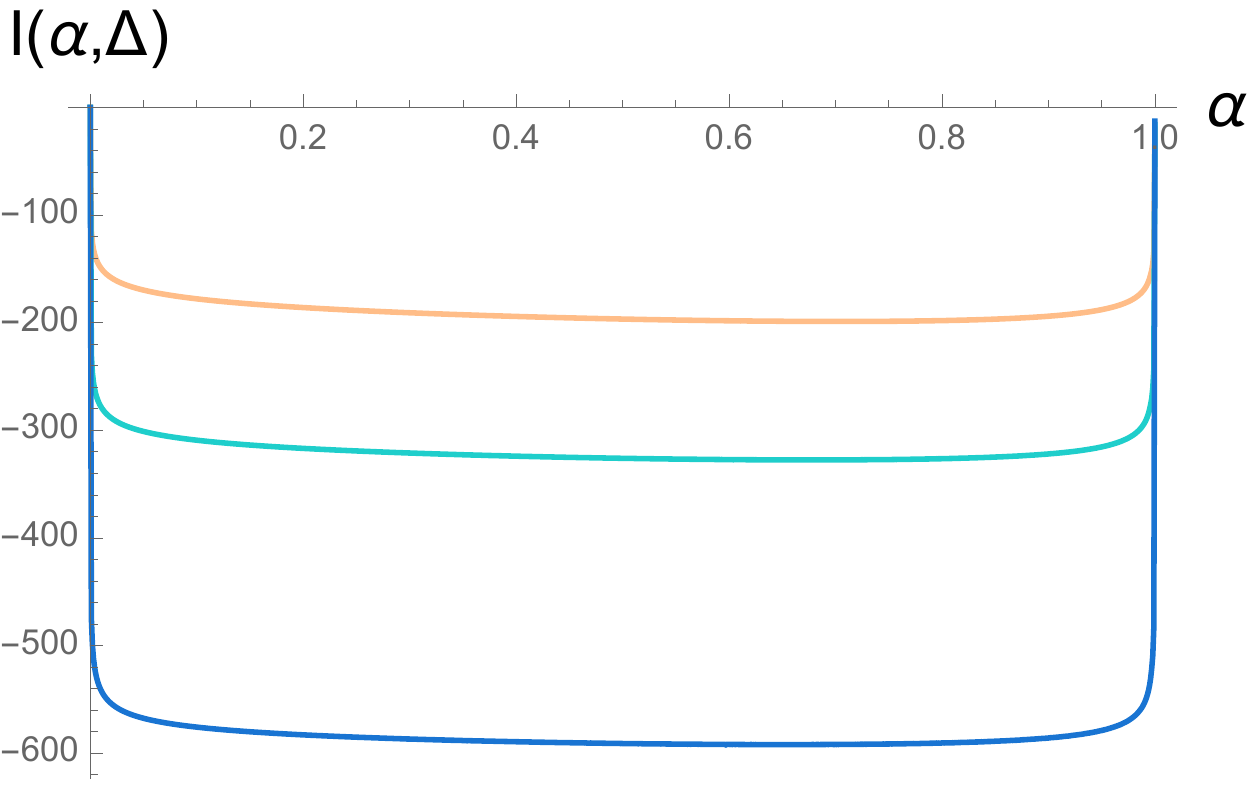}
\caption{(Left) $I(\alpha, \D)$ for $\a =0.9, 0.99, 0.999, 0.9999, 1$ from the bottom up for $1 < \D< 2$. (Right) $I(\alpha, \D)$ for $\D =1.55, 1.6, 1.65$ from the top down. In both figures $\tilde \eps=0.0001$.} 
\label{fig:Iaclose1}
\end{figure}

As we take $\tilde \eps \to 0$, the integral has the same universal divergence $\sim \tilde \eps^{3-2\D}$ of eq.~\eqref{eq:IAdSreg} for values of  conformal dimensions $\D > 3/2$. 
However as $\a$ approaches 1, the divergence in $\tilde \eps^{3-2\D}$ smoothly transits into the finite result $I(1, \D) $. We plot this behavior in figure~\ref{fig:Iaclose1}, both as a function of $\D$ for $\a$ close to 1 (left panel) and for fixed $\D>3/2$ as a function of $\a$ (right panel).   

\begin{figure}[ht]
\centering 
\includegraphics[width=.45\textwidth]{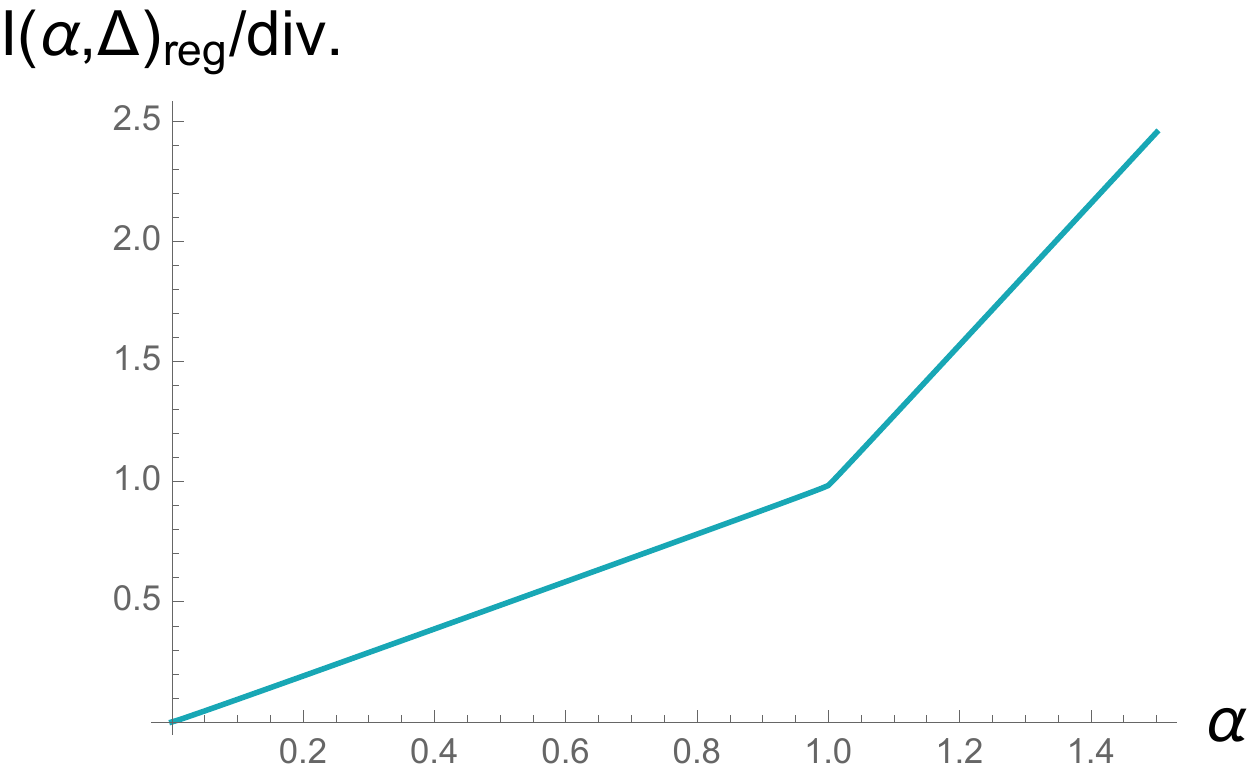}  \hfill  \includegraphics[width=.5\textwidth]{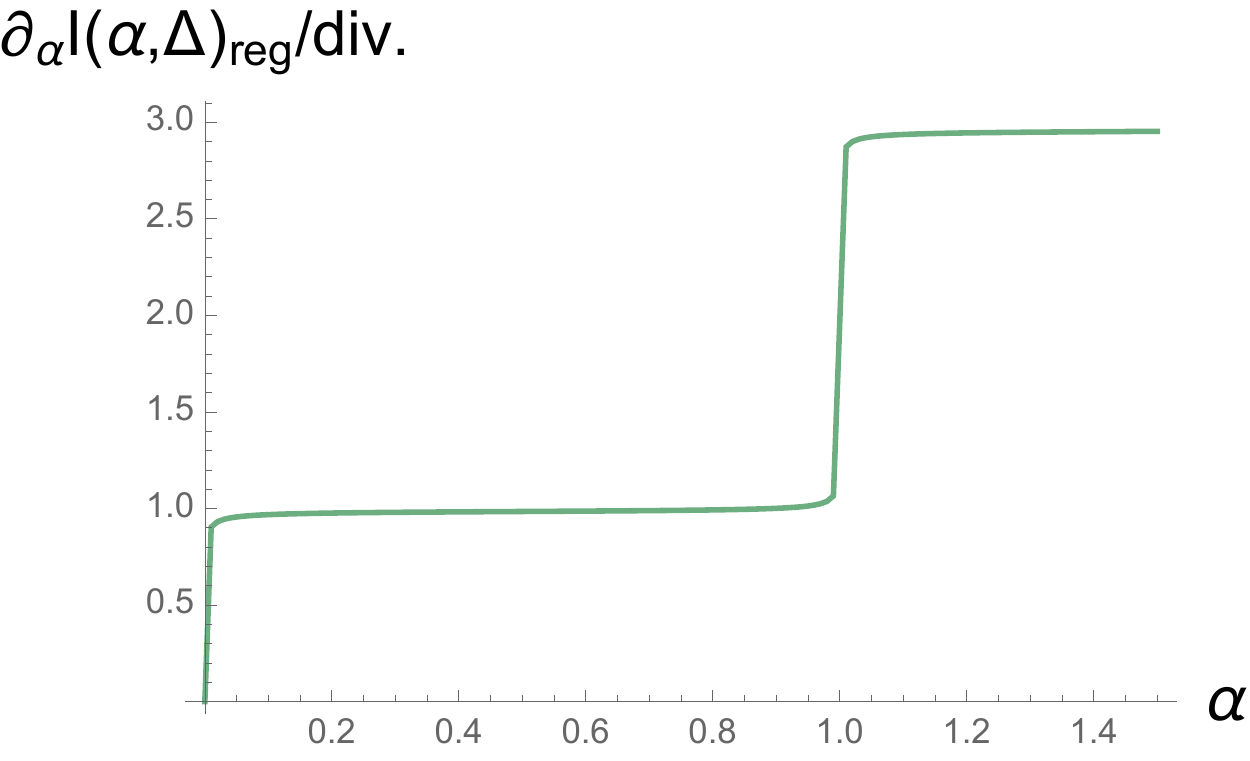}
\caption{$I(\alpha, \D)_{\rm reg}$ (left) and $ \del_\a I(\a, \D)_{\rm reg}$ (right) rescaled by the divergent term $- \frac{2^{\D+1} \pi^2 }{1-\D}\tilde \eps^{2(1-\D)} $, as a function of $\a \in [0,1.5]$. In both plots $\tilde \eps=0.001$ and $\D =5/4$.} 
\label{fig:I1primeBTZ}
\end{figure}

To evaluate the R\'enyi divergence in the limit $\a \to 1$ in the main text, we also compute the first derivative of $I(\alpha, \D) $ with respect to $\a$
\begin{align}
\del_\a I(\a, \D)_{\rm reg}&= \frac{8 \pi  }{\G(\D)} \int_0^\infty ds s^{\D-1} \int_{0}^{\infty} d x e^{- s \cosh x} \int_{0}^{2\pi \a} dp \, e^{ s \sqrt{1- \tilde \eps^2} \cos p }\nonumber\\
&= \frac{2^{3-\D} \pi^{3/2} \G(\D)}{\G\(\D+\frac 1 2\)} \int_{0}^{2\pi \a} dp\,  F\!\[\D, \D, \D+ \frac 1 2,\frac{ 1+ \sqrt{1- \tilde \eps^2} \cos p}{2}\]\,.
\end{align}
For $\a \to 1$, it gives
\be 
\del_\a I(\a, \D)_{\rm reg} \Big|_{\a=1}= 2 I(1,\D)_{\rm reg}\,,
\ee
and 
\begin{align} \label{eq:Iprime1}
\del_\a I(\a, \D) \Big|_{\a=1}&=  - \frac{2^{\D+1} \pi^2 }{1-\D}\tilde \eps^{2(1-\D)} + 2  I(1, \D)  \nonumber\\
&= - \frac{2^{\D+1} \pi^2 }{1-\D}\tilde \eps^{2(1-\D)} + \frac{ 4 \pi^{3/2} \G(\frac{1-\D}{2}) \G(\frac \D 2)^2}{\G(\D)\G(1- \frac \D 2)} \,.
\end{align}
Notice in particular, that $\del_\a I(\a, \D) |_{\a=1}$ contains the divergence $\sim \tilde  \eps^{2(1-\D)}$ for $\D>1$, even though $I(1, \D)$ is finite. As an example,  in figure~\ref{fig:I1primeBTZ} we plot up to $\a =1.5$ the functions  $I(\alpha, \D)_{\rm reg}$ and $\del_\a I(\a, \D)_{\rm reg}$ divided by the divergent term in eq.~\eqref{eq:Iprime1}.  From that, we see indeed that the leading divergence in $I(\alpha, \D)_{\rm reg}$ is always linear in $\a$, but at finite $\tilde \eps $ smoothly changes slope around $\a=1$. 

\begin{figure}[ht]
\centering 
\includegraphics[width=.5\textwidth]{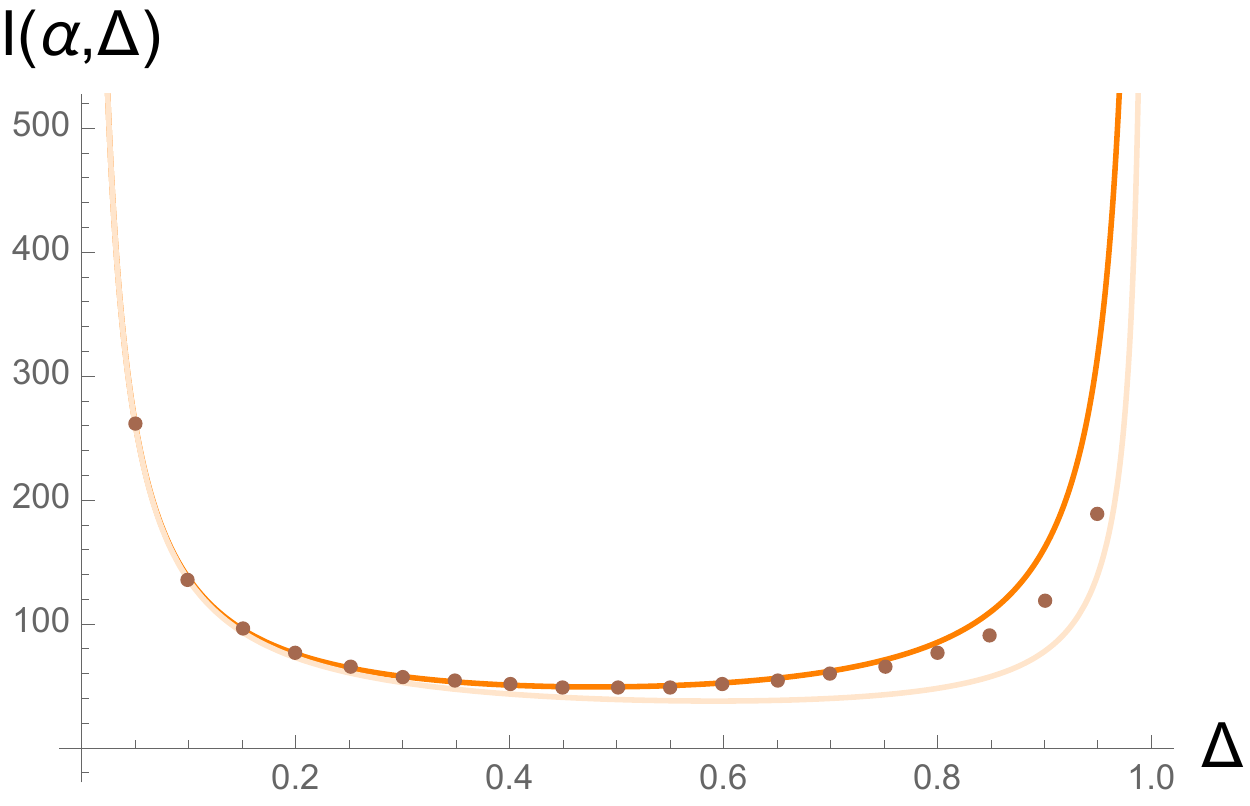}
\caption{Numerical integration of $I(\alpha, \D)$ for $\a =0.4$ and $\eps=0.0001$ compared to the series representation \eqref{eq:Iseries} truncated at $m=3000$ (brown dots). In light orange $ \a^2 I(1, \D)$, which well approximates the full result for small $\D$. The series representation we worked out strictly holds for $\D <1/2$, but it is found to well approximate the result also for slightly larger conformal dimensions. } 
\label{fig:compare}
\end{figure}

Finally notice that for the values of $\D$ for which the integral is finite as we remove the cutoff, we can also write the expression we obtained in eq.~\eqref{eq:Iareg} as 
\begin{align}
& I(\alpha, \D) =  \frac{2^{2-\D}\sqrt \pi  \G(\D)}{\G\(\D+\frac 1 2\)}  \int_{0}^{2\pi \alpha} dp \, (2\pi \a -p ) \, \sum_{m=0}^\infty \frac{[(\D)_m]^2}{\(\D +\frac 1 2\)_m} \frac{\(\cos \frac p 2\)^{2m}}{m!} \\
&=  \frac{2^{2-\D}\sqrt \pi  \G(\D)}{\G\(\D+\frac 1 2\)}  \sum_{m=0}^\infty \frac{[(\D)_m]^2}{m! 2^{2m} \(\D +\frac 1 2\)_m}   \int_{0}^{2\pi \alpha} dp (2\pi \a -p ) \left\{ \sum_{k=0}^{m-1} 2 \left(\begin{array}{c}2m \\ k\end{array}\right) \cos[(m-k)p] + \left(\begin{array}{c}2m \\ m \end{array}\right)  \right\} \,. \nonumber 
\end{align}
To exchange the integral and the series we further  restricted to $\D < 1/2$, such that the Gauss hypergeometric series is absolutely and uniformly convergent $\forall p$, and used a sum representation of $(\cos \frac p 2)^{2m}$.
Performing the integration over $p$ 
\begin{align}
& I(\alpha, \D)= \frac{2^{2-\D}\sqrt \pi  \G(\D)}{\G\(\D+\frac 1 2\)} \sum_{m=0}^\infty \frac{[(\D)_m]^2}{m! 2^{2m} \(\D +\frac 1 2\)_m}  \Bigg\{ \sum_{k=0}^{m-1} 4 \left(\begin{array}{c}2m \\ k\end{array}\right) \frac{\sin^2 [(k-m)q] }{(k-m)^2} + \left(\begin{array}{c}2m \\ m \end{array}\right) 2 \pi^2 \a^2 \Bigg\} 
\end{align}
and summing the series
\be
\sum_{m=0}^\infty \frac{[(\D)_m]^2}{m! 2^{2m} \(\D +\frac 1 2\)_m}  \left(\begin{array}{c}2m \\ m \end{array}\right)  = \frac{\G\(\frac{1-\D}{2}\)\G\(\D +\frac 1 2\)}{2^\D \G\(1-\frac \D 2\) \G\(\frac{1+\D}{2}\)^2}\,,
\ee
we obtain the following expression
\begin{align}
I(\alpha, \D) &=\a^2  \frac{ 2\pi^{3/2} \G(\frac{1-\D}{2}) \G(\frac \D 2)^2}{\G(\D)\G(1- \frac \D 2)}  + \sum_{m=0}^\infty \frac{2^{3+\D} \G(\D)^2  [(\D)_m]^3}{m! \G\(2\D +2m \)} \sum_{k=0}^{m-1} \left(\begin{array}{c}2m \\ k\end{array}\right)\frac{\sin^2[(k-m)\pi \a] }{(k-m)^2} \,. \label{eq:Iseries}
\end{align}
For $\a=1$ the coefficients of the series over $m$ all vanish and the first term gives the finite part of eq.~\eqref{eq:I1regBTZ}.  For general values of $\a$ we compare the truncated sum to the numerical integral in figure~\ref{fig:compare}. 
%

\section{Euclidean shell solution}\label{app:EVaidya}

In this appendix, we construct a new Euclidean geometry involving a shell of pressureless fluid or dust, and examine its interpretation as a trace function. Upon Wick rotation, the corresponding Lorentzian solution of 3d gravity describes a homogeneous (but infinitely thin) shell of pressureless fluid which falls into a BTZ black brane --- see figure \ref{fig:Lorentzian} below. An interesting feature of this solution will be that it takes into account the gravitational backreaction of the shell, in contrast to our perturbative calculations in section \ref{sec:quench}. The construction essentially involves joining together regions of two different (Euclidean) black brane solutions with masses, $M_{\rm in}$ and $M_{\rm out}$. In this regard, the new solution resembles the well-known Vaidya solution \cite{Vaid0,OriginalVaidya} or its extension to AdS boundary conditions \cite{VaidyaAdS}. Of course, a key difference is that these Vaidya solutions describe an infalling shell of {\it null} fluid, while our construction relies on conventional pressureless fluid, which follows timelike trajectories in the Lorentzian version.\footnote{We note that a Euclidean version of the AdS-Vaidya solution was  constructed in \cite{Balasubramanian:2012tu}, by taking the Vaidya metric to be a limit of a family of ``spacelike Vaidya'' geometries that admitted a well-defined Wick rotation to Euclidean signature. This is not the construction that we use here.} The AdS-Vaidya solution has been used extensively in holographic studies, \eg to explore features of the thermalization process in strongly coupled gauge theories \cite{AbajoArrastia:2010yt,Albash:2010mv,Balasubramanian:2010ce,Balasubramanian:2011ur,Aparicio:2011zy,Balasubramanian:2011at,
Allais:2011ys,Galante:2012pv,Caceres:2012em,Wu:2012rib,Balasubramanian:2012tu,Liu:2013iza,Liu:2013qca,Fonda:2014ula,
Callebaut:2014tva,David:2015xqa,Keranen:2015mqc}.

As explained below, in the context of the AdS/CFT correspondence, the Euclidean shell solution describes a bulk partition function which can be interpreted as computing a boundary trace function of the type 
\be
Z_{\rm bulk} \approx \tr \( \rho_{\rm out}^{\a_{\rm out}} \rho_{\rm in}^{1-\a_{\rm in}}\) \,,
\ee
where $ \rho_{\rm in},\  \rho_{\rm out}$ are two thermal density matrices at inverse temperatures $\b_{\rm in}=2\pi /\sqrt{M_{\rm in}}$ and  $\b_{\rm out} = 2\pi /\sqrt{M_{\rm out}}$. To be more precise, $\rho_{\rm out}$ is an excited state built by acting on $\rho_{\rm in}$ with a shell of CFT operators in order to produce an apparent temperature $1/\b_{\rm out}$. We discuss this interpretation in greater detail below, in section \ref{sec:interpretation}.

In section~\ref{sec:coords}, we construct coordinates adapted to geodesics moving in the radial and (Euclidean) time directions of the (Euclidean) black brane geometry.  
We then construct the Euclidean shell metric in section~\ref{sec:Vaidya} by gluing two such geometries together along the appropriate geodesics. Next, we evaluate the renormalized on-shell action of our new solution in section \ref{sec:VaidyaS}. We conclude in section~\ref{sec:interpretation} by describing the trace function interpretation of the corresponding partition function in the boundary CFT. 

\subsection{Geodesic slicing of AdS black brane}\label{sec:coords}

To begin, we recall that the Euclidean AdS$_3$  black brane is described by the following metric
\be \label{eq:BTZE}
ds^2 =(r^2- M) d\tau^2 + \frac{dr^2}{r^2-M} + r^2 dx^2\,,
\ee
with $\tau \in [- \frac \beta 2, \frac \beta 2], r \ge \sqrt M$ and $\beta = 2\pi/\sqrt M $. 
As commented above, we will construct the Euclidean shell solution by gluing together regions from two such solutions with masses $M_{\rm in}$ and $M_{\rm out}$, along space-like geodesics. In particular, we wish to consider space-like geodesics which are moving in the $(r,\tau)$-plane, and are anchored at the endpoints of intervals on the asymptotic boundary (\ie $r\to\infty$)  which are centred at $\tau=0$.  By examining Israel's junction conditions, we will show that our construction yields a solution of Einstein's equations with negative cosmological constant in presence of a pressureless perfect fluid. First, we will find new coordinates for the metric \reef{eq:BTZE}, which are adapted to the geodesics of interest.

\begin{figure}[ht]
\centering 
\includegraphics[width=.7\textwidth]{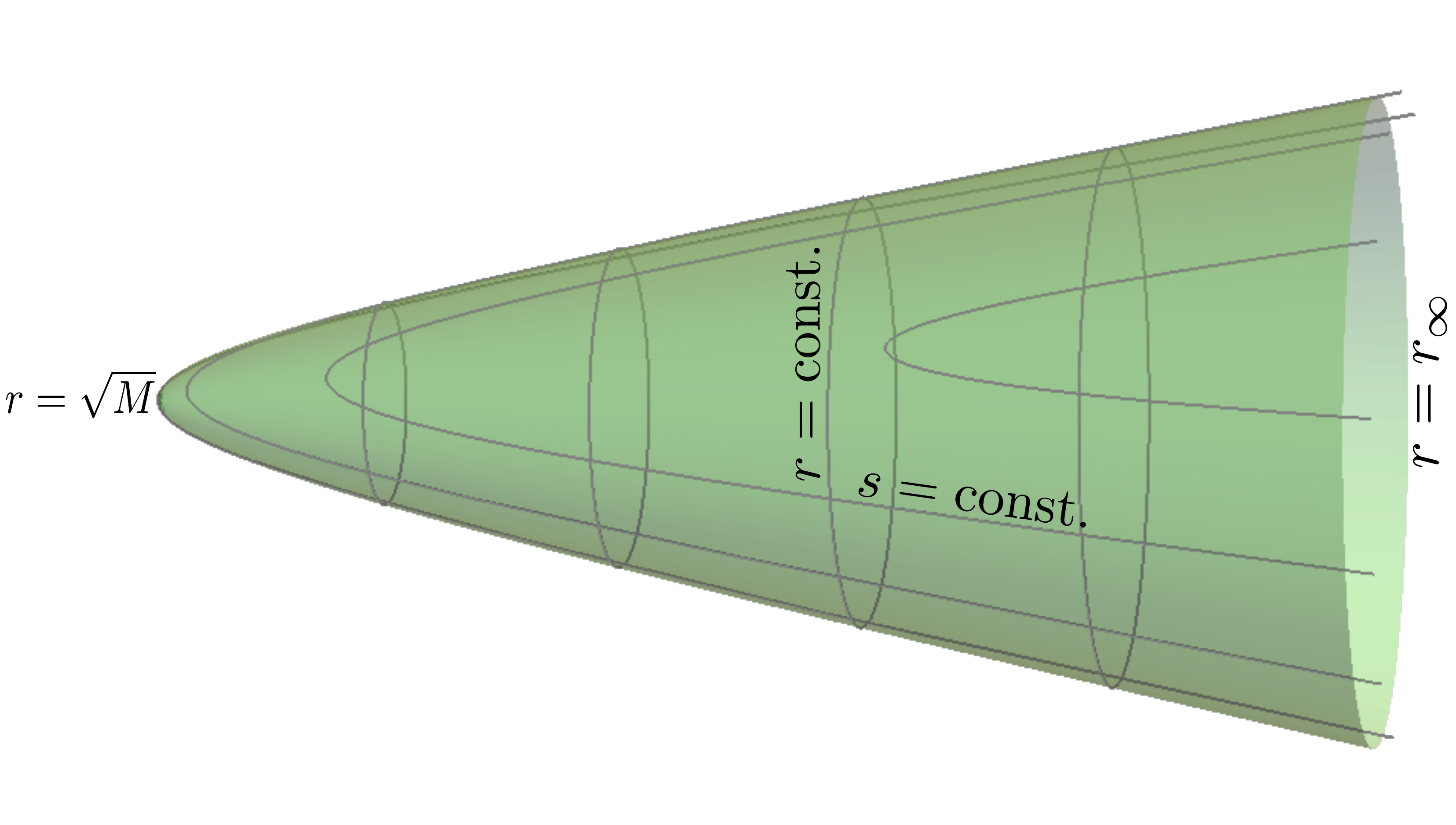} 
\caption{Surfaces of constant $r$ and $s$ in the Euclidean AdS$_3$ black brane \eqref{eq:BTZE}.} 
\label{fig:EVaidya}
\end{figure}

The geodesics anchored on the boundary at endpoints  $(- \tau_0, x_0)$ and $(\tau_0, x_0)$ are given by
\bea
\tau(\l)&=&    \frac{ 1}{\sqrt M} \arctan\(\tan (\sqrt M  \tau_0) \tanh \l \)\\
r(\l)&=& \frac{\sqrt M \cosh \l}{\sin (\sqrt M  \tau_0)}\,.
\eea
in terms of an affine parameter $\l \in (-\infty, \infty)$.  Alternatively, solving for $\l(r)$ 
we can express the geodesic as the  $\tau(r)$ curve
\be  
\tau(r) =  \pm \frac{ 1 }{\sqrt M} {\rm arctan}\(\frac{1}{\cos (\sqrt M  \tau_0)} \sqrt{\sin^2(\sqrt M  \tau_0) -\frac{M}{r^2}}\)\,, \label{eq:tauofr}
\ee 
where the minus sign branch covers the range $- \tau_0 \le \tau \le 0$ and the plus sign $ 0 \le \tau \le \tau_0$. 
For $2 \tau_0 \in [0, \pi/\sqrt M]$ these geodesic span half of the original geometry, corresponding to  $\tau \in [-\pi/(2\sqrt M), \pi/(2\sqrt M)]$,   and each geodesic can be specified in a unique way in terms of the radial value at the turning point
\be \label{eq:s0}
s_0 \equiv r(\l=0) = \frac{\sqrt M}{\sin (\sqrt M \tau_0)}\, . 
\ee
We can thus replace the time slicing with one given in terms of these geodesics  by  performing the change of coordinate $\tau \to s$ defined through
\be\label{eq:taus}
\tau =   \pm \frac{ 1}{ \sqrt M} \ {\rm arctan} \(\frac{\sqrt M }{ r} \sqrt{\frac{r^2-s^2}{s^2-M}}\)\,.
\ee
Notice that the change of coordinates is defined patchwise with the plus sign covering the patch $\tau \in [0, \pi/(2\sqrt M)]$. We draw such slicing of the geometry in figure~\ref{fig:EVaidya}. 

The new coordinate has range $s \ge \sqrt M$ and, for both signs in eq.~\eqref{eq:taus}, leads to the metric
\be \label{eq:metric_s}
ds^2 = \frac{r^2(r^2 -M)}{s^2(s^2 -M)(r^2-s^2)} \ ds^2 - \frac{2  r  }{s(r^2 -s^2)} \ ds dr +\frac{dr^2}{r^2-s^2} + r^2 dx^2\,.
\ee
Notice that surfaces at constant $s$ have an induced metric that does not depend on $M$.  

\subsection{Euclidean shell geometry}\label{sec:Vaidya}

\begin{figure}[t]
\centering 
\includegraphics[width=.65\textwidth]{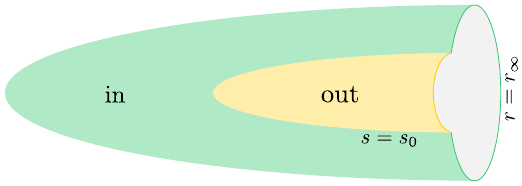} 
\caption{A pictorial representation of the Euclidean shell geometry of eq.~\eqref{eq:EVaidyaCigar}.} 
\label{fig:EVaidyaCigar}
\end{figure}

We have parametrized the metric \eqref{eq:BTZE} in terms of a new set of  coordinates $(s,r,x)$. 
In analogy to the Lorentzian case, we then cut-and-paste two black branes with different masses along a $s=s_0=\text{constant}$ surface and define a Euclidean shell geometry with
\be \label{eq:EVaidyaCigar}
ds^2 = \frac{r^2( r^2 - M(s))} {s^2 (r^2-s^2 ) (s^2 -M(s))} \  ds^2 -  \frac{2  r  }{s(r^2 -s^2)} \ ds \,dr +\frac{dr^2}{r^2-s^2} + r^2 dx^2\, ,
\ee
where
\be
M(s) = M_{\rm in} + \theta(s-s_0) (M_{\rm out} -M_{\rm in})\, .   \label{mofs}
\ee
The resulting geometry is a cigar in which a geodesic shell at $s_0$ separates an inside region where the geometry matches the black brane geometry \reef{eq:metric_s} with mass $M_{\rm in}$,  and an outside region with the local metric given in terms of the black brane geometry \reef{eq:metric_s} with mass $M_{\rm out}$. This is schematically represented in figure~\ref{fig:EVaidyaCigar}. 
Notice that  by construction the geometry is defined for a shell location $s_0 \in [{\rm Max} (\sqrt{ M_{\rm in}},\sqrt{ M_{\rm out}}), \infty)$.\\

The metric defined in this way satisfies Einstein's equations with a stress tensor with only non-vanishing component
\be
T^{rr} = - \frac{ s(r^2-s^2) M'(s)}{2 r^2}  = -\frac{ s(r^2-s^2)  (M_{\rm out}-M_{\rm in})}{2 r^2}  \delta(s-s_0) \, .
\ee
This is the stress tensor of a pressureless perfect fluid 
\be
T^{\mu\nu}  = \rho \ u^\mu u^\nu 
\ee
with 
\bea
\rho &=& - \frac{s M'(s)}{2 r^2} = -  \frac{ s(M_{\rm out}-M_{\rm in})  }{2 r^2} \delta(s-s_0) 
\eea
and $u^{\mu}$ the three-velocity associated to the geodesic with  $s =s_0$.  We note that for the solution to have a proper continuation to Lorentzian signature, we would require $M_{\rm out}>M_{\rm in}$ (\ie this constraint ensures that the Lorentzian shell has positive energy).
 One can also check explicitly that the stress tensor is conserved, $\nabla_{\mu} T^{\mu\nu} = 0$, 
and that Israel's junction conditions \cite{Israel:1966rt} are satisfied.\footnote{Indeed the induced metric and extrinsic curvature are well defined as one approaches the spacelike  surface $s=s_0$ from the two sides of the shell, and the induced metric is continuos across the surface.  The discontinuity in the extrinsic curvature
 \begin{align}
 \Delta[\delta^i_{~j} K-K^i_{~j} ]  &\equiv \lim_{\eps\to 0 }\(  \delta^i_{~j} K-K^i_{~j}  ]_{s_{0} - \eps} -    [\delta^i_{~j} K-K^i_{~j} ]_{s_{0}+ \eps}  \) \nonumber  \\
 &=  \delta^{i}_{~r}\delta_{j}^{r} \ \frac{  \sqrt{s_0^2  - M_{\rm out}}  -   \sqrt{s_0^2  - M_{\rm in}} }{r }
 \end{align}
matches the surface stress tensor $S^i_{~j} $, that is the pull back $T^i_{~j}$ of the stress tensor integrated over a small region around the surface 
 $s=s_0$
 \begin{align}
 S^i_{~j}  &= \lim_{\eps\to0} \int_{s_0 + \eps}^{s_0 - \eps} T^i_{~j} ~d\hat n_{(s)} = \delta^{i}_{~r}\delta_{j}^{r} \lim_{\eps\to0}\int_{s_0 + \eps}^{s_0 - \eps}  \frac{ M'(s)}{2r \sqrt{s^2 -M(s)}} ds   \nonumber \\
 & =  \delta^{i}_{~r}\delta_{j}^{r} \ \frac{  \sqrt{s_0^2  - M_{\rm out}}  -   \sqrt{s_0^2  - M_{\rm in}} }{r } \,.
 \end{align}
 }\\

\begin{figure}[ht]
\centering 
\includegraphics[width=.8\textwidth]{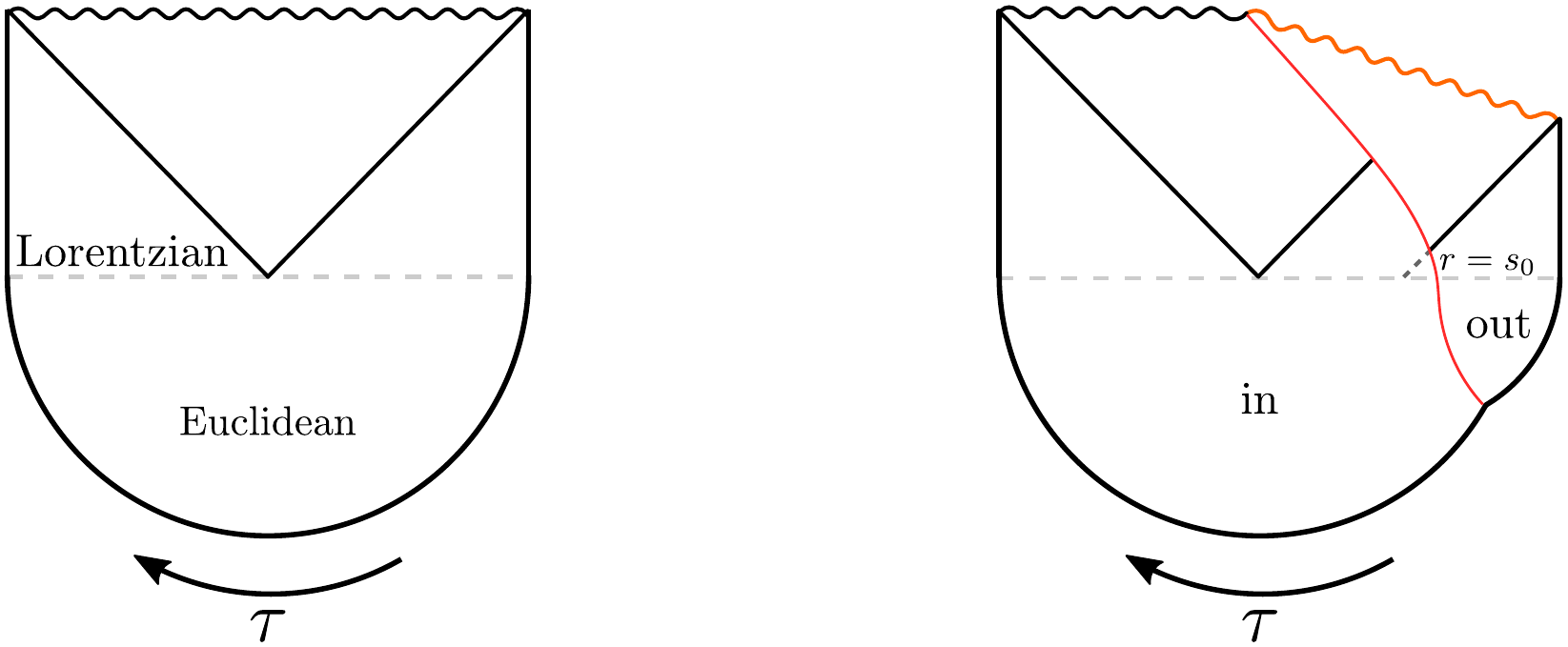} 
\caption{Representation of the analytic continuation to Lorentzian of a single Euclidean black hole geometry (left) and of the Euclidean shell geometry constructed here (right).  In the latter case, the continued geometry describes the backreaction to the presence of the homogeneous shell of dust that falls into a black brane geometry from a radial location $r= s_0$.  }
\label{fig:Lorentzian}
\end{figure}

We are primarily interested in the Euclidean construction per se, but  one can also work out the analytic continuation of this geometry to Lorentzian signature. In the case of a  single Euclidean black hole geometry,  cutting it at the surface of time reflection symmetry (\ie along $\tau=0$ and $\tau=\beta_{\rm in}/2$), the solution can be continued to the maximally extended Lorentzian solution. Here our Euclidean shell solution can be continued in an analogous way, and  the resulting geometry will describe a homogeneous shell of pressureless dust (non-interacting particles) that falls into a black brane from a radial location $r= s_0$ (see figure \ref{fig:Lorentzian}).
%

\subsection{Renormalized on-shell action}\label{sec:VaidyaS}

To evaluate the on-shell renormalized action associated to the Euclidean shell geometry constructed in the previous section, we consider separately the two disconnected geometric patches inside and outside the shell, plus a perfect fluid shell
\be
S = S^\out + S^\ins  + S^\shell \, . 
\ee

The on-shell action for a perfect fluid is proportional to its pressure \cite{Brown:1992kc}
\be
S^\shell  = \int d^3x\sqrt{g}~P 
\ee
and thus vanishes in the present case of a shell of pressureless dust. 

The complete Euclidean bulk action to be evaluated on-shell for the region  $\mathcal{M}$ outside the geodesic shell is  
\bea \label{eq:Aout}
S^{\rm out}  &=&  S^{\rm bulk}  + S^{\rm GHY} + S^{\rm corner} \\
&=&- \frac{1}{16 \pi G_N}  \int_{\mathcal{M}} d^3 x \sqrt{g} ~(R-2\Lambda)  -\frac{1}{8 \pi G_N}  \int_{\del\mathcal{M}} d^2 x \sqrt{\gamma}K  -\frac{1}{8 \pi G_N}  \int_{\del\del \mathcal{M}} dx  \sqrt{h} \Theta \, ,  \nonumber 
\eea
with  $\Lambda = - 1$. The  Gibbons-Hawking-York term  has in this case two contributions, as $\del \mathcal{M}$ is the union of the surface extending at $r=r_\infty$ (with $r_\infty$ being a IR-bulk regulator) and of the surface $s=s_0$.  The extrinsic curvature $K_{\mu\nu} = \nabla_{(\mu}\hat n_{\nu)}$, with  $\hat n$ the outgoing normal vector to the boundary $\del \mathcal{M}$, jumps at the union of this two surfaces. This gives origin to the corner terms part of the action, that is evaluated at the non-smooth boundaries  $\del\del \mathcal{M}$ of  $\del \mathcal{M}$. 
The corner action is obtained in terms of 
\be \label{eq:theta}
\Theta = \arccos |\hat n_{(s_0)} \cdot \hat n_{(r_{\infty})}|\
\ee
with $  \hat n_{(r_{\infty})}$ and $\hat n_{(s_0)}$ the unit outgoing normal repectively to   the  $r_\infty$ and   $s_0$ surfaces \cite{Hayward:1993my,Brill:1994mb}. 
The  sign of the angle depends on whether the normals are converging, $\Theta>0$, or diverging $\Theta <0$. 

As we work with different patches of the shell geometry, which  locally are each  equal to a Euclidean black brane,  we find it more convenient to work in the  $(\tau, r, x)$  coordinates of the metric \eqref{eq:BTZE}. 
Also, all integrals  extend in the  homogeneous $x$-direction. Its measure can be simply factored out and expressed in terms of an  IR regulator $L$.

The (radially IR-regulated) region $\mathcal{M}$  corresponds to $r_\infty>r> r_{s_0}(\tau)$ and  $-\tau_\infty >\tau >\tau_{\infty}$. Here $r_{s_0}(\tau) $ parametrizes the surface described by the geodesic with opening $2 \tau_0$  on the boundary and turning radius $s_0$, as obtained inverting   eq.~\eqref{eq:tauofr} and using eq.~\eqref{eq:s0}
\be \label{eq:s0surf}
r_{s_0}(\tau) =   \frac{\sqrt M \cos (\sqrt M \tau) }{\sqrt{s_0^2 - \sin^2 (\sqrt M \tau)}}  \, .
\ee
Instead $\tau_\infty$ is the time coordinate along such a  geodesic corresponding to the regulated radial position $r_\infty$
\be \label{eq:tauinfty}
\tau_{\infty} =     \frac{1 }{\sqrt M} \ {\rm arcsin} \(\frac{\sqrt M}{s_0} \sqrt{\frac{r_\infty^2-s_0^2}{r_\infty^2- M}}\)  \, . 
\ee
The action associated to the bulk $\mathcal{M}$ is then
\begin{align}
S_{\rm reg}^{\rm bulk} &= \frac{L}{16 \pi G_N}  \left[ 4 (r^2_\infty - M)  \tau_\infty  +  \sqrt M \cot (\sqrt M \tau_0) \log\( \frac{  \sin^2 \( \sqrt M(  \tau_0-\tau_\infty)\)}{ \sin ^2\( \sqrt M( \tau_0 + \tau_\infty)\)}  \)   \right] \,. 
 \end{align}

The GHY term has a contribution evaluated at the AdS boundary $r=r_\infty$, to which the time integral provides only the measure $2 \tau_\infty$ of the time interval at that radial location.  The outgoing normal at the AdS boundary $r=r_\infty$ is
\be
 \hat n_{(r_{\infty})} =  \sqrt{g_{rr}} \ dr = \frac{1}{ \sqrt{r_{\infty}^2 - M}} \ dr\, ,
\ee
which leads to  
\begin{align}
S_{\rm reg}^{\rm GHY(r_{\infty})}   &= -\frac{L}{16 \pi G_N}  \left[  4 (2r^2_\infty - M)  \tau_\infty  \] \, .  
 \end{align}
Both actions vanish in the limit of zero size interval $\t_\infty \to 0$ at the regulated boundary $r=r_\infty$.

The second contribution comes from the geodesic shell itself. 
The corresponding surface can be described in terms of  eq.~\eqref{eq:s0surf}, with 
outgoing normal
\be
\hat n_{(s_{0})} = \sqrt{r^2 - s^2_0} \ d\t  -\frac{\sqrt{ s^2_0- M } }{r^2 - M}\ dr \, , 
\ee 
and induced metric
\be
\gamma_{ij} = \left(\begin{array}{cc}\frac{1}{r^2 - s_0^2} & 0 \\0 & r^2\end{array}\right) \, ,
\ee
where  we used  eq.~\eqref{eq:s0surf} in order to eliminate $\t$. 
Computing the associated extrinsic curvature along this surface, the  contribution from the shell to the GHY action is   
\begin{align}
S_{\rm reg}^{\rm GHY(s_0)} 
&=-\frac{L}{16 \pi G_N} \[ \sqrt M \cot (\sqrt M \tau_0) \log\( \frac{  \sin^2 \( \sqrt M(  \tau_0-\tau_\infty)\)}{ \sin ^2\( \sqrt M( \tau_0 + \tau_\infty)\)}  \) \]   \, . 
\end{align}
This term  precisely cancels the logarithmic divergence in $S^{\rm bulk}$.

Finally the corner contribution  accounting for the discontinuity in the boundary region is localized at $r_\infty$ and $\pm\tau_\infty$ and extends in the homogeneous $x$ direction.  The induced metric is just $h_{xx} = g_{xx} = r_{\infty}^2$. The two normals along the $s_0$ and $r_\infty$ surfaces diverge one from the other and therefore the corresponding angle $\Theta$ defined in eq.~\eqref{eq:theta} is to be taken negative. The contribution of  a single corner  is 
\be \label{eq:onecorner}
S_{\rm {reg}}^{\rm corner (\pm \tau_\infty)} =     \frac{L}{16 \pi G_N}  \left[ 2r_{\infty} {\rm arccos} \( \sqrt{\frac{s_0^2-M}{r_\infty^2- M}}  \)\right]  \, .
\ee
and since the two corner contributions come with the same sign, the total  corner action is two times  eq.~\eqref{eq:onecorner}, which rewritten in terms of $\tau_{\infty}$ reads
\begin{align}
S_{\rm {reg}}^{\rm corners } 
&=     \frac{L}{16 \pi G_N}  \left[4 r_{\infty} {\rm arcsec} \(  \sqrt{1+ \frac{r_\infty^2}{M} \tan^2 (\sqrt M \tau_\infty) } \) \right] \, .\end{align}
It is easy to see this contribution vanishes as $\tau_\infty \to 0$, while the sum of the angles associated to the corners becomes exactly $\pi$ for a geodesic that cuts in two halves the full black hole space: $\tau_0  =\tau_\infty = \pi / (2 \sqrt M)$.  

Taking all these contributions into account, the regularized action for the geometry associated to the region outside the shell is
\begin{align}
S^{\rm out}_{\rm reg} &=  -\frac{L}{16 \pi G_N}  \Bigg\{4 r^2_\infty \tau_\infty  -   4 r_{\infty} {\rm arcsec} \[  \sqrt{1+ \frac{r_\infty^2}{M} \tan^2 (\sqrt M \tau_\infty) } \] \Bigg\} \, .  
\end{align}
The regulated action of a full black brane instead is
\begin{align}
S_{\rm reg}^{\rm BH} &=  - \frac{1}{16 \pi G_N}  \int_{\mathcal{M}} d^3 x \sqrt{g} ~(R-2\Lambda)  -\frac{1}{8 \pi G_N}  \int_{\del \mathcal{M}} d^2 x \sqrt{\gamma}K = - \frac{L}{16 \pi G_N} \frac{4\pi}{\sqrt M} r_\infty^2\,,
\end{align}
and thus the action associated to the region inside a geodesic shell is obtained from the result above as
\begin{align}
S^{\rm in}_{\rm reg} &=S_{\rm reg}^{\rm BH} - S^{\rm out}_{\rm reg}\\
&= -\frac{L}{16 \pi G_N}  \Bigg\{4 r_\infty^2 \(\frac{ \pi}{ \sqrt M} -  \tau_\infty \) +   4 r_{\infty} {\rm arcsec} \[  \sqrt{1+ \frac{r_\infty^2}{M} \tan^2 (\sqrt M \tau_\infty) } \] \Bigg\}\,. \nonumber 
\end{align}
Some comments are in order, for $\tau_\infty =0$ $S^{\rm in}_{\rm reg} = - \frac{L}{16 \pi G_N} \frac{4\pi}{\sqrt M} r_\infty^2$ gives the full black brane action. On the other hand, when the shell divides exactly in two halves the black brane geometry, that is for $\tau_\infty = \pi/(2\sqrt M)$, $S^{\rm out}_{\rm reg}$  and $S^{\rm in}_{\rm reg}$  do not coincide, but have corner contributions that differ by  $2 \pi$ 
\be
S^{\rm out}_{\rm reg}\(\tau_\infty = \frac{ \pi}{2\sqrt M}\) = S^{\rm in}_{\rm reg} \(\tau_\infty =  \frac{ \pi}{2\sqrt M}\) + 2 r_\infty \,\, 2\pi. 
\ee
This discrepancy is related to the usual issue of the (non)additivity of the Euclidean gravitational action \cite{Brill:1994mb}. In the present case,   our choice can be understood  by regulating the junction between the two black brane geometries by introducing a finite width shell of the pressureless fluid (\eg see \cite{newer}), as we comment below. \\

Using the results derived so far  and   taking   the inside and outside geometries to have  different masses, we can straightforwardly write the regulated on-shell action of the shell solution: 
\begin{align}
S_{\rm reg} &= S^{\rm in}_{\rm reg} +S^{\rm out}_{\rm reg}   \nonumber \\
&=    - \frac{L}{16 \pi G_N} \Bigg\{  4 r_\infty^2 \(\frac{ \pi}{ \sqrt{M_{\rm in}}} -  \tau_{\infty, {\rm in}}+  \tau_{\infty, {\rm out}} \) +     4 r_{\infty} {\rm arcsec} \[  \sqrt{1+ \frac{r_\infty^2}{M_{\rm in}} \tan^2 (\sqrt{M_{\rm in}} \tau_{\infty, {\rm in}}) } \]  \nonumber\\
& \qquad  -   4 r_{\infty} {\rm arcsec} \[  \sqrt{1+ \frac{r_\infty^2}{M_{\rm out}} \tan^2 (\sqrt{M_{\rm out}}\tau_{\infty, {\rm out}}) } \] \Bigg\}  \, .  
\end{align}
To renormalize the quadratic divergences, we use the standard bulk counterterm 
\begin{align}
S_{\rm ct}& =    \frac{1}{16 \pi G_N} \int_{r=r_\infty} d^2 x \ 2 \sqrt{\gamma}  \\ 
&=  \frac{L}{16 \pi G_N}  \[ 4    r_\infty\sqrt{r^2_\infty - M_{\rm out}}  \tau_{\infty, {\rm out}} + 4   r_\infty\sqrt{r^2_\infty - M_{\rm in}} \(\frac{ \pi}{ \sqrt{M_{\rm in}}} -  \tau_{\infty, {\rm in}}\) \]  \nonumber 
\end{align}
and obtain  
\begin{align}
&S_{\rm ren}  =   S_{\rm reg}  + S_{\rm ct} \\
&= \lim_{r_\infty \to \infty} - \frac{L}{16 \pi G_N} \Bigg\{ 2 M_{\rm out} \tau_{\infty, {\rm out}} + 2M_{\rm in}\(\frac{ \pi}{ \sqrt{M_{\rm in}}} -  \tau_{\infty, {\rm in}} \)   \nonumber\\
&   +    4 r_{\infty} {\rm arcsec} \[  \sqrt{1+ \frac{r_\infty^2}{M_{\rm in}} \tan^2 (\sqrt{M_{\rm in}} \tau_{\infty, {\rm in}}) } \]  -  4 r_{\infty} {\rm arcsec} \[  \sqrt{1+ \frac{r_\infty^2}{M_{\rm out}} \tan^2 (\sqrt{M_{\rm out}} \tau_{\infty, {\rm out}}) } \]  \Bigg\}\,. \nonumber 
\end{align}
While the result is finite as we remove the cutoff, we here have not explicitly evaluated the limit of the trigonometric functions, as to have an expression with a well defined limit when the boundary time interval is sent to zero.  
Eliminating $\tau_{\infty, {\rm in}}$  and $\tau_{\infty, {\rm out}}$  in favour of $s_0$ using eq.~\eqref{eq:tauinfty}, the complete renormalized action then  gives\footnote{ \label{foot:thick}
When $M(s)$ describes a thick shell of width $2\d$ centered around $s_0$,  the geometry \eqref{eq:EVaidyaCigar} continuously interpolates between the two values of the mass function across the shell. In this case, there are no corner terms in the action and the GHY term only has a contribution from the AdS boundary, which yields
\begin{align}
S^{\rm GHY}   = \frac{L}{8\pi G_N}   \int ds \left\{ \frac{2 r_\infty}{s} \frac{M(s) - 2r_\infty^2}{\sqrt{(r_\infty^2-s^2)(s^2 -M(s))}} + \frac{r_\infty\sqrt{r_\infty^2 - s^2} M'(s)}{(r_\infty^2 -M(s))\sqrt{s^2 -M(s)}} \right\}  \,.
\end{align}
In the thin shell limit $\delta \to 0 $, the only non-vanishing contribution from the region where the shell is present is given by the term proportional to $M'(s)$, which can be evaluated with an integration by parts, exactly reproducing the corner contributions in eq.~\eqref{eq:SrenShell},
\begin{align}
    \int_{s_0 -\d}^{s_0+\d} ds  \frac{r_\infty\sqrt{r_\infty^2 - s^2} M'(s)}{(r_\infty^2 -M(s))\sqrt{s^2 -M(s)}} \, \to \,  2 r_\infty \left\{  {\rm arccos} \sqrt{\frac{s_0^2 - M_{\rm out}}{r_\infty^2 -M_{\rm out}}} -   {\rm arccos} \sqrt{\frac{s_0^2 - M_{\rm in}}{r_\infty^2 -M_{\rm in}}} \right\} \ \, . 
\end{align}
 }
%
\begin{align}
&S_{\rm ren} =   S_{\rm reg}  + S_{\rm ct} \nonumber \\
&=  \lim_{r_\infty \to \infty} -\frac{L}{16 \pi G_N} \Bigg\{ 2\sqrt{M_{\rm out}} \, {\rm arcsin} \(\frac{\sqrt{M_{\rm out}}}{s_0} \sqrt{\frac{r_\infty^2-s_0^2}{r_\infty^2-M_{\rm out}}}\)  -   4 r_{\infty} {\rm arccos} \[  \sqrt{\frac{s_0^2-M_{\rm out}}{r_\infty^2- M_{\rm out}}} \] \nonumber\\
& + 2\sqrt{M_{\rm in}} \[\pi  -   {\rm arcsin} \(\frac{\sqrt{M_{\rm in}}}{s_0} \sqrt{\frac{r_\infty^2-s_0^2}{r_\infty^2- M_{\rm in}}}\)\] 
+   4 r_{\infty} {\rm arccos} \[  \sqrt{\frac{s_0^2-M_{\rm in}}{r_\infty^2- M_{\rm in}}}  \]  \Bigg\} \, . \label{eq:SrenShell}
\end{align}
%
  
\subsection{Trace function interpretation}\label{sec:interpretation}

We would here like to interpret the geometric construction of the previous sections in terms of the boundary CFT. For that we find it instructive to briefly recall what is the CFT dual of the familiar AdS$_3$-Vaidya solution, interpolating between vacuum AdS and a black hole geometry across the location of an infalling shell of null dust. The CFT dual state was worked out in \cite{Anous:2016kss} and it is obtained through the insertion of a discretized shell of local operators at some $t=t_0$ in the vacuum state. Although the system remains in a pure state throughout,  at times after the shell is injected, it appears for most observables to be a thermal state, with an effective temperature fixed by the energy injected \cite{AbajoArrastia:2010yt,Albash:2010mv,Balasubramanian:2010ce,Balasubramanian:2011ur}. 
Implicitly, this description assumes that an external operator is turned on for an instant at $t=t_0$ to excite the vacuum state (\ie we have a quantum quench where the CFT Hamiltonian is momentarily deformed). Alternatively, we can think that we have an excited state that is evolved by the undeformed Hamiltonian \cite{multi1}. However, in this case, the correct bulk solution would involve a null shell which emerges from a white hole in the past, reflects off of the asymptotic boundary at $t=t_0$ and then collapses to form a black hole in the future. In the boundary theory, we would have an excited state where the UV degrees freedom seem to be thermalized but then the excitation momentarily coheres at $t=t_0$ before dissipating into an apparent thermal ensemble again. 

Our interpretation of the Euclidean shell solution is similar in spirit.  From the boundary point of view, we can think of it as a CFT on the thermal cylinder with appropriate insertions, which support the homogeneous shell of non-interacting particles in the bulk. As in AdS-Vaidya  case, we can think that the insertions have effectively modified the temperature past the shell.  However, rather than including the boundary insertions, we can think purely in terms of the excited CFT state. In particular, here we have excitations on top of the thermal field double (TFD) state with temperature $1/\beta_{\rm in}$, which produce an apparently thermal state with a temperature $1/\beta_{\rm out}$. The bulk shell is not reflected at the asymptotic boundary but this reflects the fact that the corresponding boundary partition function ties together two different states, the original TFD and the excited state.

Following this reasoning, we interpret the boundary partition function associated to our Euclidean shell geometry as computing a trace of the form
\be \label{eq:democraticTr}
\log \tr \(\rho_{\rm out}^{\a_{\rm out}} \rho_{\rm in}^{1-\a_{\rm in}}\) = - S_{\rm ren} \,,
\ee
by identifying 
\be
\a_{\rm out} \b_{\rm out} \equiv 2\,\tau_{\rm out}\,, \qquad  \a_{\rm in} \b_{\rm in}  \equiv 2\,\tau_{\rm in}\,, 
\ee
and where $\b_{\rm in}=2\pi /\sqrt{M_{\rm in}}$ and   $\b_{\rm out} = 2\pi /\sqrt{M_{\rm out}}$. The matrix $\rho_{\rm in}$ is a thermal density matrix at inverse temperature $\b_{\rm in} $, while $\rho_{\rm out}$ is an excited state at an apparent inverse temperature $\b_{\rm out} $. 
The exponents $\a_{\rm out}, \a_{\rm in}$ are not independent in this geometric construction, but related by the constraint 
\be
s_0 = \frac{\sqrt{M_{\rm out}}}{\sin (\pi \a_{\rm out})} =  \frac{\sqrt{M_{\rm in}}}{\sin (\pi \a_{\rm in})}\,.
\ee
Observe however that, since $\rho_{\rm in}$ is a thermal density matrix, the trace \eqref{eq:democraticTr}  is of the type of those studied in sec.~\ref{sec:Xeno} for general reference states. In fact, we can write explicitly
\be \label{eq:democraticR}
\log \tr \(\rho_{\rm out}^{\a_{\rm out}} \rho_{\rm in}^{1-\a_{\rm in}}\) = \log \tr \(\rho_{\rm out}^{\a_{\rm out}} \rho_{R}^{1-\a_{\rm out}}\)   \,,
\ee
introducing the reference inverse temperature
\bea
 \b_R &=& \frac{1-\a_{\rm in}}{1-\a_{\rm out}} \, \b_{\rm in}   \\
&=& \frac{1- \frac 1 \pi {\rm arcsin} \(\sqrt{\frac{M_{\rm in}}{M_{\rm out}}} \sin(\pi \a_{\rm out})\)}{1-\a_{\rm out}} \,  \frac{2\pi}{\sqrt{M_{\rm in}}} \,. \label{eq:betaR}
\eea

As $s_0 \to r_\infty$, both  $\a_{\rm out}$ and  $\a_{\rm in}$ go to zero, and we recover  
\be
\log \tr \rho_{\rm in} = \frac{c \, \pi L}{6 \b_{\rm in}} \,.
\ee
As $s_0$ decreases, $\a_{\rm out}$ and  $\a_{\rm in}$ increase. Since our Euclidean shell geometry is only well defined for $s_0 \in \[{\rm Max}(\sqrt{M_{\rm in}},\sqrt{M_{\rm out}}), r_\infty\]$, we find that the range of $\a_{\rm out}, \a_{\rm in}$ we can cover is 
\be
\left\{\begin{array}{lll}
\a_{\rm out}  \in \[0, \frac 1 2\]\,,   & \qquad \a_{\rm in} \in \[0, \frac 1 \pi {\rm arcsin} \frac{\b_{\rm out}}{\b_{\rm in}} \]\,, & \qquad\mbox{for } M_{\rm out} \ge M_{\rm in} \\
\a_{\rm out}  \in \[0, \frac 1 \pi {\rm arcsin} \frac{\b_{\rm in}}{\b_{\rm out}}  \]\,,  & \qquad  \a_{\rm in} \in \[0, \frac 1 2\]\,, &\qquad  \mbox{for } M_{\rm out} \le M_{\rm in}
\end{array}\right. \,.
\ee
Even if the values $\a_{\rm out} = \a_{\rm in} = 1$ are formally not included in the geometric construction of the trace function \eqref{eq:democraticTr}, we still assume that
\be
\log \tr \rho_{\rm out} =  \frac{c \, \pi L}{6 \b_{\rm out}}\,,
\ee
consistently with the symmetry of our construction that would leave us with the full \emph{out} geometry when the \emph{in} geometry has shrunk to zero size. 
 
\begin{figure}[t]
\centering 
\includegraphics[width=.45\textwidth]{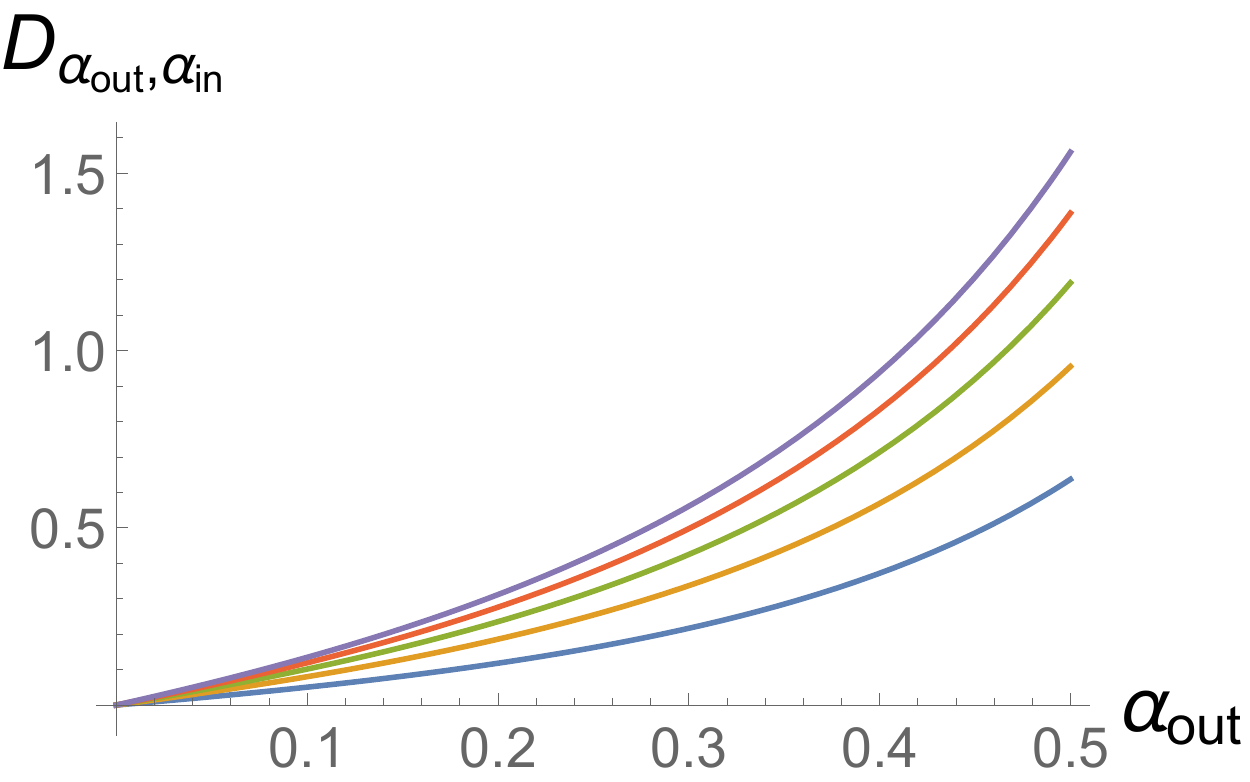} \hfill \includegraphics[width=.45\textwidth]{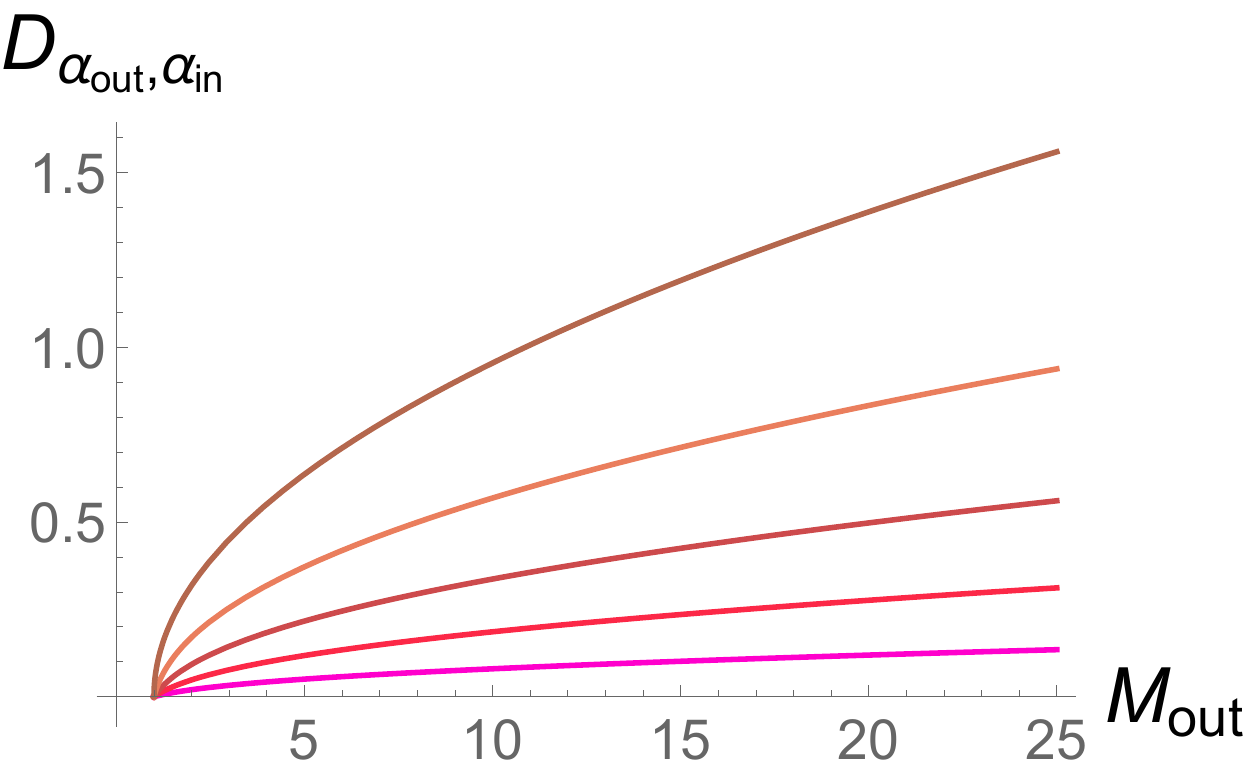} 
\caption{(Left) $D_{\a_{\rm out}, \a_{\rm in}}$ for $M_{\rm in} =1, M_{\rm out}= 5,10,15,20,25$ from the bottom-up. (Right) $D_{\a_{\rm out}, \a_{\rm in}}$ as a function of $M_{\rm out}$ for $M_{\rm in} =1$ and $\a_{\rm out}= 0.1, 0.2, 0.3, 0.4, 0.5$ from the bottom-up. We rescaled the prefactor $c/3$.} 
\label{fig:Da_outin}
\end{figure}

Normalizing the density matrices accordingly,  we define 
\begin{align}
&D_{\a_{\rm out},\a_{\rm in}} (\rho_{\rm out}\Vert\rho_{\rm in}) \equiv
  \frac{1}{\a_{\rm out}-1} 
  \log \frac{\tr \(\rho_{\rm out}^{\a_{\rm out}} \rho_{\rm in}^{1-\a_{\rm in}}\) }{\(\tr \rho_{\rm out}\)^{\a_{\rm out}}  \(\tr \rho_{\rm in}\)^{1-\a_{\rm in}}} \label{eq:Daoutain} \\
&=  \lim_{r_\infty \to \infty} 
\frac{c \, r_{\infty}}{ 6\pi(\a_{\rm out}-1)}   \left\{ {\rm arccos} \[  \sqrt{\frac{s_0^2-M_{\rm in}}{r_\infty^2- M_{\rm in}}}  - {\rm arccos} \[  \sqrt{\frac{s_0^2-M_{\rm out}}{r_\infty^2- M_{\rm out}}} \]  \]  \right\} \nonumber\\
& \approx 
 \frac{c}{3}\frac{1}{\a_{\rm out}-1} 
 \left\{ \frac{\cot(\pi \a_{\rm out})}{\b_{\rm out}}  - \frac{ \cot(\pi \a_{\rm in}) }{\b_{\rm in}}\right\}\,. 
\end{align}
Upon normalization, the trace function $D_{\a_{\rm out}, \a_{\rm in}}$ is therefore fully given by the corner terms in the action. It is identically vanishing for $M_{\rm in} = M_{\rm out}$, for which indeed  $\rho_{\rm in} = \rho_{\rm out}$ and $\a_{\rm in}=\a_{\rm out}$. We plot it in figure~\ref{fig:Da_outin} for $M_{\rm out}  \ge M_{\rm in}$, both as a function of $\a_{\rm out} \in [0,1/2]$ and $M_{\rm out}$. Notice  this is the parameter range for which we expect a physically sensible Lorentzian continuation of the geometry with a shell of positive energy density.

\begin{figure}[h]
\centering 
\includegraphics[width=.7\textwidth]{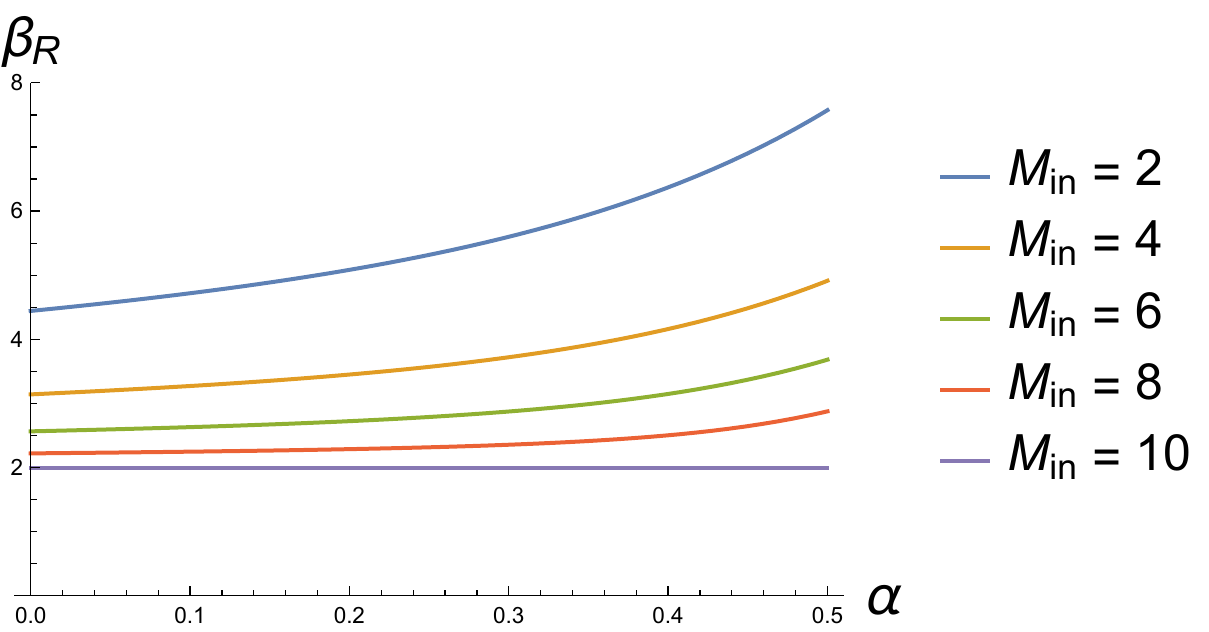} 
\caption{Reference inverse temperatures $\b_R$ that can be accessed with this Euclidean geometric construction as a function of $\a$ for $M_{\rm out}=10$ and varying $M_{\rm in} \le M_{\rm out}$.} 
\label{fig:betaR}
\end{figure}

In order to study the monotonicity constraints
\be \label{eq:monotR}
D_{\a} (\rho \Vert\rho_{R}) \ge D_{\a} (\rho' \Vert\rho_{R})
\ee
with respect to a thermal reference state $\rho_R$, we need to consider the quantity
\begin{align}
D_{\a} (\rho\Vert\rho_{R})
&= D_{\a_{\rm out},\a_{\rm in}} (\rho_{\rm out}\Vert\rho_{\rm in}) \label{eq:translate} \\
&= \frac{c}{6\pi} \frac{\sqrt{M_{\rm out}}}{\a-1} \left\{ \cot(\pi \a) - \sqrt{\frac{1}{\sin^2 (\pi \a)} - \frac{M_{\rm in}}{M_{\rm out}}} \right\}\,,
\end{align}
where we identify $\a= \a_{\rm out}$, $\rho = \rho_{\rm out}$ and $\b_R = 2\pi/ \sqrt{M_{\rm R}}$ 
is given by eq.~\eqref{eq:betaR} in terms of $\a, M_{\rm in}, M_{\rm out}$. The excited state $\rho =\rho_{\rm out}$ is specified in this geometric construction by both the mass of the background geometry $M_{\rm in}$ and that of the shell $M_{\rm out} - M_{\rm in}$, or equivalently by the pair $(M_{\rm in}, M_{\rm out})$. If we want to keep  fixed the excited state, for each given $\a$ we thus have a different value of reference temperature fixed by eq.~\eqref{eq:betaR}. This construction alone therefore does not allow to explore the full $(\a,\b_R)$ parameter space of constraints, but only fixed trajectories in that plane, as those plotted in figure~\ref{fig:betaR}.  

\begin{figure}[t]
\centering 
\includegraphics[width=.45\textwidth]{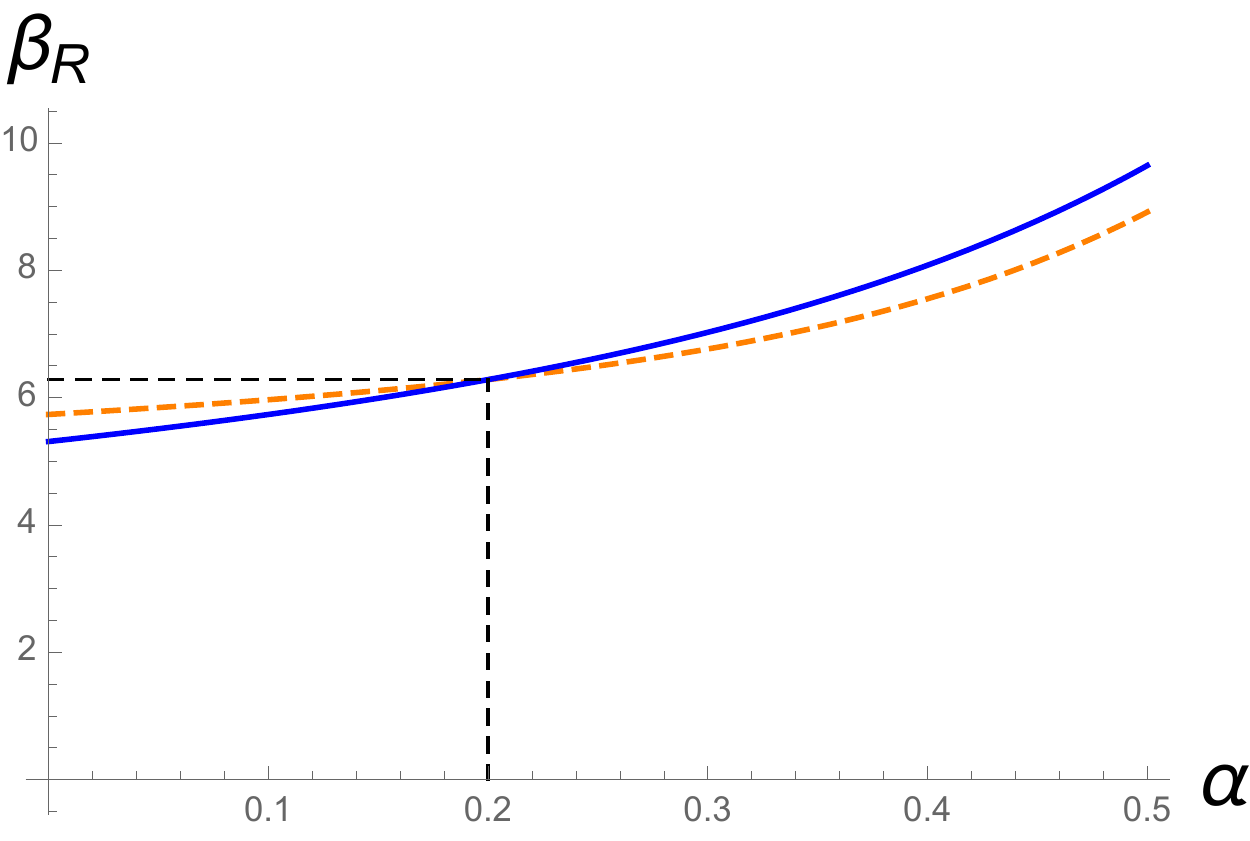} \hfill \includegraphics[width=.45\textwidth]{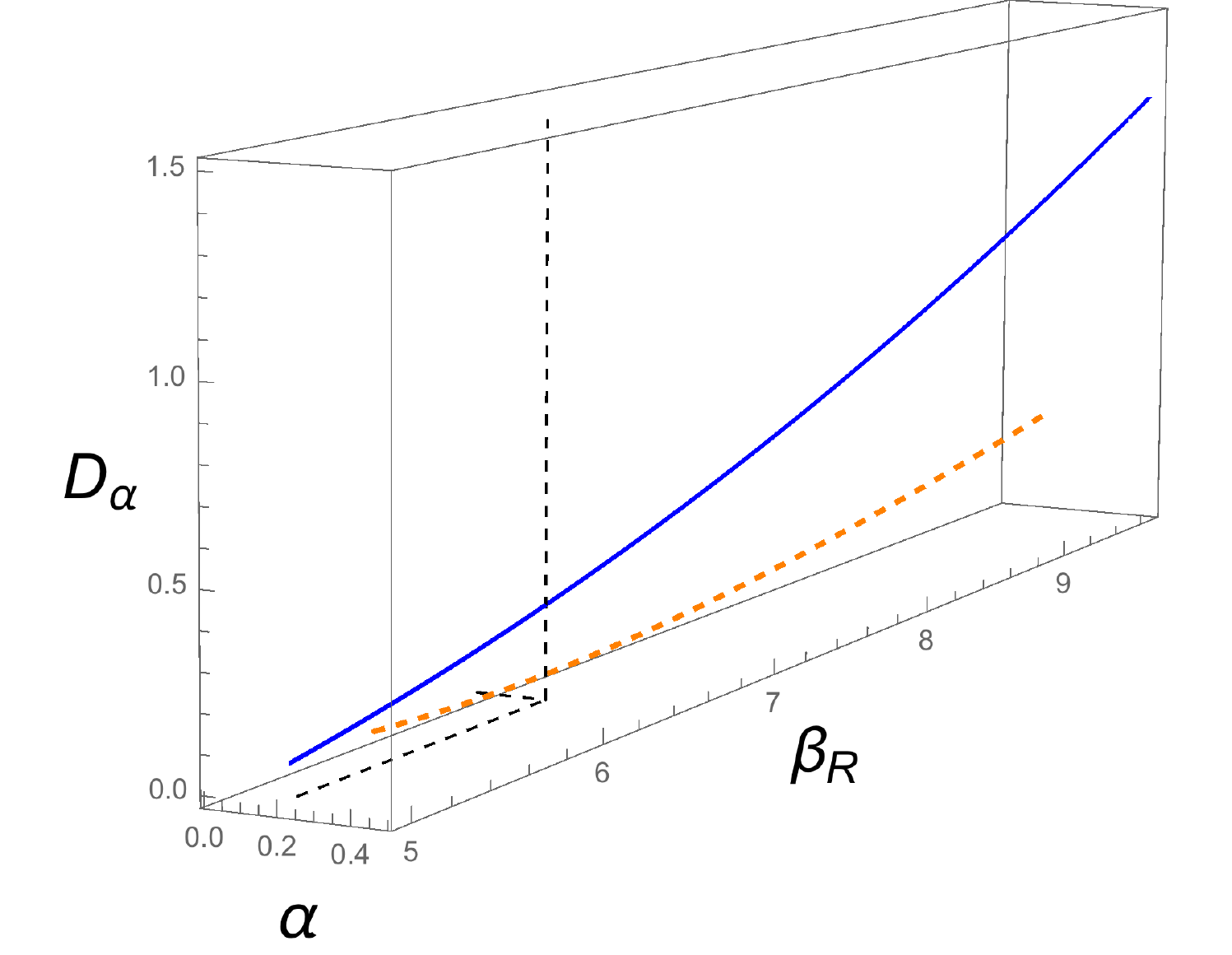} 
\caption{(Left) Reference inverse temperature $\b_R$ for $(M_{\rm in} , M_{\rm out})\approx (1.2, 2.9), (1.4,17.3)$ respectively in orange dashed and blue, as a function of $\a$. The two curves cross at $(\b_R, \a) = (2\pi,0.2)$.  (Right) $D_\a(\rho || \rho_R)$ for the two excited states. The monotonicity constraint \eqref{eq:monotR} at $(\b_R, \a) = (2\pi,0.2)$ therefore forbids transitions from the state $(M_{\rm in} , M_{\rm out})\approx (1.2, 2.9)$ to that with $(M_{\rm in} , M_{\rm out})\approx (1.4,17.3)$. We rescaled the prefactor $c/3$.} 
\label{fig:DabetaR}
\end{figure}

Still, varying the excited state $(M_{\rm in}, M_{\rm out})$, we can study some of the second laws \eqref{eq:monotR}. As an example, we plot in figure~\ref{fig:DabetaR} (right panel) the R\'enyi divergence \eqref{eq:translate} for two different excited states $(M_{\rm in} , M_{\rm out})\approx (1.2, 2.9), (1.4,17.3)$. The two curves can be compared at the point $(\b_R, \a) = (2\pi,0.2)$ (see left panel), and the corresponding monotonicity constraint  \eqref{eq:monotR} rules out transitions from $(M_{\rm in} , M_{\rm out})\approx (1.2, 2.9)$ to $(M_{\rm in} , M_{\rm out})\approx (1.4,17.3)$. 

In general, at each point in the $(\b_R, \a)$-plane there will be an infinite number of curves that cross, associated to all pairs $(M_{\rm in} , M_{\rm out})$ that solve eq.~\eqref{eq:betaR}, and thus an infinite number of excited states that can be compared. The constraints \eqref{eq:monotR} will define an ordering for all of them. 
It is however important to stress once more that within this geometric interpretation, we are not able to compare the relative strength of the various constraints, as we cannot vary $\a$ and $\b_R$ independently. Further, we are unable to take the limit $\a\to 1$, which would allow us to access the usual thermodynamic constraints.
\newline

Above, exploiting the fact that $\rho_{\rm in}$ is a thermal state, we were able to recast the quantity \eqref{eq:Daoutain} in the form of a R\'enyi divergence. We would here like to briefly discuss a more general property that holds for a quantity of the form \eqref{eq:Daoutain}.  
According to Lieb's concavity theorem in fact the trace function 
\be
f_{\a_{\rm out}, \a_{\rm in}}(\rho_{\rm out}\Vert\rho_{\rm in}) \equiv  \tr \( \rho_{\rm out}^{\a_{\rm out}} \rho_{\rm in}^{1-\a_{\rm in}} \)
\ee
is jointly concave for $\a_{\rm in}, \a_{\rm out} \in [0,1]$, $\a_{\rm out}  \le \a_{\rm in}$  \cite{Lieb}. Joint concavity means that for any two pairs of normalized density matrices $(\rho_{\rm out},\rho_{\rm in})$, $(\rho'_{\rm out},\rho'_{\rm in})$ and any $0< \kappa < 1$:
\begin{align}
f_{\a_{\rm out}, \a_{\rm in}}\Big((1-\kappa) \rho_{\rm out} + \kappa \rho'_{\rm out} \Vert& (1-\kappa) \rho_{\rm in} + \kappa \rho'_{\rm in}\Big) \ge \nonumber\\
& (1-\kappa)f_{\a_{\rm out}, \a_{\rm in}}(\rho_{\rm out}\Vert\rho_{\rm in}) + \kappa f_{\a_{\rm out}, \a_{\rm in}}(\rho'_{\rm out}\Vert\rho'_{\rm in}) \,.
\label{eq:jconc}
\end{align}
Equivalently, the quantity with opposite sign is jointly convex. Since the logarithm is monotonic and operator concave on $(0,\infty)$, the quantity, which we defined in eq.~\eqref{eq:Daoutain}, is jointly convex. In our construction we would expect joint convexity to hold for $M_{\rm out}  \le M_{\rm in}$, that is the range of masses where the geometric construction does not have a physically sensible Lorentzian continuation.  Join convexity is a useful property to have if optimising this quantity.  Together with unitary invariance and invariance under tensor products, joint convexity implies the data processing inequality (see for instance  Thm. 5.16 in \cite{Wolf}). In the case of the R\'enyi divergence, these properties allow one to prove its monotonicity under CPTP maps, but do not hold for generic $\a_{\rm out}, \a_{\rm in}$. Take for instance $\rho_{\rm out}=\rho_{\rm in}$ a pure state, so that the quantity of eq.~\eqref{eq:Daoutain} is zero. Then tracing out a share of each state will make this quantity increase unless $\a_{\rm out}= \a_{\rm in}$ (just as the R\'enyi entropy defined in eq.~\eqref{eq:renyi-entropy}). 

Note that joint concavity (convexity) implies concavity (convexity) for each of the arguments of the above function, and furthermore implies that $\sup_{\a_{\rm in}}- f_{\a_{\rm out}, \a_{\rm in}}$ is concave, which suggests that this quantity can translate into entropic like quantities under a suitable restriction of the reference set of states.
Exploring the properties of $f_{\a_{\rm out}, \a_{\rm in}}$ 
when the range of parameters are restricted is a potentially interesting line of research.



\begin{thebibliography}{150}

\bibitem{JB1}
  J.~D.~Bekenstein,
  ``Black holes and the second law,''
  Lett.\ Nuovo Cim.\ {\bf 4} (1972) 737.
  
\bibitem{Bekenstein:1973ur}
  J.~D.~Bekenstein,
  ``Black holes and entropy,''
  Phys.\ Rev.\ D {\bf 7} (1973) 2333.

\bibitem{Hawking:1974rv}
  S.~W.~Hawking,
  ``Black hole explosions,''
  Nature {\bf 248} (1974) 30.

\bibitem{Hawking:1974sw}
  S.~W.~Hawking,
  ``Particle Creation by Black Holes,''
  Commun.\ Math.\ Phys.\  {\bf 43} (1975) 199
   Erratum: [Commun.\ Math.\ Phys.\  {\bf 46} (1976) 206].

\bibitem{area}
  S.~W.~Hawking, ``Black holes in general relativity," Commun.\ Math.\ Phys.\ {\bf 25} (1972) 152.
  
\bibitem{Hawking:1973uf}
  S.~W.~Hawking and G.~F.~R.~Ellis,
  ``The Large Scale Structure of Space-Time,''  (Cambridge
University Press, Cambridge, 1973).

\bibitem{ruch1978mixing}
E.~Ruch, R.~Schranner, and T.~H. Seligman, ``The mixing distance,''   J.\ Chem.\ Phys. {\bf 69} (1978), no.~1 386--392.
%
\bibitem{janzing2000thermodynamic}
D.~{Janzing}, P.~{Wocjan}, R.~{Zeier}, R.~{Geiss}, and T.~{Beth}, 
``Thermodynamic Cost of Reliability and Low Temperatures: Tightening
  Landauer's Principle and the Second Law,''  Int.\ J.\ Theor.\ Phys. {\bf
  39} (2000), no.~12 2717--2753 [\arXiv{quant-ph/0002048}].

\bibitem{uniqueinfo}
M.~Horodecki, P.~Horodecki, and J.~Oppenheim, ``Reversible transformations
  from pure to mixed states and the unique measure of information,''  Phys.\
  Rev.\ A {\bf 67} (2003) 062104   [\arXiv{quant-ph/0212019}].
 
\bibitem{dahlsten2011inadequacy}
O.~Dahlsten, R.~Renner, E.~Rieper, and V.~Vedral, ``Inadequacy of von
  Neumann entropy for characterizing extractable work'',  New J. Phys.  {\bf 13} (2011) 053015 	
  [\arXiv{arXiv:0908.0424} [quant-ph]].

\bibitem{del2011thermodynamic}
L.~Del~Rio, J.~{\AA}berg, R.~Renner, O.~Dahlsten, and V.~Vedral, ``The
  thermodynamic meaning of negative entropy,''  Nature {\bf 474} (2011),
  no.~7349 61--63 	[\arXiv{arXiv:1009.1630} [quant-ph]].

\bibitem{horodecki2013fundamental}
M.~{Horodecki} and J.~{Oppenheim}, ``Fundamental limitations for quantum
  and nanoscale thermodynamics,''   Nat. Commun. {\bf 4} (2013) 
  [\arXiv{arXiv:1111.3834} [quant-ph]].

\bibitem{aaberg2013truly}
J.~{\AA}berg, ``Truly work-like work extraction via a single-shot
  analysis,''  Nat. Commun. {\bf 4} (2013) 1925,
[\arXiv{arXiv:1110.6121} [quant-ph]].
 
\bibitem{brandao2013second}
F.~G.~S.~L. {Brandao}, M.~{Horodecki}, N.~H.~Y. {Ng}, J.~{Oppenheim}, and
  S.~{Wehner}, ``The second laws of quantum thermodynamics,''   Proc.
  Natl. Acad. Sci. {\bf 112} (2015), no.~11 3275--3279  
[\arXiv{arXiv:1305.5278} [quant-ph]].

\bibitem{lostaglio2015description}
M.~Lostaglio, D.~Jennings, and T.~Rudolph, ``Description of quantum
  coherence in thermodynamic processes requires constraints beyond free
  energy,''   Nat. Commun. {\bf 6} (2015) 6383 [\arXiv{arXiv:1405.2188} [quant-ph]].

\bibitem{faist2015minimal}
P.~Faist, F.~Dupuis, J.~Oppenheim, and R.~Renner, 
``The minimal work cost of information processing,'' 
Nat. Commun. {\bf 6}  (2015)  [\arXiv{arXiv:1211.1037} [quant-ph]]

\bibitem{egloff2015measure}
D.~Egloff, O.~C. Dahlsten, R.~Renner, and V.~Vedral, ``A measure of  majorization emerging from single-shot statistical mechanics,'' 
New J. Phys. {\bf 17}  (2015) 073001 [\arXiv{arXiv:1207.0434} [quant-ph]]

\bibitem{cwiklinski2015limitations}
P.~{\'C}wikli{\'n}ski, M.~Studzi{\'n}ski, M.~Horodecki, and J.~Oppenheim,
  ``Limitations on the evolution of quantum coherences: Towards fully quantum  second laws of thermodynamics,'' 
  Phys. Rev. Lett. {\bf 115}, 210403 (2015) [\arXiv{arXiv:1405.5029} [quant-ph]]

\bibitem{lostaglio2014quantum}
M.~Lostaglio, K.~Korzekwa, D.~Jennings, and T.~Rudolph, ``Quantum coherence,  time-translation symmetry, and thermodynamics,'' 
Phys. Rev. X {\bf 5}, 021001 (2015) [\arXiv{arXiv:1410.4572} [quant-ph]]
  
\bibitem{petz1986quasi}
D.~Petz, ``Quasi-entropies for finite quantum systems,'' { Reports on
  mathematical physics} {\bf 23} (1986), no.~1 57--65.

\bibitem{HiaiMPB2010-f-divergences}
F.~{Hiai}, M.~{Mosonyi}, D.~{Petz}, and C.~{B{\'e}ny}, ``Quantum
  f-DIVERGENCES and Error Correction,'' { Rev. Math. Phys.} {\bf 23} (2011)
  691--747 [\arXiv{arXiv:1008.2529} [math-ph]].
   
\bibitem{Muller-LennertDSFT2013-Renyi}
M.~{M{\"u}ller-Lennert}, F.~{Dupuis}, O.~{Szehr}, S.~{Fehr}, and
  M.~{Tomamichel}, ``On quantum Renyi entropies: a new definition and some
  properties,''  	J. Math. Phys. {\bf 54} 122203 (2013) 
 [\arXiv{arXiv:1306.3142} [quant-ph]].
 

\bibitem{WildeWY2013-strong-converse}
M.~M. {Wilde}, A.~{Winter}, and D.~{Yang}, ``Strong converse for the
  classical capacity of entanglement-breaking and Hadamard channels,''
 Commun. Math. Phys. 331 (2014) 593 
[\arXiv{arXiv:1306.1586} [quant-ph]].

\bibitem{Aharony:1999ti}
  O.~Aharony, S.~S.~Gubser, J.~M.~Maldacena, H.~Ooguri and Y.~Oz,
  ``Large N field theories, string theory and gravity,''
  Phys.\ Rept.\  {\bf 323} (2000) 183
  [\arXiv{hep-th/9905111}].

\bibitem{Kiritsis}
E.~Kiritsis, 
  ``String Theory in a Nutshell,'' 
9787510058110 World Publishing Corporation (2007).

\bibitem{Ammon}
M.~Ammon and J.~Erdmenger, 
``Gauge/Gravity Duality: Foundations and Applications,'' 
Cambridge: Cambridge University Press  (2015). 
 
\bibitem{Engelhardt:2017aux}
  N.~Engelhardt and A.~C.~Wall,
  ``Decoding the Apparent Horizon: A Coarse-Grained Holographic Entropy,''
  \arXiv{arXiv:1706.02038} [hep-th].


\bibitem{Hayward:1993wb}
  S.~A.~Hayward,
  ``General laws of black hole dynamics,''
  Phys.\ Rev.\ D {\bf 49} (1994) 6467.
 
\bibitem{Ashtekar:2002ag}
  A.~Ashtekar and B.~Krishnan,
  ``Dynamical horizons: Energy, angular momentum, fluxes and balance laws,''
  Phys.\ Rev.\ Lett.\  {\bf 89} (2002) 261101
  [\arXiv{gr-qc/0207080}].

\bibitem{Ashtekar:2003hk}
  A.~Ashtekar and B.~Krishnan,
  ``Dynamical horizons and their properties,''
  Phys.\ Rev.\ D {\bf 68} (2003) 104030
  doi:10.1103/PhysRevD.68.104030
  [\arXiv{gr-qc/0308033}].

\bibitem{Gourgoulhon:2005ng}
  E.~Gourgoulhon and J.~L.~Jaramillo,
  ``A 3+1 perspective on null hypersurfaces and isolated horizons,''
  Phys.\ Rept.\  {\bf 423} (2006) 159
  [\arXiv{gr-qc/0503113}].

\bibitem{Bousso:2015mqa}
  R.~Bousso and N.~Engelhardt,
  ``New Area Law in General Relativity,''
  Phys.\ Rev.\ Lett.\  {\bf 115} (2015) no.8,  081301
  [\arXiv{arXiv:1504.07627} [hep-th]].
 
\bibitem{Bousso:2015qqa}
  R.~Bousso and N.~Engelhardt,
  ``Proof of a New Area Law in General Relativity,''
  Phys.\ Rev.\ D {\bf 92} (2015) no.4,  044031
  [\arXiv{arXiv:1504.07660} [gr-qc]].
 
\bibitem{Sanches:2016pga}
  F.~Sanches and S.~J.~Weinberg,
  ``Refinement of the Bousso-Engelhardt Area Law,''
  Phys.\ Rev.\ D {\bf 94} (2016) no.2,  021502
  [\arXiv{arXiv:1604.04919} [hep-th]].
 
\bibitem{jarzynski1997nonequilibrium}
C.~Jarzynski, ``Nonequilibrium equality for free energy differences,'' {
  Phys. Rev. Lett.} {\bf 78} (1997) 2690--2693 
  [\arXiv{cond-mat/9610209}].

\bibitem{crooks1999entropy}
G.~E. {Crooks}, ``Entropy production fluctuation theorem and the
  nonequilibrium work relation for free energy differences,''  Phys. Rev. E  {\bf
  60} (1999) 2721--2726  [\arXiv{cond-mat/9901352}]
 
\bibitem{srednicki1994chaos}
M.~Srednicki, ``Chaos and quantum thermalization,'' { Phys. Rev. E}
  {\bf 50} (1994) 888  	[\arXiv{cond-mat/9403051}].

\bibitem{deutsch1991quantum}
J.~M. Deutsch, ``Quantum statistical mechanics in a closed system,'' {
  Phys.\ Rev.\ A} {\bf 43} (1991), no.~4 2046.

\bibitem{horodecki_are_2002}
M.~Horodecki, J.~Oppenheim, and R.~Horodecki, ``Are the laws of entanglement
  theory thermodynamical?,'' { Phys. Rev. Lett.} {\bf 89} (2002)
  240403 [\arXiv{quant-ph/0207177}].

\bibitem{horodecki2013quantumness}
M.~{Horodecki} and J.~{Oppenheim}, ``(quantumness in the Context Of)
  Resource Theories,'' { Int. J. Mod. Phys. B} {\bf 27} (2013) 45019,
  [\arXiv{arXiv:1209.2162} [quant-ph]]
 
\bibitem{brandao2015reversible}
F.~G. Brand{\~a}o and G.~Gour, ``Reversible framework for quantum resource
  theories,'' { Phys. Rev. Lett.} {\bf 115} (2015), no.~7 070503 [\arXiv{arXiv:1502.03149} [quant-ph]].

\bibitem{Neumann}
J.~von Neumann, ``Mathematische Grundlagen der Quantenmechanic,''
\newblock Springer, Berlin, 1932.

\bibitem{MuellerOppenheim}
A.~Bernamonti, F.~Galli, M.~Mueller, R.~C.~Myers and J.~Oppenheim, et. al, to appear.

\bibitem{JaksicOPP2012-entropy}
V.~Jaksic, Y.~Ogata, Y.~Pautrat, and C.-A. Pillet,``Quantum Theory from
  Small to Large Scales: Lecture Notes of the Les Houches Summer School,''
  vol.~95, ch.~Entropic fluctuations in quantum statistical mechanics. An
  Introduction.
\newblock Oxford University Press, 2010.

\bibitem{frank2013monotonicity}
R.~L. Frank and E.~H. Lieb, ``Monotonicity of a relative r{\'e}nyi entropy,''
J. Math. Phys. {\bf 54} (2013), no.~12 122201 [\arXiv{arXiv:1306.5358} [math-ph]].

\bibitem{beigi2013sandwiched}
S.~Beigi, ``Sandwiched r{\'e}nyi divergence satisfies data processing
  inequality,'' J. Math. Phys. {\bf 54} (2013), no.~12
  122202 	[\arXiv{arXiv:1306.5920} [quant-ph]].

\bibitem{Erven}
 T. van Erven and P. Harremos, 
 ``R\'enyi Divergence and Kullback-Leibler Divergence,'' 
 in IEEE Transactions on Information Theory, vol. {\bf 60}, no. 7, pp. 3797-3820, July 2014,
[\arXiv{arXiv:1206.2459} [cs.IT]].

\bibitem{Lieb} 
Elliott~H.~Lieb, ``Convex trace functions and the Wigner-Yanase-Dyson conjecture,''  In Advances in Mathematics, Volume 11, Issue 3, 1973, Pages 267-288, ISSN 0001-8708.

\bibitem{card1}
  P.~Calabrese and J.~L.~Cardy,
  ``Time-dependence of correlation functions following a quantum quench,''
  Phys.\ Rev.\ Lett.\  {\bf 96} (2006) 136801
  [\arXiv{cond-mat/0601225}].
 
\bibitem{card2}
  P.~Calabrese and J.~Cardy,
  ``Quantum Quenches in Extended Systems,''
  J.\ Stat.\ Mech.\  {\bf 0706} (2007) P06008
  [\arXiv{arXiv:0704.1880} [cond-mat.stat-mech]].

\bibitem{card3}
  S.~Sotiriadis and J.~Cardy,
  ``Quantum quench in interacting field theory: A Self-consistent approximation,''
  Phys.\ Rev.\ B {\bf 81} (2010) 134305
  [\arXiv{arXiv:1002.0167} [quant-ph]].

\bibitem{abrupt1}
  S.~R.~Das, D.~A.~Galante and R.~C.~Myers,
  ``Universal scaling in fast quantum quenches in conformal field theories,''
  Phys.\ Rev.\ Lett.\  {\bf 112} (2014) 171601
  [\arXiv{arXiv:1401.0560} [hep-th]].

\bibitem{abrupt2}
  S.~R.~Das, D.~A.~Galante and R.~C.~Myers,
  ``Smooth and fast versus instantaneous quenches in quantum field theory,''
  JHEP {\bf 1508} (2015) 073
  [\arXiv{arXiv:1505.05224} [hep-th]].

\bibitem{abrupt3}
  A.~Buchel, R.~C.~Myers and A.~van Niekerk,
  ``Universality of Abrupt Holographic Quenches,''
  Phys.\ Rev.\ Lett.\  {\bf 111} (2013) 201602
  [\arXiv{arXiv:1307.4740} [hep-th]].

\bibitem{BF1}
  P.~Breitenlohner and D.~Z.~Freedman,
  ``Stability in Gauged Extended Supergravity,''
  Annals Phys.\  {\bf 144} (1982) 249.
 
\bibitem{BF2}
  L.~Mezincescu and P.~K.~Townsend,
  ``Stability at a Local Maximum in Higher Dimensional Anti-de Sitter Space and Applications to Supergravity,''
  Annals Phys.\  {\bf 160} (1985) 406.

\bibitem{Witten:1998zw}
  E.~Witten,
  ``Anti-de Sitter space, thermal phase transition, and confinement in gauge theories,''
  Adv.\ Theor.\ Math.\ Phys.\  {\bf 2} (1998) 505
  [\arXiv{hep-th/9803131}].

\bibitem{Maldacena:2001kr}
  J.~M.~Maldacena,
  ``Eternal black holes in anti-de Sitter,''
  JHEP {\bf 0304} (2003) 021
  [\arXiv{hep-th/0106112}].

\bibitem{deHaro:2000vlm}
  S.~de Haro, S.~N.~Solodukhin and K.~Skenderis,
  ``Holographic reconstruction of space-time and renormalization in the AdS / CFT correspondence,''
  Commun.\ Math.\ Phys.\  {\bf 217} (2001) 595
  [\arXiv{hep-th/0002230}].
   
\bibitem{Skenderis:2002wp}
  K.~Skenderis,
  ``Lecture notes on holographic renormalization,''
  Class.\ Quant.\ Grav.\  {\bf 19} (2002) 5849
  [\arXiv{hep-th/0209067}].
 
\bibitem{Detournay:2014fva}
  S.~Detournay, D.~Grumiller, F.~Sch\"oller and J.~Sim\'on,
  ``Variational principle and one-point functions in three-dimensional flat space Einstein gravity,''
  Phys.\ Rev.\ D {\bf 89} (2014) no.8,  084061
  [\arXiv{arXiv:1402.3687} [hep-th]].

\bibitem{Compere:2008us}
  G.~Compere and D.~Marolf,
  ``Setting the boundary free in AdS/CFT,''
  Class.\ Quant.\ Grav.\  {\bf 25} (2008) 195014
  [\arXiv{arXiv:0805.1902} [hep-th]].

\bibitem{Andrade:2011nh}
  T.~Andrade and C.~F.~Uhlemann,
  ``Beyond the unitarity bound in AdS/CFT$_{(A)dS}$,''
  JHEP {\bf 1201} (2012) 123
  [\arXiv{arXiv:1111.2553} [hep-th]].
  
\bibitem{Andrade:2011dg}
  T.~Andrade and D.~Marolf,
  ``AdS/CFT beyond the unitarity bound,''
  JHEP {\bf 1201} (2012) 049
  [\arXiv{arXiv:1105.6337} [hep-th]].

\bibitem{Casini:2016rwj}
  H.~Casini, D.~A.~Galante and R.~C.~Myers,
  ``Comments on Jacobson's "Entanglement equilibrium and the Einstein equation",''
  \arXiv{arXiv:1601.00528} [hep-th].

\bibitem{RR-phd}
R.~Renner, ``Security of Quantum Key Distribution,'' PhD thesis, ETH, Zurich, 2005 [\arXiv{quant-ph/0512258}].

\bibitem{Lashkari:2014yva}
  N.~Lashkari,
  ``Relative Entropies in Conformal Field Theory,''
  Phys.\ Rev.\ Lett.\  {\bf 113} (2014) 051602
  [\arXiv{arXiv:1404.3216} [hep-th]].

\bibitem{Lashkari:2015dia}
  N.~Lashkari,
  ``Modular Hamiltonian for Excited States in Conformal Field Theory,''
  Phys.\ Rev.\ Lett.\  {\bf 117} (2016) no.4,  041601
  [\arXiv{arXiv:1508.03506} [hep-th]].

\bibitem{Ruggiero:2016khg}
  P.~Ruggiero and P.~Calabrese,
  ``Relative Entanglement Entropies in 1+1-dimensional conformal field theories,''
  JHEP {\bf 1702} (2017) 039
  [\arXiv{arXiv:1612.00659} [hep-th]].

\bibitem{Nozaki:2014hna}
  M.~Nozaki, T.~Numasawa and T.~Takayanagi,
  ``Quantum Entanglement of Local Operators in Conformal Field Theories,''
  Phys.\ Rev.\ Lett.\  {\bf 112} (2014) 111602
  \arXiv{[arXiv:1401.0539} [hep-th]].

\bibitem{Asplund:2014coa}
  C.~T.~Asplund, A.~Bernamonti, F.~Galli and T.~Hartman,
  ``Holographic Entanglement Entropy from 2d CFT: Heavy States and Local Quenches,''
  JHEP {\bf 1502} (2015) 171
  [\arXiv{arXiv:1410.1392} [hep-th]].

\bibitem{Caputa:2014eta}
  P.~Caputa, J.~Simon, A.~Stikonas and T.~Takayanagi,
  ``Quantum Entanglement of Localized Excited States at Finite Temperature,''
  JHEP {\bf 1501} (2015) 102
  [\arXiv{arXiv:1410.2287} [hep-th]].
 
\bibitem{Marolf:2017kvq}
  D.~Marolf, O.~Parrikar, C.~Rabideau, A.~I.~Rad and M.~Van Raamsdonk,
  ``From Euclidean Sources to Lorentzian Spacetimes in Holographic Conformal Field Theories,''
  \arXiv{arXiv:1709.10101} [hep-th].
 
\bibitem{alicki_entanglement_2013}
R.~Alicki and M.~Fannes,
``Entanglement boost for extractable work from ensembles of quantum batteries,''
 Phys. Rev. E \textbf{87}, 042123 (2013) 
 [\arXiv{arXiv:1211.1209} [quant-ph]]

\bibitem{sparaciari2017resource}
C.~Sparaciari, J.~Oppenheim, and T.~Fritz,
``Resource theory for work and heat'',  Phys. Rev. A, {\bf 96}, 052112 (2017)
[\arXiv{arXiv:1607.01302} [quant-ph]] 
  
\bibitem{RT0}
  S.~Ryu and T.~Takayanagi,
  ``Holographic derivation of entanglement entropy from AdS/CFT,''
  Phys.\ Rev.\ Lett.\  {\bf 96} (2006) 181602
  [\arXiv{hep-th/0603001}].
  
\bibitem{RT1}
  S.~Ryu and T.~Takayanagi,
  ``Aspects of Holographic Entanglement Entropy,''
  JHEP {\bf 0608} (2006) 045
  [\arXiv{hep-th/0605073}].

\bibitem{Hung1}
  L.~Y.~Hung, R.~C.~Myers, M.~Smolkin and A.~Yale,
  ``Holographic Calculations of Renyi Entropy,''
  JHEP {\bf 1112} (2011) 047
  [\arXiv{arXiv:1110.1084} [hep-th]].

 \bibitem{juan}
 A.~Lewkowycz and J.~Maldacena,
  ``Generalized gravitational entropy,''
  JHEP {\bf 1308} (2013) 090
  [\arXiv{arXiv:1304.4926} [hep-th]].
  
\bibitem{Dong:2016hjy}
  X.~Dong, A.~Lewkowycz and M.~Rangamani,
  ``Deriving covariant holographic entanglement,''
  JHEP {\bf 1611} (2016) 028
  [\arXiv{arXiv:1607.07506} [hep-th]].

\bibitem{Dong:2016fnf}
  X.~Dong,
  ``The Gravity Dual of Renyi Entropy,''
  Nature Commun.\  {\bf 7} (2016) 12472
  [\arXiv{arXiv:1601.06788} [hep-th]].
 
\bibitem{rcm}
R.~C.~Myers, unpublished.

\bibitem{Bekenstein:1974ax}
  J.~D.~Bekenstein,
  ``Generalized second law of thermodynamics in black hole physics,''
  Phys.\ Rev.\ D {\bf 9} (1974) 3292.

\bibitem{Wall:2009wm}
  A.~C.~Wall,
  ``Ten Proofs of the Generalized Second Law,''
  JHEP {\bf 0906} (2009) 021
  [\arXiv{arXiv:0901.3865} [gr-qc]].

\bibitem{Wall:2010cj}
  A.~C.~Wall,
  ``A Proof of the generalized second law for rapidly-evolving Rindler horizons,''
  Phys.\ Rev.\ D {\bf 82} (2010) 124019
  [\arXiv{arXiv:1007.1493} [gr-qc]].

\bibitem{Wall:2011hj}
  A.~C.~Wall,
  ``A proof of the generalized second law for rapidly changing fields and arbitrary horizon slices,''
  Phys.\ Rev.\ D {\bf 85} (2012) 104049
   Erratum: [Phys.\ Rev.\ D {\bf 87} (2013) no.6,  069904]
  [\arXiv{arXiv:1105.3445} gr-qc]].

\bibitem{Bianchi:2012br}
  E.~Bianchi,
  ``Horizon entanglement entropy and universality of the graviton coupling,''
  \arXiv{arXiv:1211.0522} [gr-qc].


\bibitem{Balasubramanian:2012tu}
  V.~Balasubramanian, A.~Bernamonti, B.~Craps, V.~Ker\"anen, E.~Keski-Vakkuri, B.~Muller, L.~Thorlacius and J.~Vanhoof,
  ``Thermalization of the spectral function in strongly coupled two dimensional conformal field theories,''
  JHEP {\bf 1304} (2013) 069
  [\arXiv{arXiv:1212.6066} [hep-th]].

\bibitem{AbajoArrastia:2010yt}
  J.~Abajo-Arrastia, J.~Aparicio and E.~Lopez,
  ``Holographic Evolution of Entanglement Entropy,''
  JHEP {\bf 1011} (2010) 149
  [\arXiv{arXiv:1006.4090} [hep-th]].
 
\bibitem{Albash:2010mv}
  T.~Albash and C.~V.~Johnson,
  ``Evolution of Holographic Entanglement Entropy after Thermal and Electromagnetic Quenches,''
  New J.\ Phys.\  {\bf 13} (2011) 045017
  [\arXiv{arXiv:1008.3027} [hep-th]].
 
\bibitem{Balasubramanian:2010ce}
  V.~Balasubramanian {\it et al.},
  ``Thermalization of Strongly Coupled Field Theories,''
  Phys.\ Rev.\ Lett.\  {\bf 106} (2011) 191601
  [\arXiv{arXiv:1012.4753} [hep-th]].

\bibitem{Balasubramanian:2011ur}
  V.~Balasubramanian {\it et al.},
  ``Holographic Thermalization,''
  Phys.\ Rev.\ D {\bf 84} (2011) 026010
  [\arXiv{arXiv:1103.2683} [hep-th]].
  
\bibitem{Aparicio:2011zy}
  J.~Aparicio and E.~Lopez,
  ``Evolution of Two-Point Functions from Holography,''
  JHEP {\bf 1112} (2011) 082
  [\arXiv{arXiv:1109.3571} [hep-th]].

\bibitem{Balasubramanian:2011at}
  V.~Balasubramanian, A.~Bernamonti, N.~Copland, B.~Craps and F.~Galli,
  ``Thermalization of mutual and tripartite information in strongly coupled two dimensional conformal field theories,''
  Phys.\ Rev.\ D {\bf 84} (2011) 105017
  [\arXiv{arXiv:1110.0488} [hep-th]].

\bibitem{Allais:2011ys}
  A.~Allais and E.~Tonni,
  ``Holographic evolution of the mutual information,''
  JHEP {\bf 1201} (2012) 102
  [\arXiv{arXiv:1110.1607}[hep-th]].

\bibitem{Galante:2012pv}
  D.~Galante and M.~Schvellinger,
  ``Thermalization with a chemical potential from AdS spaces,''
  JHEP {\bf 1207} (2012) 096
  [\arXiv{arXiv:1205.1548} [hep-th]].

\bibitem{Caceres:2012em}
  E.~Caceres and A.~Kundu,
  ``Holographic Thermalization with Chemical Potential,''
  JHEP {\bf 1209} (2012) 055
  [\arXiv{arXiv:1205.2354} [hep-th]].

\bibitem{Wu:2012rib}
  B.~Wu,
  ``On holographic thermalization and gravitational collapse of massless scalar fields,''
  JHEP {\bf 1210} (2012) 133
  [\arXiv{arXiv:1208.1393} [hep-th]].
 
\bibitem{Liu:2013iza}
  H.~Liu and S.~J.~Suh,
  ``Entanglement Tsunami: Universal Scaling in Holographic Thermalization,''
  Phys.\ Rev.\ Lett.\  {\bf 112} (2014) 011601
  [\arXiv{arXiv:1305.7244}[hep-th]].
  
\bibitem{Liu:2013qca}
  H.~Liu and S.~J.~Suh,
  ``Entanglement growth during thermalization in holographic systems,''
  Phys.\ Rev.\ D {\bf 89} (2014) no.6,  066012
  [\arXiv{arXiv:1311.1200} [hep-th]].

\bibitem{Alishahiha:2014cwa}
  M.~Alishahiha, A.~Faraji Astaneh and M.~R.~Mohammadi Mozaffar,
  ``Thermalization in backgrounds with hyperscaling violating factor,''
  Phys.\ Rev.\ D {\bf 90} (2014) no.4,  046004
  [\arXiv{arXiv:1401.2807} [hep-th]].

\bibitem{Fonda:2014ula}
  P.~Fonda, L.~Franti, V.~KerŠnen, E.~Keski-Vakkuri, L.~Thorlacius and E.~Tonni,
  ``Holographic thermalization with Lifshitz scaling and hyperscaling violation,''
  JHEP {\bf 1408} (2014) 051
  [\arXiv{arXiv:1401.6088} [hep-th]].
 
\bibitem{Callebaut:2014tva}
  N.~Callebaut, B.~Craps, F.~Galli, D.~C.~Thompson, J.~Vanhoof, J.~Zaanen and H.~b.~Zhang,
  ``Holographic Quenches and Fermionic Spectral Functions,''
  JHEP {\bf 1410} (2014) 172
  [\arXiv{arXiv:1407.5975} [hep-th]].

\bibitem{David:2015xqa}
  J.~R.~David and S.~Khetrapal,
  ``Thermalization of Green functions and quasinormal modes,''
  JHEP {\bf 1507} (2015) 041
  [\arXiv{arXiv:1504.04439} [hep-th]].

\bibitem{Keranen:2015mqc}
  V.~Keranen and P.~Kleinert,
  ``Thermalization of Wightman functions in AdS/CFT and quasinormal modes,''
  Phys.\ Rev.\ D {\bf 94} (2016) no.2,  026010
  [\arXiv{arXiv:1511.08187}[hep-th]]. 
  
\bibitem{Anous:2016kss}
  T.~Anous, T.~Hartman, A.~Rovai and J.~Sonner,
  ``Black Hole Collapse in the 1/c Expansion,''
  JHEP {\bf 1607} (2016) 123
  [\arXiv{arXiv:1603.04856} [hep-th]].



\bibitem{Chesler:2008hg}
  P.~M.~Chesler and L.~G.~Yaffe,
  ``Horizon formation and far-from-equilibrium isotropization in supersymmetric Yang-Mills plasma,''
  Phys.\ Rev.\ Lett.\  {\bf 102} (2009) 211601
  [\arXiv{arXiv:0812.2053} [hep-th]].
  
\bibitem{Heller:2011ju}
  M.~P.~Heller, R.~A.~Janik and P.~Witaszczyk,
  ``The characteristics of thermalization of boost-invariant plasma from holography,''
  Phys.\ Rev.\ Lett.\  {\bf 108} (2012) 201602
  [\arXiv{arXiv:1103.3452} [hep-th]].

\bibitem{Basu:2011ft}
  P.~Basu and S.~R.~Das,
  ``Quantum Quench across a Holographic Critical Point,''
  JHEP {\bf 1201} (2012) 103
  [\arXiv{arXiv:1109.3909} [hep-th]].   

\bibitem{Erdmenger:2012xu}
  J.~Erdmenger and S.~Lin,
  ``Thermalization from gauge/gravity duality: Evolution of singularities in unequal time correlators,''
  JHEP {\bf 1210} (2012) 028
  [\arXiv{arXiv:1205.6873} [hep-th]].

\bibitem{Buch1}
  A.~Buchel, L.~Lehner and R.~C.~Myers,
  ``Thermal quenches in N=2* plasmas,''
  JHEP {\bf 1208} (2012) 049
  [\arXiv{arXiv:1206.6785} [hep-th]].

\bibitem{Buch2}
  A.~Buchel, L.~Lehner, R.~C.~Myers and A.~van Niekerk,
``Quantum quenches of holographic plasmas,''
  JHEP {\bf 1305} (2013) 067
  [\arXiv{arXiv:1302.2924} [hep-th]].



\bibitem{Hubeny:2013hz}
  V.~E.~Hubeny, M.~Rangamani and E.~Tonni,
  ``Thermalization of Causal Holographic Information,''
  JHEP {\bf 1305} (2013) 136
  [\arXiv{arXiv:1302.0853} [hep-th]].
  
\bibitem{Nozaki:2013wia}
  M.~Nozaki, T.~Numasawa and T.~Takayanagi,
  ``Holographic Local Quenches and Entanglement Density,''
  JHEP {\bf 1305} (2013) 080
  [\arXiv{arXiv:1302.5703} [hep-th]].

\bibitem{Hartman:2013qma}
  T.~Hartman and J.~Maldacena,
  ``Time Evolution of Entanglement Entropy from Black Hole Interiors,''
  JHEP {\bf 1305} (2013) 014
  [\arXiv{arXiv:1303.1080} [hep-th]].

\bibitem{Balasubramanian:2013oga}
  V.~Balasubramanian {\it et al.},
  ``Inhomogeneous holographic thermalization,''
  JHEP {\bf 1310} (2013) 082
  [\arXiv{arXiv:1307.7086} [hep-th]].
  
\bibitem{Balasubramanian:2013rva}
  V.~Balasubramanian {\it et al.},
  ``Inhomogeneous Thermalization in Strongly Coupled Field Theories,''
  Phys.\ Rev.\ Lett.\  {\bf 111} (2013) 231602
  [\arXiv{arXiv:1307.1487} [hep-th]].

\bibitem{Asplund:2013zba}
  C.~T.~Asplund and A.~Bernamonti,
  ``Mutual information after a local quench in conformal field theory,''
  Phys.\ Rev.\ D {\bf 89} (2014) no.6,  066015
  [\arXiv{arXiv:1311.4173} [hep-th]].

\bibitem{Craps:2013iaa}
  B.~Craps, E.~Kiritsis, C.~Rosen, A.~Taliotis, J.~Vanhoof and H.~b.~Zhang,
  ``Gravitational collapse and thermalization in the hard wall model,''
  JHEP {\bf 1402} (2014) 120
  [\arXiv{arXiv:1311.7560} [hep-th]].
  
\bibitem{Hubeny:2013dea}
  V.~E.~Hubeny and H.~Maxfield,
  ``Holographic probes of collapsing black holes,''
  JHEP {\bf 1403} (2014) 097
  [\arXiv{arXiv:1312.6887} [hep-th]].
  
\bibitem{Buchel:2014gta}
  A.~Buchel, R.~C.~Myers and A.~van Niekerk,
  ``Nonlocal probes of thermalization in holographic quenches with spectral methods,''
  JHEP {\bf 1502} (2015) 017
   Erratum: [JHEP {\bf 1507} (2015) 137]
  [\arXiv{arXiv:1410.6201} [hep-th]].

\bibitem{Ishii:2015gia}
  T.~Ishii, E.~Kiritsis and C.~Rosen,
  ``Thermalization in a Holographic Confining Gauge Theory,''
  JHEP {\bf 1508} (2015) 008
  [\arXiv{arXiv:1503.07766} [hep-th]].
  
\bibitem{Asplund:2015eha}
  C.~T.~Asplund, A.~Bernamonti, F.~Galli and T.~Hartman,
  ``Entanglement Scrambling in 2d Conformal Field Theory,''
  JHEP {\bf 1509} (2015) 110
  [\arXiv{arXiv:1506.03772} [hep-th]].

\bibitem{Ziogas:2015aja}
  V.~Ziogas,
  ``Holographic mutual information in global Vaidya-BTZ spacetime,''
  JHEP {\bf 1509} (2015) 114
  [\arXiv{arXiv:1507.00306} [hep-th]].

\bibitem{Kundu:2016cgh}
  S.~Kundu and J.~F.~Pedraza,
  ``Spread of entanglement for small subsystems in holographic CFTs,''
  Phys.\ Rev.\ D {\bf 95} (2017) no.8,  086008
  [\arXiv{arXiv:1602.05934} [hep-th]].
  
\bibitem{Mezei:2016zxg}
  M.~Mezei,
  ``On entanglement spreading from holography,''
  JHEP {\bf 1705} (2017) 064
  [\arXiv{arXiv:1612.00082} [hep-th]].

\bibitem{Jahn:2017xsg}
  A.~Jahn and T.~Takayanagi,
  ``Holographic entanglement entropy of local quenches in AdS4/CFT3: a finite-element approach,''
  J.\ Phys.\ A {\bf 51} (2018) no.1,  015401
  [\arXiv{arXiv:1705.04705} [hep-th]].  

\bibitem{Myers:2017sxr}
  R.~C.~Myers, M.~Rozali and B.~Way,
  ``Holographic Quenches in a Confined Phase,''
  J.\ Phys.\ A {\bf 50} (2017) no.49,  494002
  [\arXiv{arXiv:1706.02438} [hep-th]].  

\bibitem{Arefeva:2017pho}
  I.~Y.~Aref'eva, M.~A.~Khramtsov and M.~D.~Tikhanovskaya,
  ``Thermalization after holographic bilocal quench,''
  JHEP {\bf 1709} (2017) 115
  [\arXiv{arXiv:1706.07390} [hep-th]].  
  
\bibitem{Bagrov:2017tqn}
  A.~Bagrov, B.~Craps, F.~Galli, V.~KerŠnen, E.~Keski-Vakkuri and J.~Zaanen,
  ``Holography and thermalization in optical pump-probe spectroscopy,''
  \arXiv{arXiv:1708.08279} [hep-th].

\bibitem{VaidyaAdS}
  A.~Wang and Y.~Wu,
  ``Generalized Vaidya solutions,''
  Gen.\ Rel.\ Grav.\  {\bf 31} (1999) 107
    [\href{https://arxiv.org/abs/gr-qc/9803038}{gr-qc/9803038}].

\bibitem{Faulkner:2013ica}
  T.~Faulkner, M.~Guica, T.~Hartman, R.~C.~Myers and M.~Van Raamsdonk,
  ``Gravitation from Entanglement in Holographic CFTs,''
  JHEP {\bf 1403} (2014) 051
  [\arXiv{arXiv:1312.7856} [hep-th]].
  
\bibitem{Lashkari:2016idm}
  N.~Lashkari, J.~Lin, H.~Ooguri, B.~Stoica and M.~Van Raamsdonk,
  ``Gravitational positive energy theorems from information inequalities,''
  PTEP {\bf 2016} (2016) no.12,  12C109
  [\arXiv{arXiv:1605.01075} [hep-th]].

\bibitem{Vaid0}
 P.~C.~Vaidya, ``The External Field of a Radiating Star in General Relativity," Curr. Sci. {\bf 12}
(1943) 183.

\bibitem{OriginalVaidya}
  P.~C.~Vaidya,
  ``The gravitational field of a radiating star,''
 Indian Acad. Sci. (Math. Sci.) (1951) {\bf 33}: 264.
 
\bibitem{Israel:1966rt}
  W.~Israel,
  ``Singular hypersurfaces and thin shells in general relativity,''
  Nuovo Cim.\ B {\bf 44S10} (1966) 1
   [Nuovo Cim.\ B {\bf 44} (1966) 1]
   Erratum: [Nuovo Cim.\ B {\bf 48} (1967) 463].
 
\bibitem{Brown:1992kc}
  J.~D.~Brown,
  ``Action functionals for relativistic perfect fluids,''
  Class.\ Quant.\ Grav.\  {\bf 10} (1993) 1579
  [\arXiv{gr-qc/9304026}].

\bibitem{Hayward:1993my}
  G.~Hayward,
  ``Gravitational action for space-times with nonsmooth boundaries,''
  Phys.\ Rev.\ D {\bf 47} (1993) 3275.
  
\bibitem{Brill:1994mb}
  D.~Brill and G.~Hayward,
   ``Is the gravitational action additive?,''
  Phys.\ Rev.\ D {\bf 50} (1994) 4914
  [\arXiv{gr-qc/9403018}].
    
\bibitem{newer}
S.~Chapman, H.~Marrochio and R.~C.~Myers,
  ``Holographic Complexity in Vaidya Spacetimes,''
  in preparation
  [arXiv:1803.nnnnn [hep-th]].
 
\bibitem{multi1}
  S.~H.~Shenker and D.~Stanford,
  ``Multiple Shocks,''
  JHEP {\bf 1412} (2014) 046
  [\arXiv{arXiv:1312.3296} [hep-th]].

\bibitem{Wolf} 
M.~M.~Wolf, ``Quantum channels \& operations: Guided tour, 2012.''  

%
%
%

\end{thebibliography}
\end{document}